\documentclass[preprint,amsmath,amssymb,aps]{revtex4-2}
\usepackage{graphicx}
\usepackage{bm}
\usepackage{txfonts}
\usepackage{hyperref}
\usepackage{caption}
\date{\today}
\newcommand{\be}{\begin{eqnarray}}
\newcommand{\ee}{\end{eqnarray}}

\newcommand{\bfk}{{\bf k}_{\perp}}

\usepackage{xcolor}
\begin{document}
\title{Leading twist T-even TMDs for the spin-1 heavy vector mesons}

\author{Satyajit Puhan}
\email{puhansatyajit@gmail.com}
\affiliation{Department of Physics, Dr. B.R. Ambedkar National
	Institute of Technology, Jalandhar, 144008, India}

\author{Harleen Dahiya}
\email{dahiyah@nitj.ac.in}
\affiliation{Department of Physics, Dr. B.R. Ambedkar National
	Institute of Technology, Jalandhar, 144008, India}

\date{\today}%
\begin{abstract}
We have presented the leading twist quark transverse momentum-dependent parton distribution functions (TMDs) for the spin-1 heavy vector mesons $J/\psi$-meson and $\Upsilon$-meson using the overlap of the light-front wave functions. We have computed their TMDs in the light-front holographic model (LFHM) as well as  the light-front quark model (LFQM) and further compared the results with the Bethe-Salpeter  (BSE) model. We have discussed the behavior of  the TMDs with respect to momentum fraction carried by active quark ($x$) and the transverse quark momenta ($k_\perp$) in both the models. We have also calculated the $k_\perp$ moments of the quark in both the models and have compared the results with the BSE model. The  predictions of  LFQM are found to be in accord with the BSE model. Further, we have analyzed the leading twist parton distribution functions (PDFs) for both the heavy mesons in both the models and the results are found to be in accord with the basic light-front quantization (BLFQ) and BSE model.

 \vspace{0.1cm}
\end{abstract}
%
\maketitle
%
%
\section{Introduction\label{secintro}}
Hadrons, which are strongly interacting and relativistic bound states of quarks and gluons, have a complex structure. An important goal of both theoretical and experimental physicists is to understand the hadron structure in terms of the theory of strong interactions: quantum chromodynamics (QCD) \cite{qcd,qcd1,qcd2} and the hadronic matrix elements of quark-gluon field operators. Even though the parton distribution functions (PDFs) \cite{pdf,pdf1,pdf2,pdf3}, extracted from the deep  inelastic scattering (DIS) \cite{dis} experiments, provide information about the longitudinal momentum fraction $(x)$ distribution of the quarks, they do not specify the spatial location, transverse momentum distribution and spin densities of the quarks inside the hadrons. Further, the three-dimensional transverse momentum-dependent parton distributions (TMDs) \cite{tmd,tmd1,tmd2,tmd3} and the generalized parton distributions (GPDs) \cite{gpd,gpd2,gpd3} play an important role in describing a complete internal structure of the hadron. Both TMDs and GPDs are widely investigated experimentally as well as theoretically in these recent years. GPDs are a function of longitudinal momentum fraction of parton ($x$) as well as momentum transferred fraction ($\Delta$) between initial and final state of a hadron. It contains information about spatial distribution of the partons inside a hadron which cannot be accessed through TMDs. GPDs can be accessed through the deeply virtual Compton scattering (DVCS) \cite{dvcs} and deeply virtual meson production (DVMP) \cite{dvmp,dvmp1} processes. On the other hand, TMDs contain information about both the longitudinal momentum fraction ($x$) and transverse momentum distribution ($k_\perp$) of the partons inside a hadron. TMDs also provide nonperturbative information about parton structure of hadrons, particularly angular momentum, spin-orbit correlation and polarization degrees of freedom for the partons inside a hadron \cite{non}. The extended form of co-linear PDFs gives rise to the three-dimensional TMDs. Experimentally, TMDs can be extracted from  deep inelastic scattering (DIS) processes at high energy \cite{dis}, Drell-Yan processes \cite{drell,drell1,drell2,drell3,drell4}, semi-inclusive deep inelastic scattering (SIDIS) processes \cite{sidis,sidis1,sidis2}, and $Z^0/W^{\pm}$ production \cite{z1,z2,z3}.

A wide range of theoretical studies are going on to study the TMDs of hadrons. For the case of nucleons which are spin-$\frac{1}{2}$ particles, TMDs have been studied at the leading twist in different QCD-inspired models like light-front constituent quark model (LFCQM) \cite{lfcm}, quark-diquark model \cite{lfcm1}, chiral quark soliton model \cite{lfcm2}, bag model \cite{lfcm3}  and  lattice QCD \cite{lattice}. A few higher twist TMDs calculations for the nucleons have also been done \cite{shu,shu1}. There are in total eight TMDs for the case of spin-$\frac{1}{2}$ particles out of which six are T-even at leading twist \cite{spin0.5}. TMDs have also been calculated for the case of spin-$0$ light mesons  at the leading twist \cite{spt,spt1,spt2} and also for spin-$1$ light mesons at the leading twist \cite{bse,rho}. There are only two TMDs at leading twist for the case of spin-$0$ mesons \cite{spin0} whereas there are eighteen TMDs for the case of spin-$1$ vector mesons
at lower twist  out of which nine are T-even \cite{spin1}. Three tensor polarized TMDs $f_{1LL}(x,{\textbf{k}^2_\perp})$, $f_{1LT}(x, {\textbf{k}^2_\perp})$ and $f_{1TT}(x,{\textbf{k}^2_\perp})$ come into the picture in the case of spin-$1$ particles. These TMDs occur due to the tensor polarizations of the target and are absent in case of spin-$0$ pseudoscalar mesons and spin-$\frac{1}{2}$ nucleons. For the case of spin-$1$ hadron, there is an extra tensor PDF $f_{1LL}(x)$ when compared with the spin-$0,\frac{1}{2}$ targets. 

At present, there is no experimental data available for extracting the spin-$1$ meson TMDs but a very few
theoretical predictions are available. Spin-$1$ TMDs and TMD fragmentation functions have been predicted at the leading twist as well as  at higher twist with positivity bounds on them \cite{spin1}. Detailed investigations on the $\rho$ meson T-even TMDs along with PDFs have been carried out in the Nambu-Jona-Lasinio (NJL) model \cite{NJL}, light-front quark model (LFQM) \cite{rho}, light-front holographic model (LFHM) \cite{rho} and BSE model \cite{bse}. Photon TMDs have been studied in the basic light-front quantization (BLFQ) method \cite{blfq}. All these models successfully encode the TMDs, PDFs and TMD fragmentation functions for the light vector mesons. There are however very few investigations on the heavy vector mesons like $J/\psi$ and $\Upsilon$. In the BLFQ method, the study of the heavy vector mesons PDFs have been done for the case of quark being unpolarized \cite{blfqpdf}. A detailed study of all the PDFs for $J/\psi$ has been performed using the light-front wave functions (LFWFs) method \cite{lfwf,Li_2017}. An overview of the TMDs and PDFs for spin-$1$ mesons  has been presented in BSE model \cite{bse}, however, there is no mention of $f_{1LL}(x)$ PDF, $f_{1LL}(x, {\textbf{k}^2_\perp})$ TMDs for both $J/\psi$ and $\Upsilon$ particles in this work. The $J/\psi$ and $\Upsilon$ mesons get their masses from the current quark mass through the Higgs mechanism. The parton's motion inside $J/\psi$ is slower as compared to the other vector mesons making the motion non-relativistic. The motion of parton in $\Upsilon$ is even slower than $J/\psi$ meson hence providing a benchmark for TMDs in the non-relativistic limit. Following the success of the LFHM as well as  LFQM in predicting the results for  $\rho$-mesons and in light of the above developments, a detailed study of  $J/\psi$ and $\Upsilon$ mesons would be interesting in order to understand the three-dimensional structure of  the heavy vector mesons.

In this work, we have mainly targeted the quark distributions of $J/\psi$ and $\Upsilon$ heavy vector mesons using the light-front (LF) dynamics based LFHM and LFQM. The LFQM is a non-perturbative approach framework describing the structure and dynamics of hadrons as well as the behavior of quarks within them \cite{lfqm,lfqm1,lfqm2,lfqm3,lfqm4,lfqm5,lfqm6,lfqm7}. LFQM is relativistic and gauge invariant in nature. The advantage of LFQM is that it focuses on the valence quarks of the hadrons and the valence quarks are the primary constituents responsible for the overall structure and properties of hadrons. At times, LFQM can describe the hadrons in the strong interaction regime when the perturbative QCD calculations cannot. However, higher Fock-state contributions and effect of confinement, which play an important role to describe the complete dynamics of hadrons, are not included in LFQM. LFQM describes the structure only in the lower Fock-states. Further, the LFHM goes beyond the simple picture of constituent quarks and gluons by considering the full holographic dual description. LFHM relates the higher dimensional gravitational theory to lower dimensional quantum field theory for strongly interacting systems \cite{lfhm,lfhm1,lfhm2}.

In the present work, we have proceeded through both the LFQM and LFHM to calculate the T-even TMDs for $J/\psi$ and $\Upsilon$ mesons. We have obtained the explicit forms of the quark TMDs in the overlapping form of LFWFs in both the models. We have taken the lower Fock-state for these particles i.e, $|M \rangle = \sum | q \bar{q} \rangle \psi_{q \bar{q}}$ as there is very less contribution from higher Fock-states for the heavy mesons.  In the LFQM, the tensor TMDs $f_{1LL}(x, {\textbf{k}^2_\perp})$, $f_{1LT}(x, {\textbf{k}^2_\perp})$ and $f_{1TT}(x, {\textbf{k}^2_\perp})$  come out to be zero. Whereas,  $h^{\perp}_{1T}(x, {\textbf{k}}^2_\perp)$ TMD is zero in the LFHM. We have discussed the TMDs in both models with the help of three-dimensional and two-dimensional plots with respect to $x$ and $\textbf{k}^2_{\perp}$. We have calculated average quark transverse momenta $\textbf{k}^2_{\perp}$ for both the models and compared the results with the BSE model \cite{bse} as no experimental data is available.  The PDFs have also been computed for both the models and have been compared with the results from the BLFQ \cite{blfqpdf}, designed light-front wave functions (D-LFWF) \cite{lfwf} and BSE Mode \cite{bse}.

This paper is arranged as follows. In Section \ref{tmd_defi}, we have defined the leading twist TMDs and the co-relation functions for the spin-1 hadrons at lower Fock-state. In Sections \ref{LFHme} and \ref{LFQMe}, the essential details of the  LFHM and LFQM have respectively been presented. Further in Section \ref{tmd_overlap}, the basic formalism of LFWFs has been given.  We represent  different TMDs in the form of light-front amplitude with their explicit overlap form in both the models. In Section \ref{results}, we have given the details of the numerical results with the help of three-dimensional and two-dimensional plots. We have also compared our results with other model results in this section. In Section \ref{pdF}, we have discussed the collinear PDFs. Finally,  we summarize our results in Section \ref{conclude}.

 \section{Methodology}
\subsection{Transverse momentum-dependent parton distributions}\label{tmd_defi}
The quark TMDs for spin-1 particles are defined through transverse momentum-dependent quark  correlation function expressed as \cite{drell2,NJL,tmd11,tmd12,tmd13,tmd14,tmd15}
\begin{eqnarray}
\Theta_{i j}^{(\Lambda)_{\bf \mathcal{S}}}(x, {\bf k}^2_\perp)&=&
\int \frac{{\rm d}z^- \, {\rm d}^2 {\bf z}_\perp}{(2\pi)^3} \, 
e^{\iota (k^+z^- -k_{\perp} \cdot z_{\perp})} \ {}_{\bf {\mathcal{S}}}\langle P,\Lambda | \bar \psi_{j}(0) W(0,z)
\psi_{i}(z^-,{\bf z}_\perp) | P, \Lambda \rangle_{\bf {\mathcal{S}}} \nonumber \\[0.2em]
&\equiv& \epsilon^*_{\Lambda (\mu)}(P) \ \Theta_{ij}^{\mu\nu}(x,{\bf k}_\perp) \ \epsilon_{\Lambda (\nu)}(P),
\label{phi1}
\end{eqnarray}
where $\psi_{i(j)}$ is  the flavor SU(2) quark field operator, \textit{i} and \textit{j} are the Dirac indices and $W(0,z)$ is the gauge link \cite{tmd14,tmd15}. For simplicity, we have taken the gauge link as unity to study the T-even TMDs. $P$ is the four-vector momentum of heavy vector mesons and is expressed as
\begin{eqnarray*}
  P=\left(P^+,P^-,{\bf P}_\perp \right)=\left(P^+, \frac{M_{\alpha}}{P^+},{\bf 0}_\perp\right),  
\end{eqnarray*}
  where $M_{\alpha}$ is the mass for heavy vector meson. $M_\alpha= M_{c \bar c}$ and $M_\alpha= M_{b \bar b}$ for $J/\psi$ meson and $\Upsilon$ meson respectively. $k_{\perp}$ and $k^+$ are transverse and longitudinal momentum carried by the active quark, $z$ is the position  four vector  and is expressed as, $z=(z^+,z^-,z^{\perp})$ in LF dynamics. In Eq. (\ref{phi1}), $\Theta_{ij}^{\mu\nu}$ is the polarization-independent Lorentz tensor matrix, $\epsilon_{\mu(\nu)}$ is the polarization four vector and the $| P, \Lambda \rangle_{\bf {\mathcal{S}}}$ state indicates that the spin projection of the target hadron in  $S=(S_{L},S_{T})$ direction with helicities $\Lambda=\pm 1,0$. At the leading twist, there are total nine T-even TMDs for  $J/\psi$ and $\Upsilon$ which are expressed with unpolarized ($U$), transversely polarized ($T$) and longitudinally polarized ($L$) target as
\begin{eqnarray}
\epsilon^{*}_{\Lambda(\mu)} (P)~\langle \gamma^+ \rangle^{\mu \nu}_{\mathcal{S}} (x,{\bf k}^2_\perp)~\epsilon_{\Lambda(\nu)}(P) &=&
f_1(x,\vec{k}^2_{\perp}) + S_{LL}f_{1LL}(x,\vec{k}^2_{\perp})\nonumber\\
&&+ \frac{\vec{S}_{LT}\cdot \vec{k}_{\perp}}{M_\alpha}\, f_{1LT}(x, \vec{k}_{\perp}^2)
+ \frac{ \vec{S}_{TT} \cdot \vec{k}_{\perp}}{M_\alpha^2}  f_{1TT}(x, \vec{k}_{\perp}^2),
\label{f1} \\
\epsilon^{*}_{\Lambda(\mu)} (P)~\langle \gamma^+ \gamma_5 \rangle^{\mu \nu}_{\mathcal{S}} (x,{\bf k}^2_\perp)~\epsilon_{\Lambda(\nu)}(P)&=&
\mathcal{S}_L\,g_{1L}(x,{\bf k}^2_\perp)+ \frac{{\bf k}_\perp \cdot {\mathcal{S}}_\perp}{M_\alpha} g_{1T}(x,{\bf k}^2_\perp),
\label{f2} \\
\epsilon^{*}_{\Lambda(\mu)} (P)~\langle \gamma^+ \gamma^i \gamma_5 \rangle^{\mu \nu}_{\mathcal{S}} (x,{\bf k}^2_\perp)~\epsilon_{\Lambda(\nu)}(P)
&=& \mathcal{S}_\perp^ih_1(x,{\bf k}^2_\perp)+\mathcal{S}_L\frac{k_\perp^i}{M_\alpha}h_{1L}^\perp(x,{\bf k}^2_\perp) \nonumber\\
&&+\frac{1}{2\,M_\alpha^2}\left(2\,k_\perp^i {\bf k}_\perp \cdot {\mathcal{S}}_\perp - \mathcal{S}_\perp^i~{\bf k}_\perp^2\right) h_{1T}^{\perp}(x,{\bf k}_\perp^2),
\label{f3}
\end{eqnarray}
with 
\begin{eqnarray*}
\mathcal{S}_{LL}&=&\left(3 \Lambda^2-2 \right)\left(\frac{1}{6}-\frac{1}{2} \mathcal{S}_L^2\right),\\
\mathcal{S}^i_{LT}&=&\left(3 \Lambda^2-2 \right)\mathcal{S}_L \mathcal{S}^i_\perp,\\
\mathcal{S}_{TT}^{ij}&=&\left(3 \Lambda^2-2 \right)(\mathcal{S}_\perp^i \mathcal{S}_\perp^j-\frac{1}{2}\mathcal{S}_\perp^2~ \delta^{ij})\,.
\end{eqnarray*}
The Lorenz tensor is defined as
\begin{eqnarray*}
\epsilon^{*}_{\Lambda(\mu)} (P)~\langle \Gamma\rangle^{\mu \nu}(x,{\bf k}^2_\perp)~\epsilon_{\Lambda(\nu)}(P)=\frac{1}{2} {\rm Tr}_{D}\left(\Gamma \Theta^{(\Lambda)_{\bf \mathcal{S}}}(x, {\bf k}^2_\perp)\right).
\label{ap}
\end{eqnarray*}
In the above equations $\mathcal{S}^{i(j)}_\perp$
symbolize the transverse polarization of the target meson in the directions $i(j)$ or $x(y)$.
The Dirac matrices structure  $\Gamma$ for leading twist are $\gamma^+$, $\gamma^+\gamma_5$ and $ \gamma^+\gamma^i\gamma_5$ with $\textit{i}=(1,2)$. The function $\textit{f}$, $\textit{g}$ and $\textit{h}$ denote the unpolarized, longitudally polarized and transversly polarized quark within the hadron. The subscript $1$ in the functions $\textit{f}$, $\textit{g}$, $\textit{h}$ (Eqs. (\ref{f1})-(\ref{f3})) denotes the leading  twist TMDs \cite{spin1}. The longitudinal and transverse hadron polarizations have been denoted as $L$ and $T$ respectively. $x$ in the above equations is the momentum fraction carried by the active quark from the target hadron i.e, $x= \frac{k^+}{P^+}$. When compared with the spin-$\frac{1}{2}$ case of nucleons, the  spin-1 hadrons have three extra  tensor polarized TMD functions $f_{1LT}$, ${S}_{LT}$ and $f_{1TT}$.

\subsection{Light-front holographic model}\label{LFHme}

 LFHM is derived from the AdS/QCD correspondence which connects  the five-dimensional Anti-de Sitter (AdS) space-time to the lower dimension QCD in the LF dynamics \cite{rho, lfhm, ads}. In the  LFHM, hadrons are described by string-like objects in the higher dimensional AdS space. The behavior of these strings corresponds to the confinement of quarks and gluons within the hadrons. The complete holographic wave function $\psi^{(1)} (x,\zeta,\varphi)$ is expressed as \cite{ads2,ads,ads1}
\begin{eqnarray}
    \psi^{(1)} (x,\zeta,\varphi)= \frac{\Phi(\zeta)}{\sqrt{2\pi \zeta}}\, X(x) \, e^{iL\varphi} ,
\label{h12}
\end{eqnarray}
where $ X(x)$ is the longitudinal mode of LFWF and can be written as $X(x) = \sqrt{x(1-x)}$ \cite{ads},
  $\varphi$ is the transverse angular dependence in light-front plane, $\zeta=X(x) b_\perp$ is a light-front variable and $b_\perp$ is the transverse quark pair separation variable in the hadrons. The dynamic part of the holographic wave function, derived from the holographic Schr\"odinger equation \cite{rho}, is expressed as
  \begin{eqnarray}
   \Phi_{nL}(\zeta) = \kappa^{1+L} \sqrt{\frac{2 n!}{(n+L)!}} \zeta^{1/2 + L} exp(\frac{-\kappa^2 \zeta^2}{2})L_n^L (\kappa^2 \zeta^2),
   \label{k2}
  \end{eqnarray}
  with meson mass spectrum
  \begin{eqnarray}
      M^2=(4n+2L+2)\kappa^2+2\kappa^2(J-1),
  \end{eqnarray}
  where $\kappa$ is the mass scale parameter and it's value is $0.894$ GeV and $1.49$ GeV for $J/\psi (c\bar c)$ \cite{Swarnkar:2015osa} and $\Upsilon (b\bar b)$ \cite{Nielsen:2018ytt} respectively. \textit{ n}, \textit{J} and \textit{ L} are the radial quantum number, total angular momentum and internal angular momentum respectively. We can get the ground state holographic wave function from Eq. (\ref{h12}) in the transverse impact-parameter space as
  \begin{eqnarray}
    \psi^{(1)}(x,\zeta^2)=\frac{\kappa}{\sqrt{\pi}}\sqrt{x(1-x)}\,{\rm exp}\left(\frac{-\kappa^2 \zeta^2}{2}\right).
    \label{k3}
  \end{eqnarray}
  By Fourier transforming Eq. (\ref{k3}), we obtain the wave function for the mesons with massless quarks in transverse momentum space
  \begin{eqnarray}
      \psi^{(1)}(x,{\bf k}^2_\perp) \propto \frac{1}{\sqrt{x(1-x)}}\,{\rm exp}\left(-\frac{{M}^2}{2 \kappa^2 }\right)\,.
      \label{k4}
\end{eqnarray}
Here $M$ is the meson invariant mass which is given as
\begin{eqnarray*}
   {M}^2=\frac{{\bf k}^2_\perp}{x(1-x)},
\end{eqnarray*}
where $k_\perp$ is the transverse quark momentum obtained by taking Fourier transform of $b_\perp$. Since we are dealing with the massive quarks, we can include the quark mass in the invariant mass form \cite{mass,mass1}. We have
\begin{eqnarray}
   {M}_{q\bar{q}^\prime}^2= \frac{{\bf k}^2_\perp}{x(1-x)} + \frac{m_q^2}{x} + \frac{m^2_{\bar{q}^\prime}}{1-x}~,
   \label{k5} 
\end{eqnarray}
where $m_{q}$ and $m_{\bar q}$ are the quark and anti-quark mass respectively. Now the holographic light-front wave function for meson bound state in Eq. (\ref{k4}) with quark masses is modified as
\begin{eqnarray}
    \psi^{(1)}(x,{\bf k}^2_\perp)\propto \frac{1}{\sqrt{x(1-x)}}\,\exp\left(-\frac{ {\bf k}^2_\perp}{2 \kappa^2 x(1-x)}\right)\,  \exp\left(-\frac{\mu_{12}^2}{2 \kappa^2}\right),
\label{k6}
\end{eqnarray}
where $\mu_{12}^2=\frac{m_q^2}{x}+\frac{m_{\bar{q}^\prime}^2}{1-x}$. The wave function in Eq. (\ref{k6}) has an additional term corresponding to quark masses when compared to Eq. (\ref{k4}). It would be important to mention here that this is not a solution of the Schr\"odinger equation for the holographic model, however, the results outcome of the model are more precise by taking the mass term. The holographic wave function in Eq. (\ref{k6}) has been used in the study of $\rho$ meson \cite{rho}, $\rho$ meson electroproduction at HERA \cite{ads2} in the decays $B\to \rho \gamma$ \cite{rho2}, $B\to K^*$ \cite{rho3}, $B\to \rho$ \cite{rho4}, $B\to K^*\mu^-\mu^+$ \cite{rho5} etc.

For heavy vector mesons $J/\psi$ $(\Upsilon)$, $m_q=(m_c,m_b)$  and  $m_{\bar q} =(m_{\bar c} ,m_{\bar b})$ denote the masses of quark and anti-quarks respectively. As $m_{q}=m_{\bar q}$, so we can rewrite Eq. (\ref{k6}) as

\begin{eqnarray}
\psi^{(1)}(x,{\bf k}^2_\perp)\propto \frac{1}{\sqrt{x(1-x)}}\,{\rm exp}\left(-\frac{ {\bf k}^2_\perp+m^2_q}{2 \kappa^2 x(1-x)}\right)\,.
\label{holographic-wf}
\end{eqnarray}  
In our work, we have neglected the small contribution coming from the massive quark mass shift following Ref. \cite{rho}. The complete wave function is obtained by adding the quark and anti-quark helicities with transverse and longitudinal spin projection in the hadrons which are expressed as 
\begin{eqnarray}
   \chi^{{(1)}L}_{h_q, h_{\bar{q}}}=\frac{1}{\sqrt{2}}\delta_{h_q, -h_{\bar{q}}} \ \ \ ; \ \ \ \chi^{{(1)}{T(\pm)}}_{h_q, h_{\bar{q}}}=\frac{1}{\sqrt{2}}\delta_{h_q \pm , h_{\bar{q}}\pm}. 
   \label{k8}
\end{eqnarray}
In Eq. (\ref{k8}), $h_q$ and $h_{\bar q}$ are the helicities of quark and anti-quark respectively. Eq. (\ref{k8}) can be written in the form of Lorentz invariant spin structure for spin-$1$ meson by using the photon quark-anti quark vertex factor
\begin{eqnarray}
    \chi^{{(1)}{L(T)}}_{h_q, h_{\bar{q}}}(x,{\bf k}^2_\perp)= \frac{\bar{u}_{h_q}(k^+,{\bf k}_\perp)}{\sqrt{x}} \, \epsilon_\Lambda \cdot \gamma \, \frac{v_{h_{\bar{q}}}(k^{\prime +},{\bf k}^{\prime}_\perp)}{\sqrt{1-x}}\,,
\label{k9}
\end{eqnarray}
where $k$ and $k^\prime$ are the quark and anti-quark four vector momentum. $x$ and $1-x$ are  the longitudinal momentum fraction carried by the quark and anti-quark from the hadron i.e, $x=\frac{k^+}{P^+}$ and $(1-x)=\frac{k^{\prime +}}{P^+}$ and the kinematics for polarization vector $\epsilon_\Lambda$ is given as follows 
\begin{eqnarray}
    \epsilon_L=\left(\frac{P^+}{M_\alpha},-\frac{M_\alpha}{P^+},0,0\right)\ \ \ ; \ \ \ \epsilon^{\pm}_T=\mp\frac{1}{\sqrt{2}}\left(0,0,1,\pm \iota\right).
\end{eqnarray}
The holographic wave function with  dynamical spin effects in the longitudinal and transverse spin projection for $J/\psi$ and $\Upsilon$-mesons  can now be written as \cite{sat,ads2}
\begin{equation}
\Psi^{{(1)}L}_{h_q, h_{\bar{q}}}(x,{\bf k}^2_\perp)= \chi^{{(1)}L}_{h_q, h_{\bar{q}}}(x,{\bf k}^2_\perp)\psi^{(1)}(x,{\bf k}^2_\perp),
\label{k10}
\end{equation} 

\begin{equation}
\Psi^{{(1)}T}_{h_q, h_{\bar{q}}}(x,{\bf k}^2_\perp)= \chi^{{(1)}T}_{h_q, h_{\bar{q}}}(x,{\bf k}^2_\perp)\psi^{(1)}(x,{\bf k}^2_\perp).
\label{k11}
\end{equation} 
The explicit form of Eqs. (\ref{k10}) and (\ref{k11}) at the model scale $\mu_{\rm LFHM}^2=0.20$ GeV$^2$ are respectively expressed as

\begin{equation}
	\Psi^{(1)L}_{h_q, h_{\bar{q}}}(x,{\bf k}^2_\perp)= \mathcal{N}_{L} \delta_{h_q, h_{\bar{q}}} \left[M_\mathcal{\alpha}^2  +  \left(\frac{m^2_q + k_\perp^2}{x (1-x)} \right)\right] \psi^{(1)} (x, \textbf{k}_\perp^2),
\label{k12}
\end{equation}
\begin{equation}
\Psi^{(1)(T \pm)}_{h_q, h_{\bar{q}}}(x,\mathbf{k}^2_{\perp}) = \mathcal{N}_{T} \left[]\pm k_\perp e^{\pm i\theta_{k_\perp}} \left(\frac{\delta_{h_q,\pm} \delta_{h_{\bar q},\mp}}{1-x} - \frac{\delta_{h_q,\mp} \delta_{h_{\bar q},\pm}}{x}\right) + \left(\frac{m_q}{x(1-x)}\right) \delta_{h_q,\pm} \delta_{h_{\bar q},\pm}\right] \psi^{(1)}(x,\mathbf{k}_\perp^2),
\label{k13}
\end{equation}
where $\mathcal{N}_{L}$ and $\mathcal{N}_{T}$ are the normalization constants and depending on the polarization of the particles they can be computed using
\begin{eqnarray}
 \sum_{h_q,h_{\bar{q}}}\int  \frac{{\rm d} x \, {\rm d}^2{\bf k}_\perp  }{2(2 \pi)^3}\, |\Psi^{{(1)}{L,(T)}}_{h_q,h_{\bar{q}}}(x,{\bf k}^2_\perp)|^2=1.
\end{eqnarray}
Our spin improved holographic wave functions differ from the ``boosted'' wave functions in the quark model which are obtained by boosting the non-relativistic Schr\"odinger wave function in the rest frame of mesons to the light-front. Our holographic wave functions are directly formulated in the light-front and are frame-independent, avoiding the ambiguities associated with a boosting prescription. These two wave functions differ by a factor of $1/\sqrt{x(1-x)}$ which is mainly responsible for better results in the present case. In our wave function, the AdS/QCD scale parameter $\kappa$  is extracted from the mass spectroscopic data, which fixes the width of the holographic Gaussian. While in boosted wave function,  width of the boosted Gaussian is a free parameter and has to be fixed by some constraint on the wave function.  The TMDs and PDFs for light vector mesons have been successfully studied with LFHM. We have extended the model for heavy vector mesons $J/\psi$ and $\Upsilon$ as very less work is available for the theoretical predictions of TMDs and PDFs for these heavy vector mesons.

\subsection{Light-front quark model}\label{LFQMe}

LFQM focuses on the valence quarks which are the primary constituents responsible for the overall structure and properties of hadrons \cite{lfqm,lfqm1,lfqm2,lfqm3,lfqm4,lfqm5,lfqm6,lfqm7}. The LFHM is basically an extension of LFQM. The momentum wave function from Brodsky-Huang-Lepage (BHL) prescription, in the LFQM, is given as \cite{Lfqm,Lfqm1,bhl,bhl1}
\begin{eqnarray}
 \psi^{(2)}(x,{\bf k}^2_\perp)= \mathcal{N} \,{ \rm exp} \left[-\frac{1}{8 \beta^2}\left(\frac{{\bf k^2_\perp}+m_q^2}{x}+\frac{{\bf k^2_\perp} +m^2_{\bar q}}{1-x}\right) \right].
 \label{l11}
\end{eqnarray}
Here $\beta$ is a model parameter and its value is $0.699$ GeV for $J/\psi$ and $1.376$ GeV for $\Upsilon$ meson  \cite{ll1}. Since we are dealing with the $c \bar c$ and $b \bar b$ systems here, the quark masses follow the relation  $m_q =m_{\bar q}$. Therefore,  Eq. (\ref{l11}) can now be expressed as
\begin{eqnarray}
 \psi^{(2)}(x,{\bf k}^2_\perp)= \mathcal{N} \,{ \rm exp} \left[-\frac{{\bf k}^2_\perp+m_q^2}{8 \beta^2 x(1-x)} \right].
 \label{l1}
\end{eqnarray}
By taking into account the spin part  on the spin projection of $J/\psi$ and $\Upsilon$ \cite{Lfqm,Lfqm1}, Eq. (\ref{l1}) becomes
\begin{eqnarray}
\Psi^{{(2)}\Lambda}_{h_q,h_{\bar{q}}}(x,{\bf k}^2_\perp)=\chi^{(2)\Lambda}_{h_q,h_{\bar{q}}} (x,{\bf k}_\perp) \psi^{(2)}(x,{\bf k}^2_\perp),
\label{l2}
\end{eqnarray}
with the normalization 
\begin{eqnarray}
\sum_{h_q,h_{\bar{q}}}  \chi^{(2){\Lambda *}}_{h_q,h_{\bar{q}}}(x,{\bf k}_\perp)\chi^{(2)\Lambda}_{h_q,h_{\bar{q}}}(x,{\bf k}_\perp) =1. 
\end{eqnarray}
The spin part of Eq. (\ref{l2}) has already been calculated for the longitudinal $(L)$ and transverse $(T)$ spin projections from instant form to the front form through Melosh-Wigner method \cite{Lfqm1}. For the longitudinal spin projection $\Lambda=L$,  
  $\chi_{h_q,h_{\bar{q}}}^{(2)\Lambda}$ is given as
\begin{eqnarray*}
\chi^{(2){L}}_{+,+}(x,{\bf k}_\perp)&=&\frac{\left(1-2x \right) {M_{\alpha}} k_L}{\left({M_{\alpha}}+2 m_q \right)\sqrt{2\left({\bf k}^2_\perp+m_q^2\right)}} \, ,\\
\chi^{(2){L}}_{+,-}(x,{\bf k}_\perp)&=&\frac{m_q ({M_{\alpha}}+2 m_q)+2{\bf k}^2_\perp}{\left({M_{\alpha}}+2 m_q \right)\sqrt{2\left({\bf k}^2_\perp+m_q^2\right)}}\, ,\\
\chi^{(2){L}}_{-,+}(x,{\bf k}_\perp)&=&\frac{m_q({M_{\alpha}}+2 m_q)+2{\bf k}^2_\perp}{\left({M_{\alpha}}+2 m_q \right)\sqrt{2\left({\bf k}^2_\perp+m^2\right)}}\, ,\\
\chi^{(2){L}}_{-,-}(x,{\bf k}_\perp)&=&-\frac{\left(1-2x \right) {M_{\alpha}} k_R}{\left({M_{\alpha}}+2 m_q \right)\sqrt{2\left({\bf k}^2_\perp+m_q^2\right)}} \, .
\end{eqnarray*}
For the transverse spin projection $\Lambda=T(+)$, we have
\begin{eqnarray*}
\chi^{(2)T(+)}_{+,+}(x,{\bf k}_\perp)&=&\frac{m_q({M_{\alpha}}+2 m)+{\bf k}^2_\perp}{\left({M_{\alpha}}+2 m_q \right)\sqrt{{\bf k}^2_\perp+m_q^2}}\, ,\\
\chi^{(2)T(+)}_{+,-}(x,{\bf k}_\perp)&=&\frac{\left(x {M_{\alpha}}+m_q\right) k_R}{\left({M_{\alpha}}+2 m_q \right)\sqrt{{\bf k}^2_\perp+m_q^2}}\, ,\\
\chi^{(2)T(+)}_{-,+}(x,{\bf k}_\perp)&=&-\frac{\left(\left(1-x\right){M_{\alpha}}+m_q \right) k_R}{\left({M_{\alpha}}+2 m_q \right)\sqrt{{\bf k}^2_\perp+m_q^2}}\, ,\\
\chi^{(2)T(+)}_{-,-}(x,{\bf k}_\perp)&=& -\frac{k_R^2}{\left({M_{\alpha}}+2 m_q \right)\sqrt{{\bf k}^2_\perp+m_q^2}}\, ,
\end{eqnarray*} 
and for $\Lambda=T(-)$, we have
\begin{eqnarray*}
  \chi^{(2)T(-)}_{+,+}(x,{\bf k}_\perp)&=& -\frac{k_L^2}{\left({M_{\alpha}}+2 m_q \right)\sqrt{{\bf k}^2_\perp+m_q^2}}\, ,\\
\chi^{(2)T(-)}_{+,-}(x,{\bf k}_\perp)&=& \frac{\left(\left(1-x\right) {M_{\alpha}}+m_q \right) k_L}{\left({M_{\alpha}}+2 m_q \right)\sqrt{{\bf k}^2_\perp+m_q^2}}\, ,\\
\chi^{(2)T(-)}_{-,+}(x,{\bf k}_\perp)&=&-\frac{\left(x {M_{\alpha}}+m_q\right) k_L}{\left({M_{\alpha}}+2 m_q \right)\sqrt{{\bf k}^2_\perp+m_q^2}}\, ,\\
\chi^{(2)T(-)}_{-,-}(x,{\bf k}_\perp)&=&\frac{m_q({M_{\alpha}}+2 m_q)+{\bf k}^2_\perp}{\left({M_{\alpha}}+2 m_q \right)\sqrt{{\bf k}^2_\perp+m_q^2}}\, ,  
\end{eqnarray*}
with 
\begin{eqnarray*}
    {M_{\alpha}}=\sqrt{\frac{{\bf k}^2_\perp+m_q^2 }{x(1-x)}} \, .
\label{l3}
\end{eqnarray*}
Now the LFWFs of Eq. (\ref{l2}) can be written explicitly for longitudinal projection $\Lambda=L$ as 
\begin{eqnarray}
\label{m1}
\Psi^{(2)L}_{+,+}(x,{\bf k}^2_\perp)&=& \frac{\left(1-2x \right) {M_{\alpha}} k_L}{\sqrt{2} \omega} \psi^{(2)}(x,{\bf k}^2_\perp)\, ,\\
\Psi^{(2)L}_{+,-}(x,{\bf k}^2_\perp)&=& \frac{m_q({M_{\alpha}}+2 m_q)+2{\bf k}^2_\perp}{\sqrt{2} \omega} \psi^{(2)}(x,{\bf k}^2_\perp)\, ,\\
\Psi^{(2)L}_{-,+}(x,{\bf k}^2_\perp)&=& \frac{m_q({M_{\alpha}}+2 m_q)+2{\bf k}^2_\perp}{\sqrt{2} \omega} \psi^{(2)}(x,{\bf k}^2_\perp)\, ,\\
\Psi^{(2)L}_{-,-}(x,{\bf k}^2_\perp)&=& \frac{\left(1-2x \right) {M_{\alpha}} k_R}{\sqrt{2} \omega} \psi^{(2)}(x,{\bf k}^2_\perp)\, .
\label{l4}
\end{eqnarray}
For transverse projection $\Lambda=T(+)$, the LFWFs are
 \begin{eqnarray}
    \Psi^{(2)T(+)}_{+,+}(x,{\bf k}^2_\perp)&=& \frac{m_q({M_{\alpha}}+2 m)+{\bf k}^2_\perp}{\omega} \psi^{(2)}(x,{\bf k}^2_\perp)\, ,\\
    \Psi^{(2)T(+)}_{+,+}(x,{\bf k}^2_\perp)&=& \frac{\left(x {M_{\alpha}}+m_q\right) k_R}{\omega} \psi^{(2)}(x,{\bf k}^2_\perp)\, ,\\
    \Psi^{(2)T(+)}_{-,+}(x,{\bf k}^2_\perp)&=& -\frac{\left(\left(1-x\right) {M_{\alpha}}+m_q \right) k_R}{\omega} \psi^{(2)}(x,{\bf k}^2_\perp)\, ,\\
    \Psi^{(2)T(+)}_{-,-}(x,{\bf k}^2_\perp)&=& -\frac{k_R^2}{\omega} \psi^{(2)}(x,{\bf k}^2_\perp)\, ,
    \label{l5}
 \end{eqnarray}
 and for $\Lambda=T(-)$, the LFWFs are
 \begin{eqnarray}
    \Psi^{(2)T(-)}_{+,+}(x,{\bf k}^2_\perp)&=& -\frac{k_L^2}{\omega} \psi^{(2)}(x,{\bf k}^2_\perp)\, ,\\
    \Psi^{(2)T(-)}_{+,+}(x,{\bf k}^2_\perp)&=& \frac{\left(\left(1-x\right) {M_{\alpha}}+m_q \right) k_L}{\omega} \psi^{(2)}(x,{\bf k}^2_\perp)\, ,\\
    \Psi^{(2)T(-)}_{-,+}(x,{\bf k}^2_\perp)&=& -\frac{\left(x {M_{\alpha}}+m_q\right) k_L}{\omega} \psi^{(2)}(x,{\bf k}^2_\perp)\, ,\\
    \Psi^{(2)T(-)}_{-,-}(x,{\bf k}^2_\perp)&=& \frac{m_q({M_{\alpha}}+2 m_q)+{\bf k}^2_\perp}{\omega} \psi^{(2)}(x,{\bf k}^2_\perp)\,.
    \label{l6}
 \end{eqnarray}
We have
 \begin{eqnarray*}
\omega=({M_{\alpha}}+2m_q)\sqrt{{\bf k}^2_\perp+m_q^2}\, .
\end{eqnarray*}
LFQM provides very reasonable results in the lower $Q^2$ region. The LFQM wave functions have been used for the study of $B_{\textit{i}{(\frac{1}{2}^+)}}\to B_{\textit{f}{(\frac{3}{2}^+)}}$ transitions \cite{LFQm}, $ \Xi_{cc} \to \Xi_c$ weak decay \cite{LFQm1}, $V' \to V''$ transition \cite{trans}, $(\pi^0, \eta, \eta') \to \gamma^* \gamma^*$ transtions \cite{trans1} etc. LFQM has also been used for calculating the form factors (FFs), distribution amplitudes (DAs) etc. \cite{da,ll1,da2,da3}.

\subsection{Overlap form of the LFWFs} \label{tmd_overlap}

The overlap form of the LFWFs in terms of the LF helicity amplitudes with quark and anti-quark helicities along with their polarizations for the heavy vector mesons for both the models can be expressed as \cite{bse,rho}
\begin{eqnarray}
A_{h^\prime_q \Lambda^\prime, h_q \Lambda}(x,{\bf k}^2_\perp)&=&\frac{1}{(2 \pi)^3} \sum_{ h_{\bar{q}}} \Psi^{(n)\Lambda^\prime *}_{h^\prime_q, h_{\bar{q}}}(x,{\bf k}^2_\perp)\,\Psi^{(n)\Lambda}_{h_q, h_{\bar{q}}}(x,{\bf k}^2_\perp)\, .
\label{O1}
\end{eqnarray}
Here $n=1$ and $2$ for the LFHM and LFQM respectively.
The explicit overlap form for all the T-even TMDs in terms of helicity amplitudes for both the models is given by \cite{bse}
\begin{eqnarray}
\label{a11}
f_{1}(x,\mathbf{k}_{\perp}^2) &=& \frac{1}{6}(A_{+0,+0}+A_{-0,-0} \nonumber \\
&&+A_{++,++}+A_{-+,-+}+A_{+-,+-}+A_{--,--}), \\
g_{1L}(x,\mathbf{k}_{\perp}^2) &=& \frac{1}{4}(A_{++,++}-A_{-+,-+}-A_{+-,+-}+A_{--,--}), \\
g_{1T}(x,\mathbf{k}_{\perp}^2) &=& \frac{M_{\alpha}}{4\sqrt{2}\mathbf{k}_{\perp}^2}\left(k_R(A_{++,+0}-A_{-+,-0}+A_{+0,+-}-A_{-0,--}) \right. \nonumber \\
&&\left. + k_L(A_{+0,++}-A_{-0,-+}+A_{+-,+0}-A_{--,-0})\right), \\
h_{1}(x,\mathbf{k}_{\perp}^2) &=& \frac{1}{4\sqrt{2}}(A_{++,-0}+A_{-0,++}+A_{+0,--}+A_{--,+0}), \\
h_{1L}^{\perp}(x,\mathbf{k}_{\perp}^2) &=& \frac{M_{\alpha}}{4\mathbf{k}_{\perp}^2}\left(k_R(A_{-+,++}-A_{--,+-}) + k_L(A_{++,-+}-A_{+-,--})\right), \\
h_{1T}^{\perp}(x,\mathbf{k}_{\perp}^2) &=& \frac{M_{\alpha}^2}{2\sqrt{2}\mathbf{k}_{\perp}^4}\left(k_R^2(A_{-+,+0}+A_{-0,+-}) + k_L^2(A_{+0,-+}+A_{+-,-0})\right), \\
f_{1LL}(x,\mathbf{k}_{\perp}^2) &=& \frac{1}{2}A_{+0,+0}+A_{-0,-0} \nonumber \\
&&-\frac{1}{4}(A_{++,++}+A_{-+,-+}+A_{+-,+-}+A_{--,--}), \\
f_{1LT}(x,\mathbf{k}_{\perp}^2) &=& \frac{M_{\alpha}}{4\sqrt{2}\mathbf{k}_{\perp}^2}\left(k_R(A_{++,+0}+A_{-+,-0}-A_{+0,+-}-A_{-0,--}) \right. \nonumber \\
&&\left. + k_L(A_{+0,++}+A_{-0,-+}-A_{+-,+0}-A_{--,-0})\right), \\
f_{1TT}(x,\mathbf{k}_{\perp}^2) &=& \frac{M_{\alpha}^2}{4\sqrt{2}\mathbf{k}_{\perp}^2}\left(k_R^2(A_{++,+-}+A_{-+,--}) + k_L^2(A_{+-,++}+A_{--,-+})\right).
\label{a12}
\end{eqnarray}

Substituting Eqs. (\ref{m1})-(\ref{l6}) in the above LF helicity amplitude equations, through Eq. (\ref{O1}) for $n=2$, the T-even TMDs for LFQM can be computed and are expressed as
\begin{eqnarray}
f_1(x,{\bf k}^2_\perp)&=&\frac{1}{3(2 \pi)^3}\bigg(\frac{1}{2}\left(3\left(m_q\left({M_\alpha}+2m_q\right)\right)^2+(1-2x)^2{M_\alpha}^2{\bf k}^2_\perp \right)+4 {\bf k}^2_\perp\left( m_q({M_\alpha}+2m_q)+{\bf k}^2_\perp\right)\nonumber\\
&&+{\bf k}^2_\perp\left(2m_q({M_\alpha}+m_q)+{M_\alpha}^2(1-2x+2x^2)\right)\bigg)\frac{| \psi^{(2)}(x,{\bf k}^2_\perp)|^2}{\omega^2}\, ,
\label{f1_LF}
\end{eqnarray}
\begin{equation}
g_{1L}(x,{\bf k}^2_\perp)=\frac{1}{2(2 \pi)^3}\big(m_q({M_\alpha}+2m_q)\left(m_q({M_\alpha}+2m_q)+2 {\bf k}^2_\perp\right)-{M_\alpha}({M_\alpha}+2m_q)(1-2x)\big)\frac{| \psi^{(2)}(x,{\bf k}^2_\perp)|^2}{\omega^2}\, ,
\label{g1_LF}
\end{equation}
\begin{equation}
g_{1T}(x,{\bf k}^2_\perp)=\frac{M_\alpha}{2 (2 \pi)^3}\left({M_\alpha}+2m_q\right)\left(m_q {M_\alpha}(1-2x)+\left(2{\bf k}^2_\perp + m_q({M_\alpha}+2m_q)\right)\right)\times\frac{| \psi^{(2)}(x,{\bf k}^2_\perp)|^2}{\omega^2}\, ,
\end{equation}
\begin{eqnarray}
h_1(x,{\bf k}^2_\perp)&=&\frac{1}{2 (2 \pi)^3}\big(\left(m_q({M_\alpha}+2m_q)+2 {\bf k}^2_\perp\right)\left(m_q({M_\alpha}+2m_q)+ {\bf k}^2_\perp\right)\nonumber\\
&&-\mathcal{M_\alpha}(x{M_\alpha}+m_q)(1-2x){\bf k}^2_\perp\big)\frac{| \psi^{(2)}(x,{\bf k}^2_\perp)|^2}{\omega^2}\, ,
\label{h1_LF}
\end{eqnarray}
\begin{eqnarray}
h^\perp_{1L}(x,{\bf k}^2_\perp)&=&-\frac{M_\alpha}{(2 \pi)^3}({M_\alpha}+2m_q)\left(m_q\left((1-x){M_\alpha}+m_q\right)+{\bf k}^2_\perp\right)\frac{| \psi^{(2)}(x,{\bf k}^2_\perp)|^2}{\omega^2}\, ,\nonumber\\
\label{h1lq}
\end{eqnarray}
\begin{equation}
h^\perp_{1T}(x,{\bf k}^2_\perp)=-\frac{M_{\alpha}^2}{(2 \pi)^3}\left({M_\alpha}\left(m_q+(1-x){M_\alpha}\right)(1-2x)+\left(2{\bf k}^2_\perp +m_q({M_\alpha}+2m_q)\right)\right)\times\frac{| \psi^{(2)}(x,{\bf k}^2_\perp)|^2}{\omega^2}\, ,
\end{equation}
\begin{eqnarray}
f_{1LL}(x,{\bf k}^2_\perp)&=& 0 \, ,
\label{f1LL_LF}
\end{eqnarray}
\begin{eqnarray}
f_{1LT}(x,{\bf k}^2_\perp)&=& 0 \, ,
\end{eqnarray}
\begin{eqnarray}
f_{1TT}(x,{\bf k}^2_\perp) &=& 0\,.\label{q1}
\end{eqnarray}
Similarly, substituting Eqs. (\ref{k12}) and (\ref{k13}) in Eqs. (\ref{a11})-(\ref{a12}) through Eq. (\ref{O1}) for $n=1$, the T-even TMDs in LFHM can be derived and are expressed as
\begin{eqnarray}
f_1(x,{\bf k}^2_\perp)&=& \frac{1}{3(2 \pi)^3}\bigg(\mathcal{N}^2_L\left( M^2_\alpha~ x(1-x)+m_q^2+{\bf k}_\perp^2 \right)^2 \frac{|\psi^{(1)}(x,{\bf k}^2_\perp)| ^2 }{{x^2 (1-x)^2}}\nonumber\\
&+& \mathcal{N}^2_T\left(m_q^2 + {\bf k}^2_\perp \left(2 x^2-2 x+1\right)\right)\frac{|\psi^{(1)}(x,{\bf k}^2_\perp)|^2}{{x^2 (1-x)^2}} \bigg)\,,\label{f1-LFH} \\
g_{1L}(x,{\bf k}^2_\perp)&=& \frac{\mathcal{N}^2_T}{2(2 \pi)^3}\left(m_q^2+{\bf k}^2_\perp\left(2 x-1\right) \right)\frac{|\psi^{(1)}(x,{\bf k}^2_\perp)| ^2}{{x^2 (1-x)^2}}\,,\label{g1l-LFH}\\
g_{1T}(x,{\bf k}^2_\perp)&=&\mathcal{N}_L \mathcal{N}_T\frac{M_\alpha}{\sqrt{2}(2 \pi)^3}\left(M^2_\alpha~ x(1-x)+m_q^2 +{\bf k}^2_\perp \right)\frac{|\psi^{(1)}(x,{\bf k}^2_\perp)|^2}{{x^2 (1-x)^2}}\,,\label{g1t-LFH}\nonumber\\
\end{eqnarray}
\begin{eqnarray}
h_1(x,{\bf k}^2_\perp)&=& \mathcal{N}_L \mathcal{N}_T\frac{m_q}{\sqrt{2}(2 \pi)^3}\left(M^2_\alpha~ x(1-x)+m_q^2 +{\bf k}^2_\perp \right)\frac{|\psi^{(1)}(x,{\bf k}^2_\perp)|^2}{{x^2 (1-x)^2}}\,, \label{h1-LFH}\nonumber\\
\end{eqnarray}
\begin{eqnarray}
h^\perp_{1L}(x,{\bf k}^2_\perp)&=& -\mathcal{N}^2_T\frac{m_q \, M_\alpha}{(2 \pi)^3}\frac{|\psi^{(1)}(x,{\bf k}^2_\perp)| ^2}{x^2 (1-x)}\,, \label{hlp-LFH}
\end{eqnarray}
\begin{eqnarray}
h^\perp_{1T}(x,{\bf k}^2_\perp)&=&0\,,\label{htp-LFH}
\end{eqnarray}
\begin{eqnarray}
f_{1LL}(x,{\bf k}^2_\perp)&=& \frac{1}{(2 \pi)^3}\bigg(\mathcal{N}^2_L\left( M^2_\alpha~ x(1-x)+m_q^2+{\bf k}_\perp^2 \right)^2 \frac{|\psi^{(1)}(x,{\bf k}^2_\perp)| ^2 }{x^2 (1-x)^2}\nonumber\\
&&-\mathcal{N}^2_T\left(m_q^2 + {\bf k}^2_\perp \left(2 x^2-2 x+1\right)\right)\frac{|\psi^{(1)}(x,{\bf k}^2_\perp)|^2}{{2 x^2 (1-x)^2}}\bigg)\,,\label{f1ll-LFH}
\end{eqnarray}
\begin{equation}
f_{1LT}(x,{\bf k}^2_\perp)=\mathcal{N}_L \mathcal{N}_T \frac{M_\alpha}{ \sqrt{2}(2 \pi)^3}\left(2x-1 \right)\left( M^2_\alpha~ x(1-x)+m_q^2+{\bf k}_\perp^2\right) \times \frac{|\psi^{(1)}(x,{\bf k}^2_\perp)|^2}{{x^2 (1-x)^2}}\,,\label{f1lt-LFH}
\end{equation}
\begin{eqnarray}
f_{1TT}(x,{\bf k}^2_\perp)&=& \mathcal{N}^2_T\frac{M^2_\alpha}{(2 \pi)^3} \frac{|\psi^{(1)}(x,{\bf k}^2_\perp)| ^2}{x (1-x)}\,.\label{f1tt-LFH}
\end{eqnarray}

 Eqs. (\ref{f1_LF})-(\ref{q1}) are the T-even TMDs in the LFQM and Eqs. (\ref{f1-LFH})-(\ref{f1tt-LFH}) are in the LFHM. Angular momentum along $z$-axis is conserved for all the TMDs.There is zero orbital angular momentum (OAM) transfer between initial and final state of the hadron for $f_1$, $g_{1L}$, $h^{\perp}_{1L}$ and $f_{1LL}$ TMDs. Eqs. (\ref{g1l-LFH}), (\ref{hlp-LFH}) and  
 (\ref{f1tt-LFH}) are the overlapping  LFWFs for $\Lambda=\pm 1$, therefore, these TMDs have the factor of $\mathcal{N}^2_T$. Eqs. (\ref{g1t-LFH}), (\ref{h1-LFH}) are (\ref{f1lt-LFH}) are a  consequence of $\Lambda=0$ to $\Lambda=\pm 1$ overlapping, therefore, they have a $\mathcal{N}_L \mathcal{N}_T$ factor. Eqs. (\ref{f1-LFH}) and (\ref{f1ll-LFH}) do not have a common factor. In the LFQM, $f_{1LL},f_{1LT}$ and $f_{1TT}$ come out to be zero whereas in the LFHM, $h^{\perp}_T$ TMD is zero. Since the higher Fock-states are suppressed in case of heavy vector mesons \cite{bse}, we have considered the lower Fock-states only for both the models which provides good results.


\subsection{Numerical results}\label{results}
\begin{figure}[ht]
	\centering
	\begin{minipage}[c]{1\textwidth}\begin{center}
			(a)\includegraphics[width=.44\textwidth]{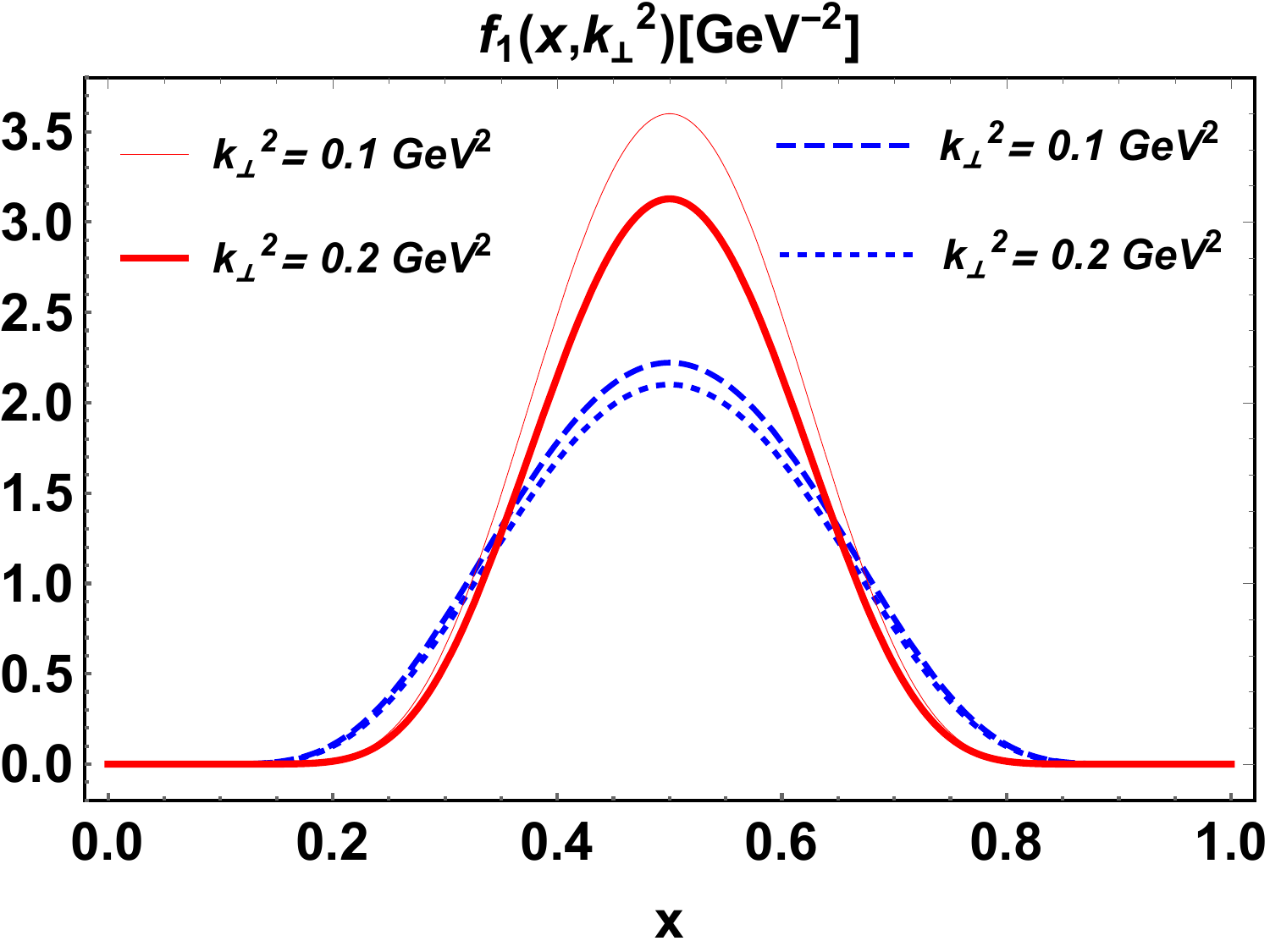}
			(b)\includegraphics[width=.445\textwidth]{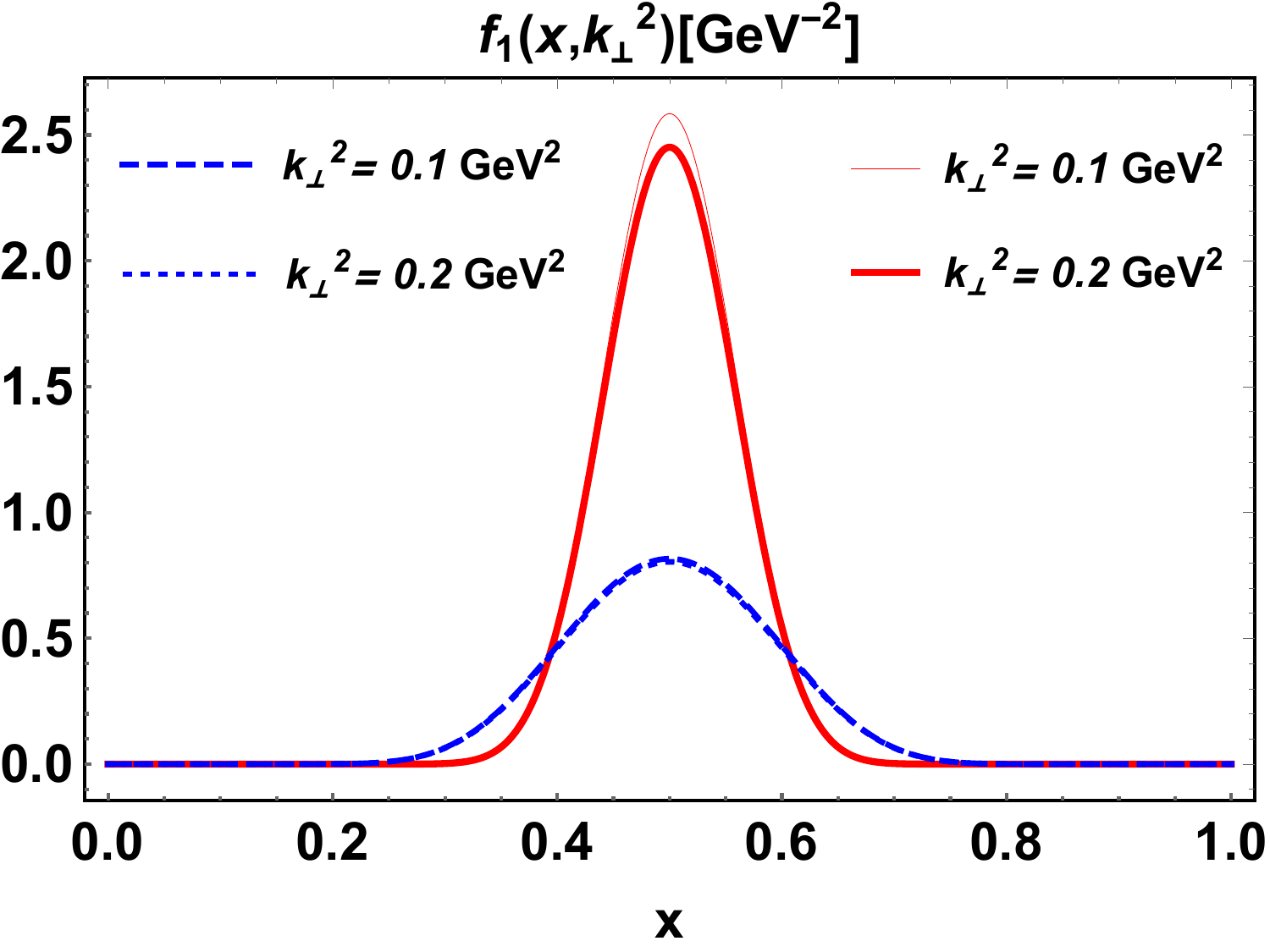}\end{center}
	\end{minipage}
	\begin{minipage}[c]{1\textwidth}\begin{center}
			(c)\includegraphics[width=.44\textwidth]{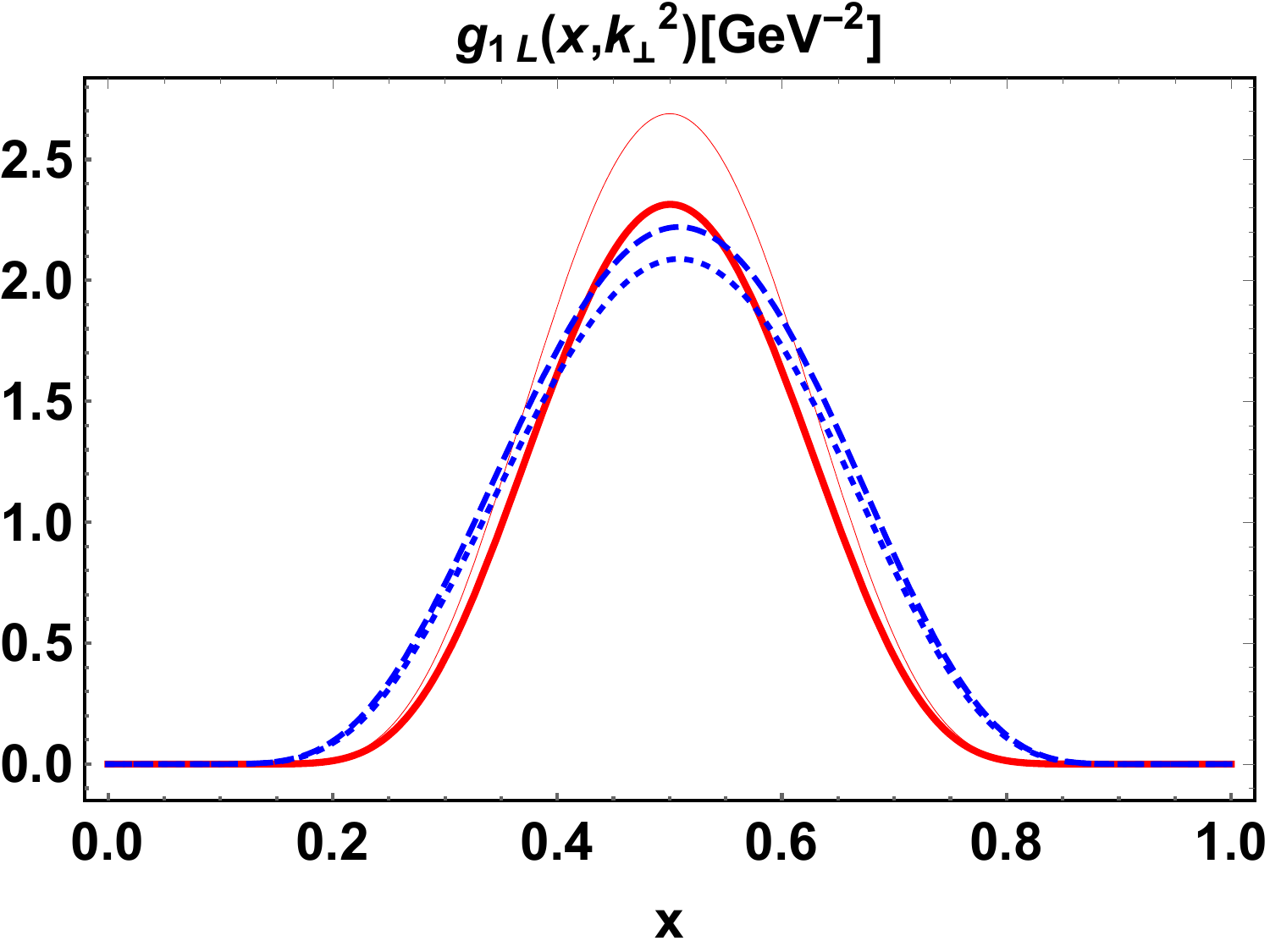}
			(d)\includegraphics[width=.445\textwidth]{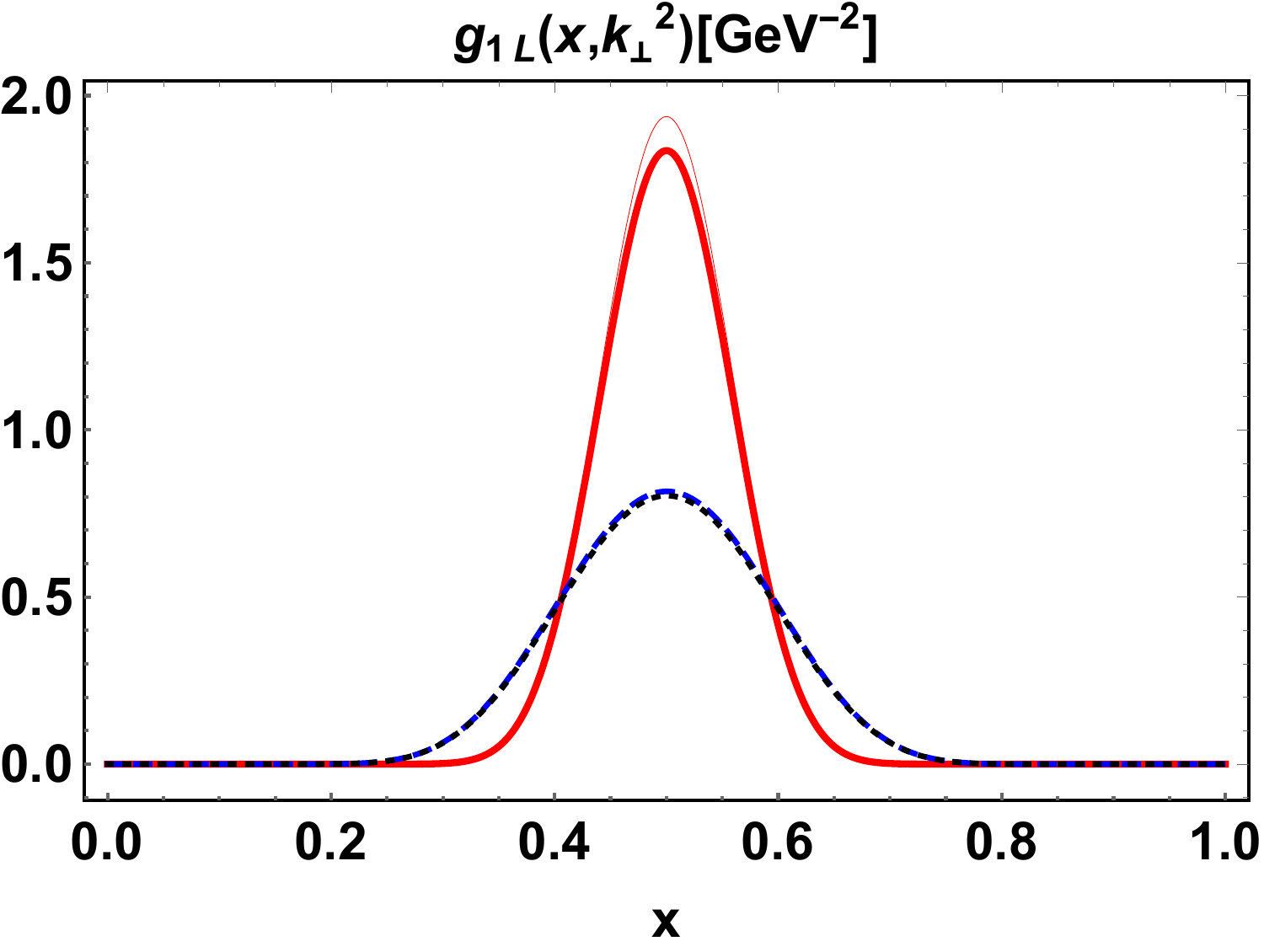}\end{center}
	\end{minipage}
	\begin{minipage}[c]{1\textwidth}\begin{center}
			(e)\includegraphics[width=.44\textwidth]{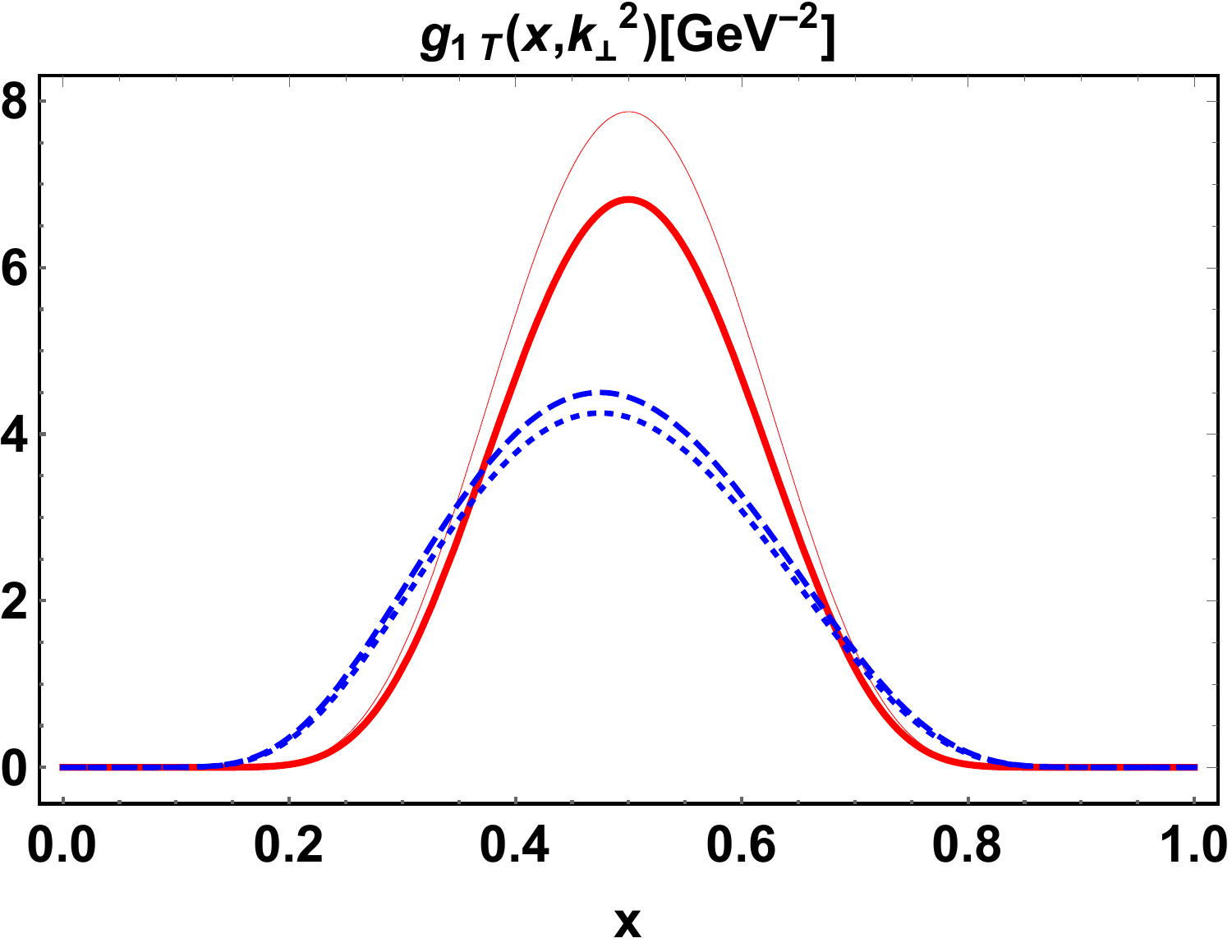}
			(f)\includegraphics[width=.445\textwidth]{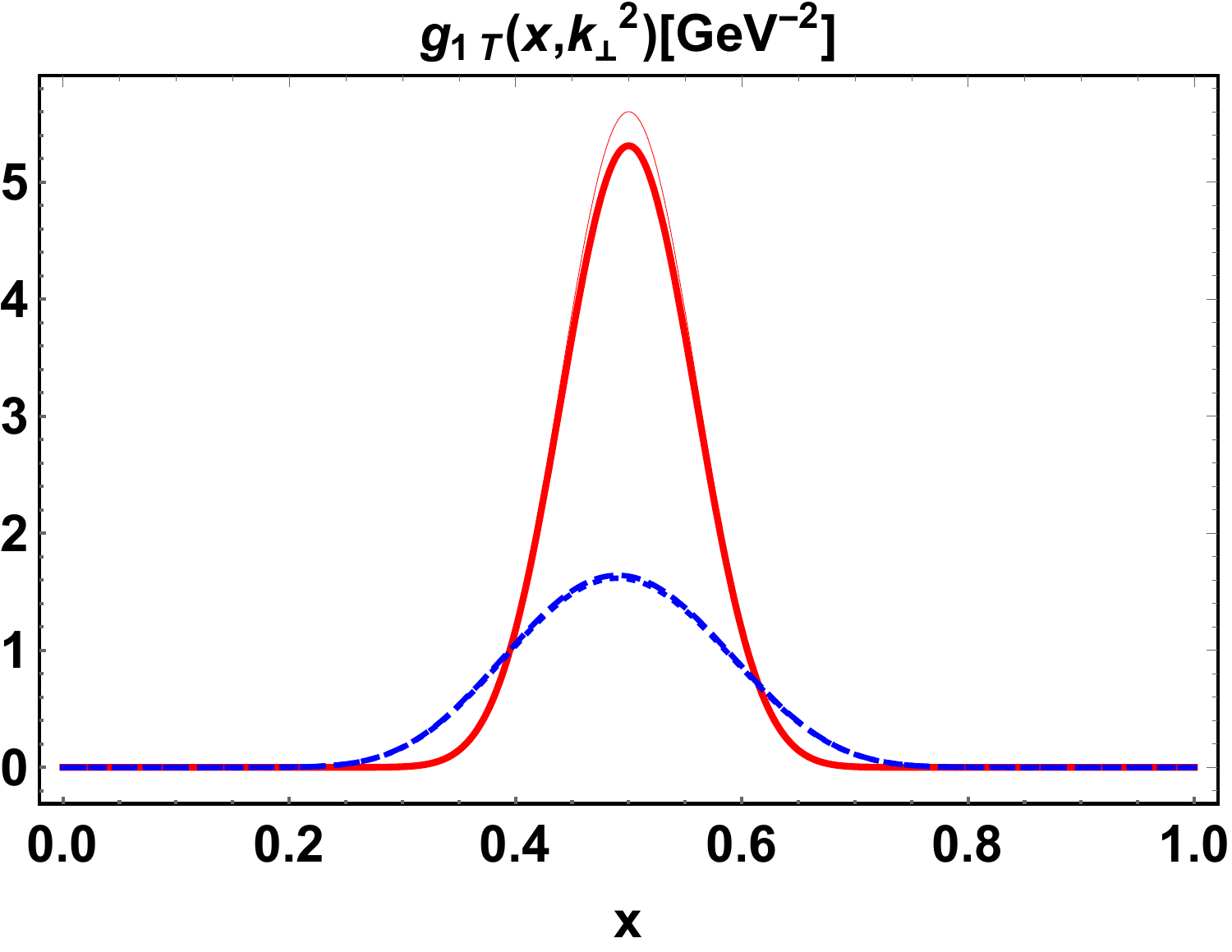}\end{center}
	\end{minipage}
	\caption{(Color online) $f_1(x,{\bf k}^2_\perp)$, $g_{1L}(x,{\bf k}^2_\perp)$ and $g_{1T}(x,{\bf k}^2_\perp)$ TMDs are plotted with respect to $x$ at different values of ${\bf k}^2_\perp$ i.e., ${\bf k}^2_\perp=0.1 {\ \rm GeV}^2$  and ${\bf k}^2_\perp=0.2 {\ \rm GeV}^2$. The solid red thick and thin curves represents the TMDs in LFHM, while blue dot and dashed curves are for LFQM. The TMDs comparison of $J/\Psi$ and $\Upsilon$ meson in the LFHM and LFQM at the model scale $\mu_{\rm LFHM}^2=0.20$ GeV$^2$ and $\mu_{\rm LFQM}^2=0.19$ GeV$^2$ in the left and right panels respectively. }
	\label{tmds}
\end{figure}

\begin{figure}[ht]
	\centering
	\begin{minipage}[c]{1\textwidth}\begin{center}
			(a)\includegraphics[width=.43\textwidth]{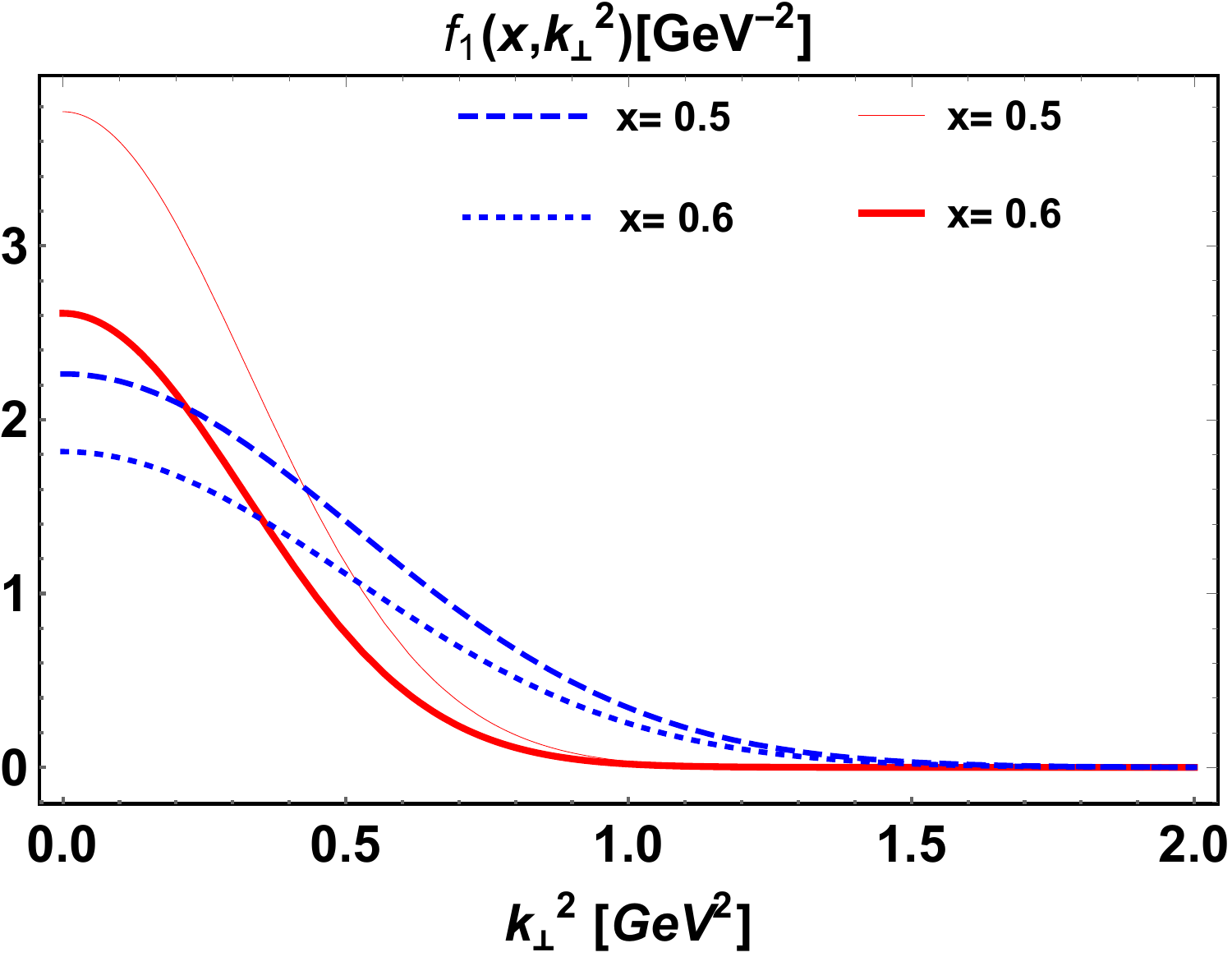}
			(b)\includegraphics[width=.445\textwidth]{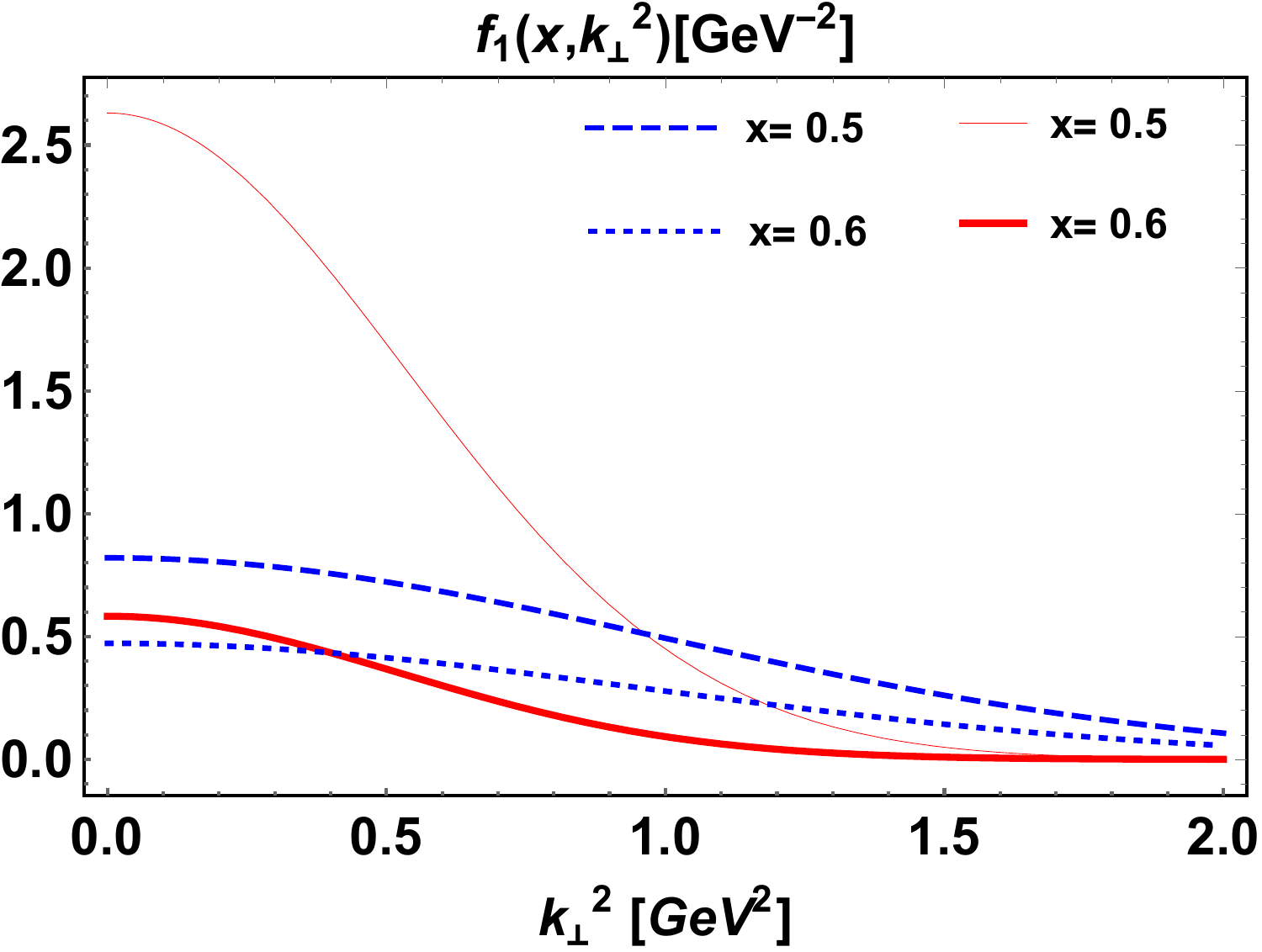}\end{center}
	\end{minipage}
	\begin{minipage}[c]{1\textwidth}\begin{center}
			(c)\includegraphics[width=.43\textwidth]{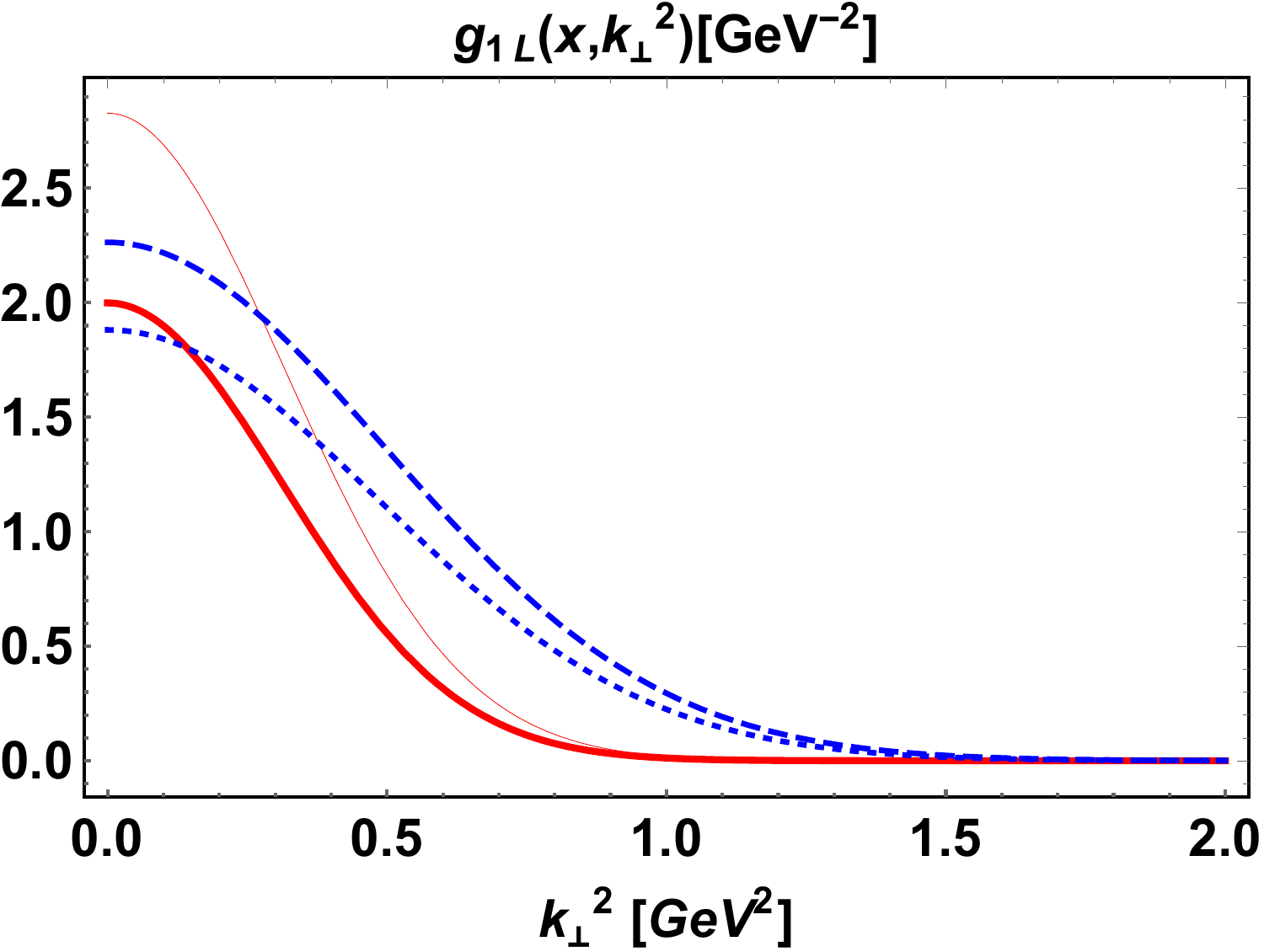}
			(d)\includegraphics[width=.445\textwidth]{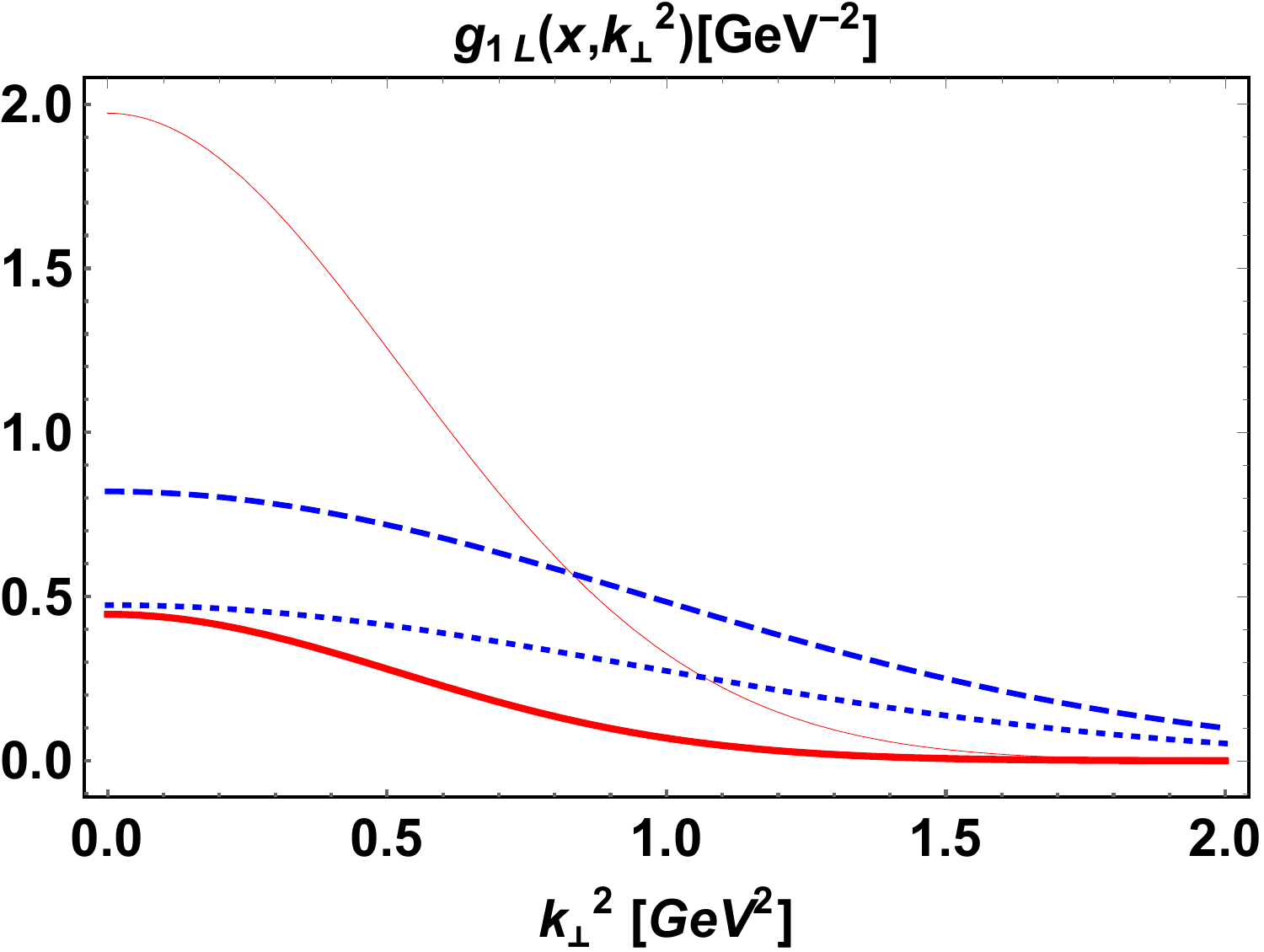}\end{center}
	\end{minipage}
	\begin{minipage}[c]{1\textwidth}\begin{center}
			(e)\includegraphics[width=.43\textwidth]{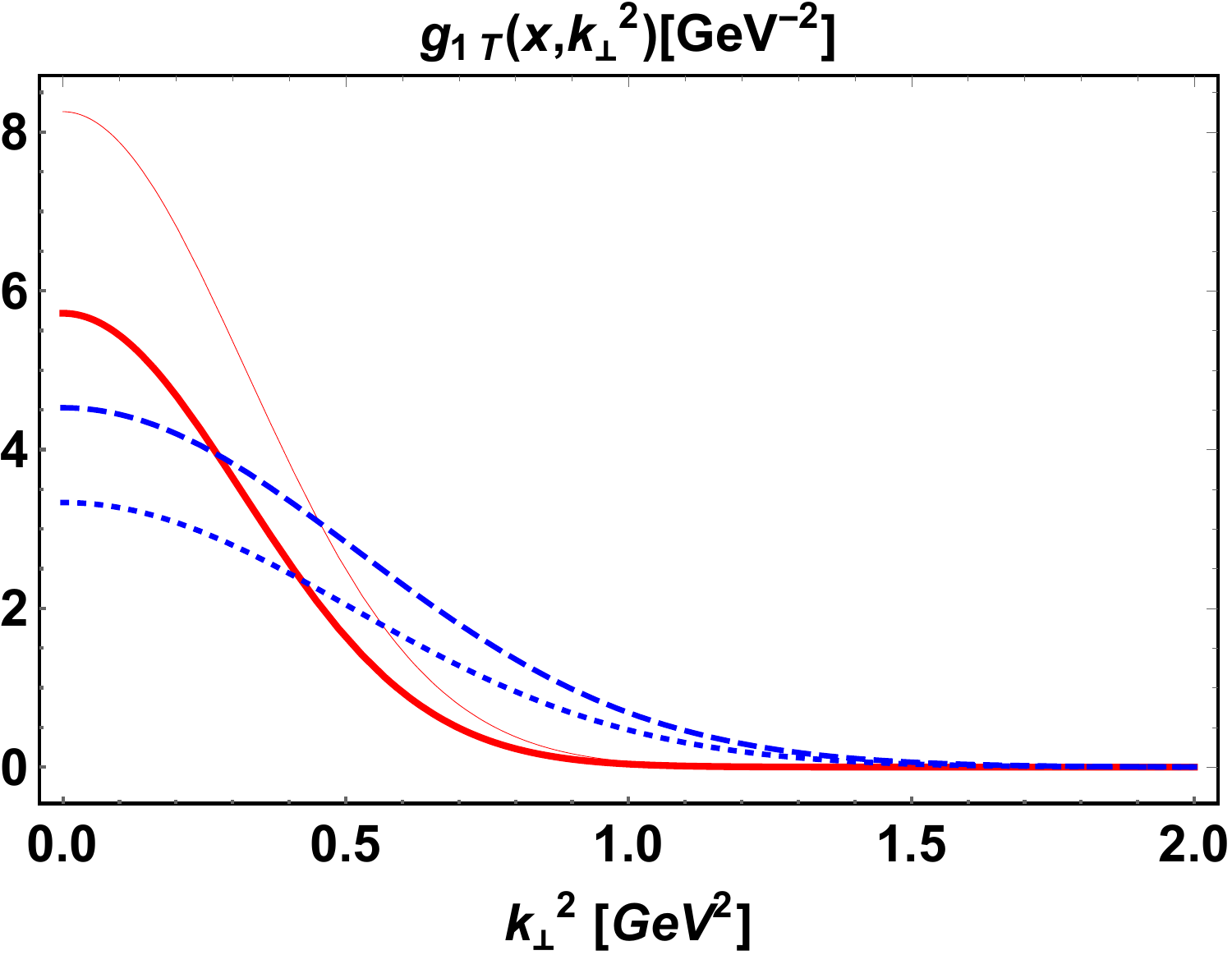}
			(f)\includegraphics[width=.445\textwidth]{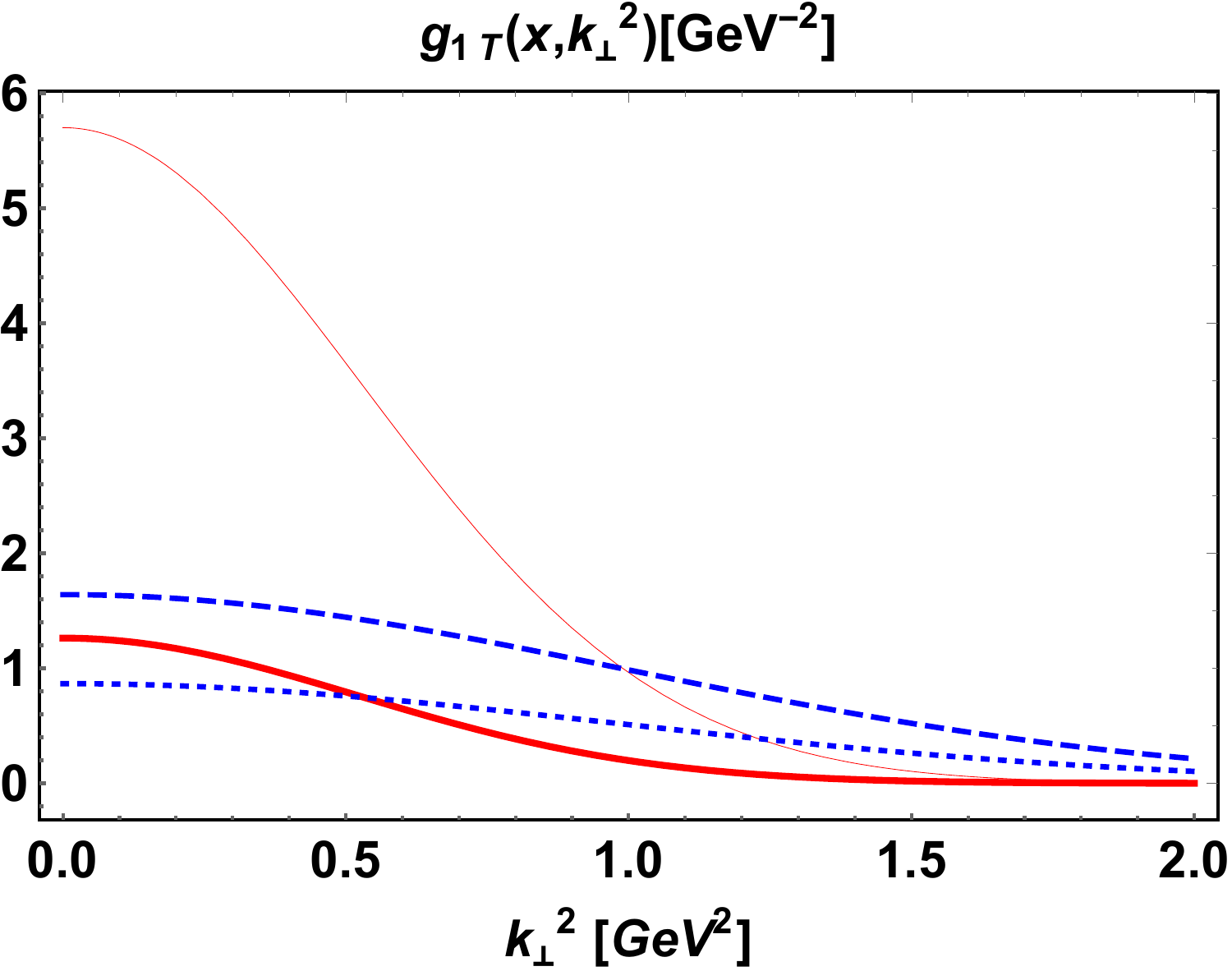}\end{center}
	\end{minipage}
	\caption{(Color online) $f_1(x,{\bf k}^2_\perp)$, $g_{1L}(x,{\bf k}^2_\perp)$ and $g_{1T}(x,{\bf k}^2_\perp)$ TMDs are plotted with respect to ${\bf k}^2_\perp$ at different values of $x$ i.e., $x=0.5$  and $x=0.6$. The solid red thick and thin curves represents the TMDs in LFHM, while blue dot and dashed curves are for LFQM. The TMDs comparison of $J/\Psi$ and $\Upsilon$ meson in the LFHM and LFQM at the model scale $\mu_{\rm LFHM}^2=0.20$ GeV$^2$ and $\mu_{\rm LFQM}^2=0.19$ GeV$^2$ in the left and right panels respectively.}
	\label{tmds2}
\end{figure}

\begin{figure}[ht]
	\begin{minipage}[c]{1\textwidth}\begin{center}
			(a)\includegraphics[width=.43\textwidth]{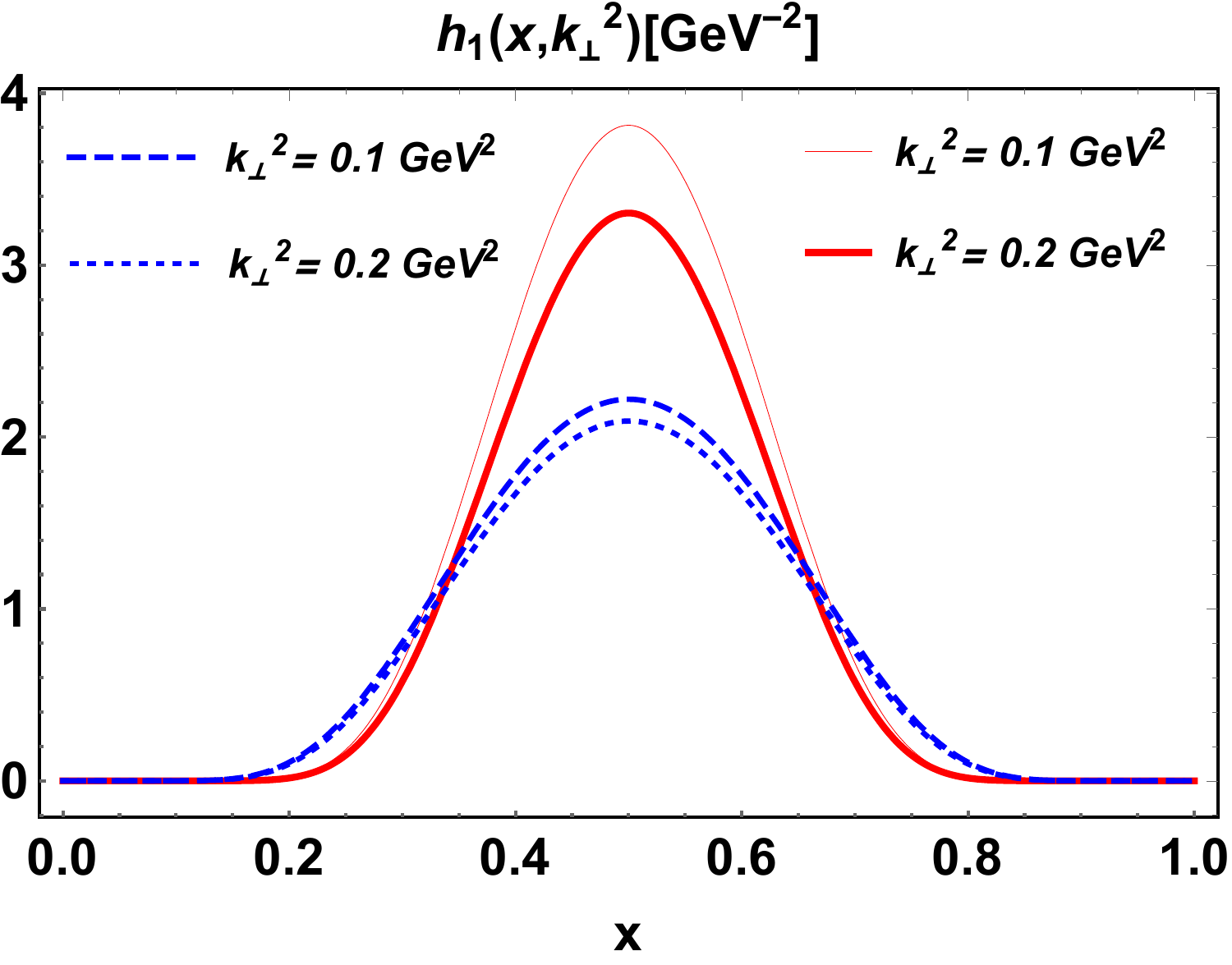}
			(b)\includegraphics[width=.445\textwidth]{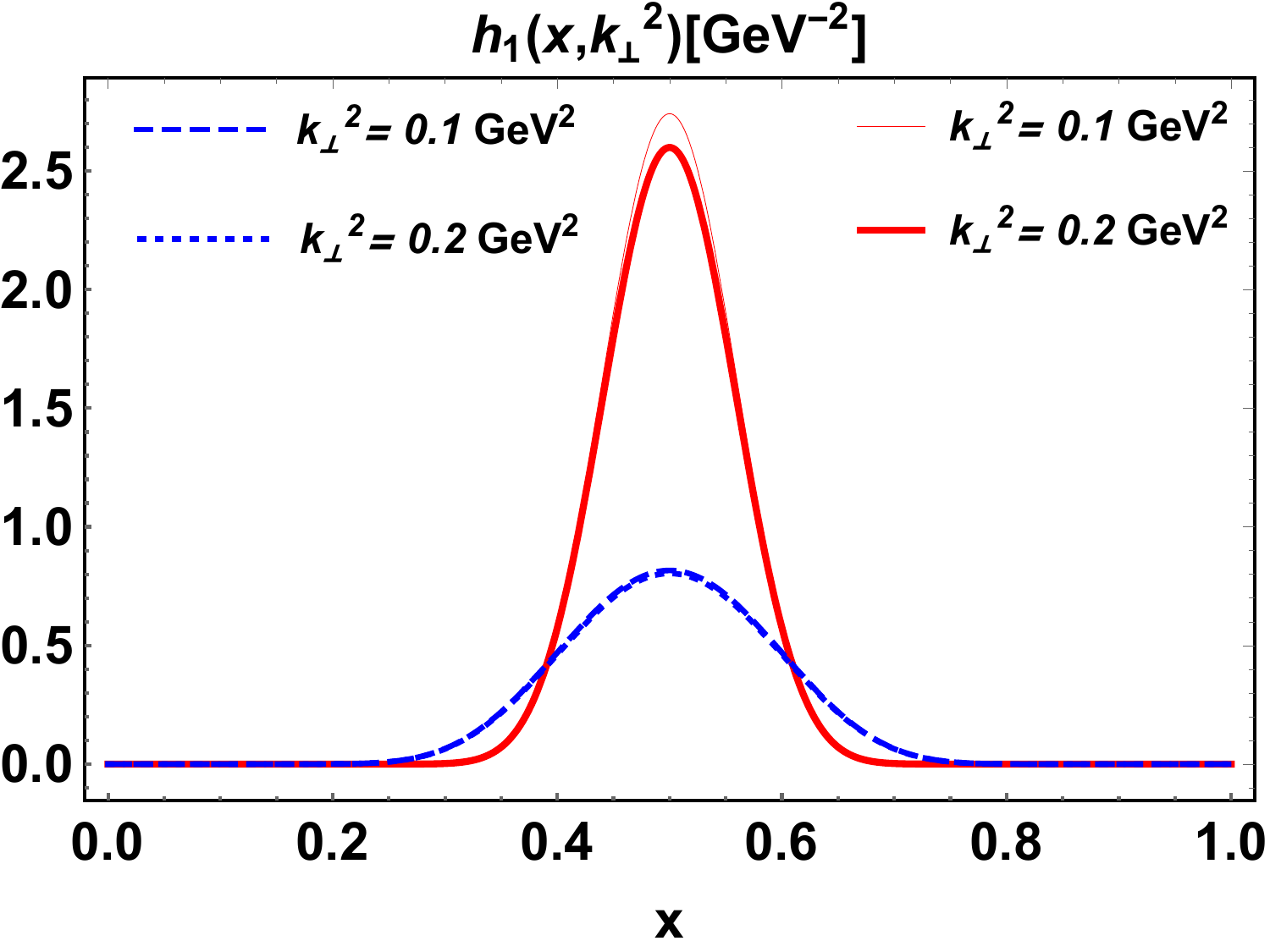}\end{center}
	\end{minipage}
	\begin{minipage}[c]{1\textwidth}\begin{center}
			(c)\includegraphics[width=.43\textwidth]{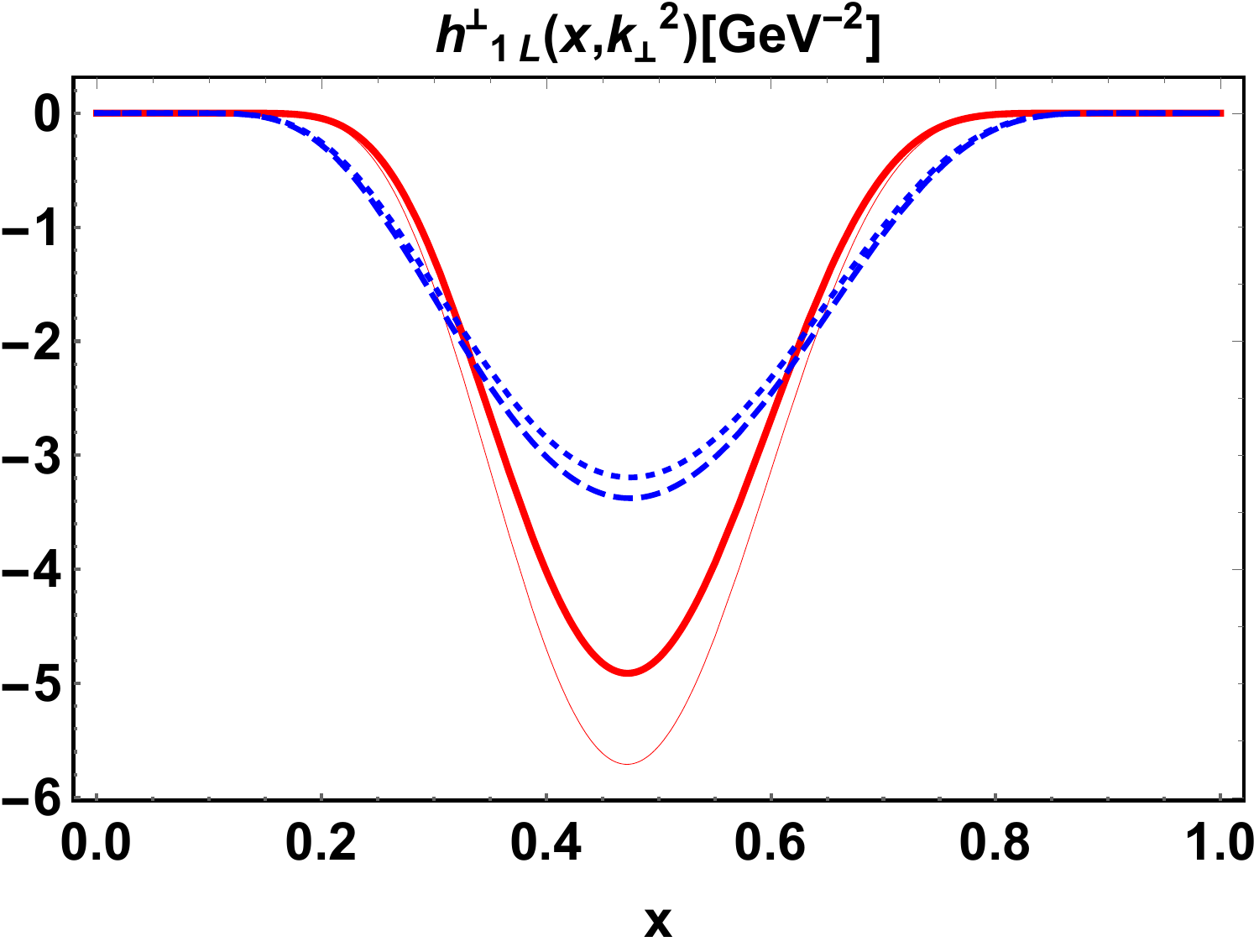}
			(d)\includegraphics[width=.445\textwidth]{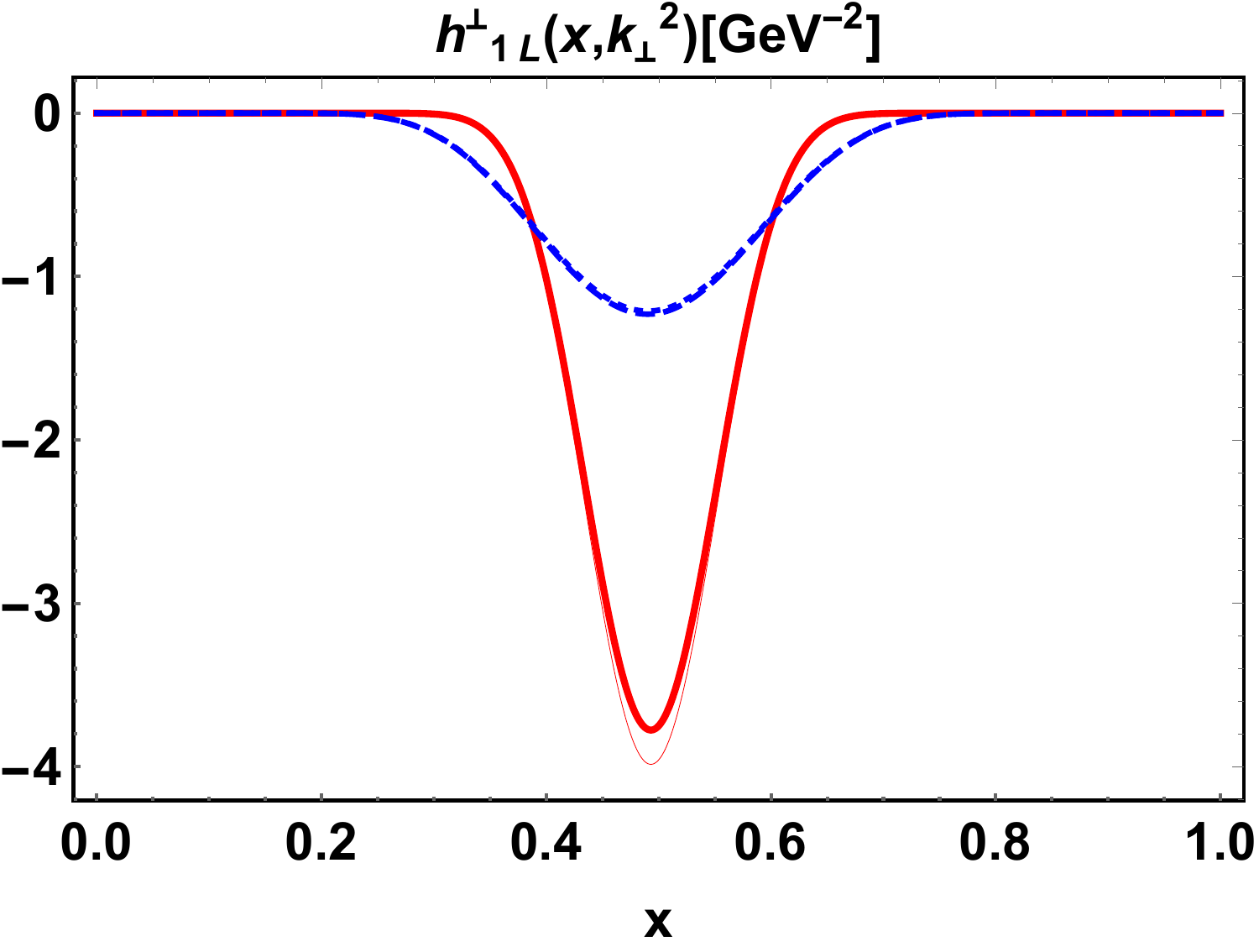}\end{center}
	\end{minipage}
	\begin{minipage}[c]{1\textwidth}\begin{center}
			(e)\includegraphics[width=.43\textwidth]{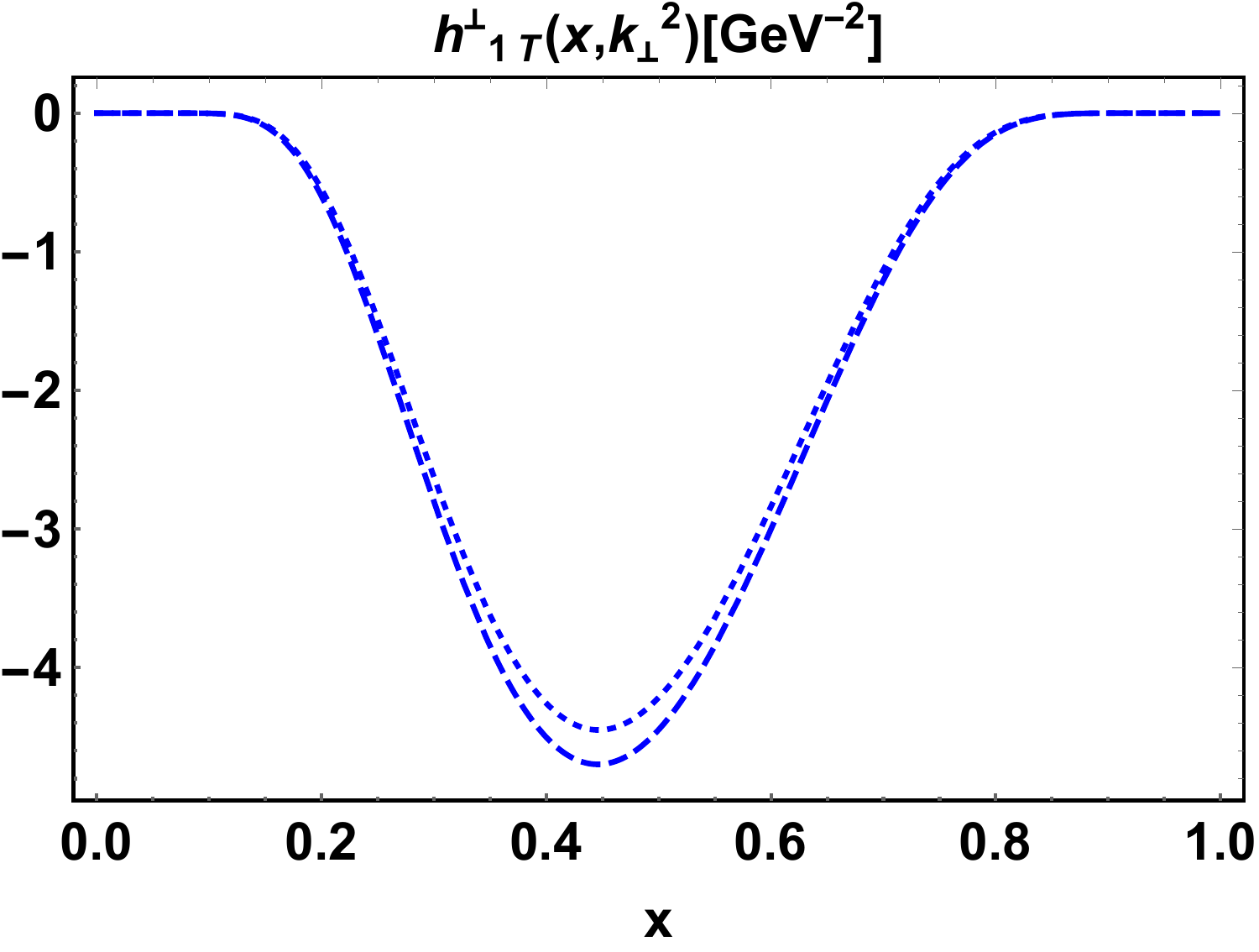}
               (f)\includegraphics[width=.45\textwidth]{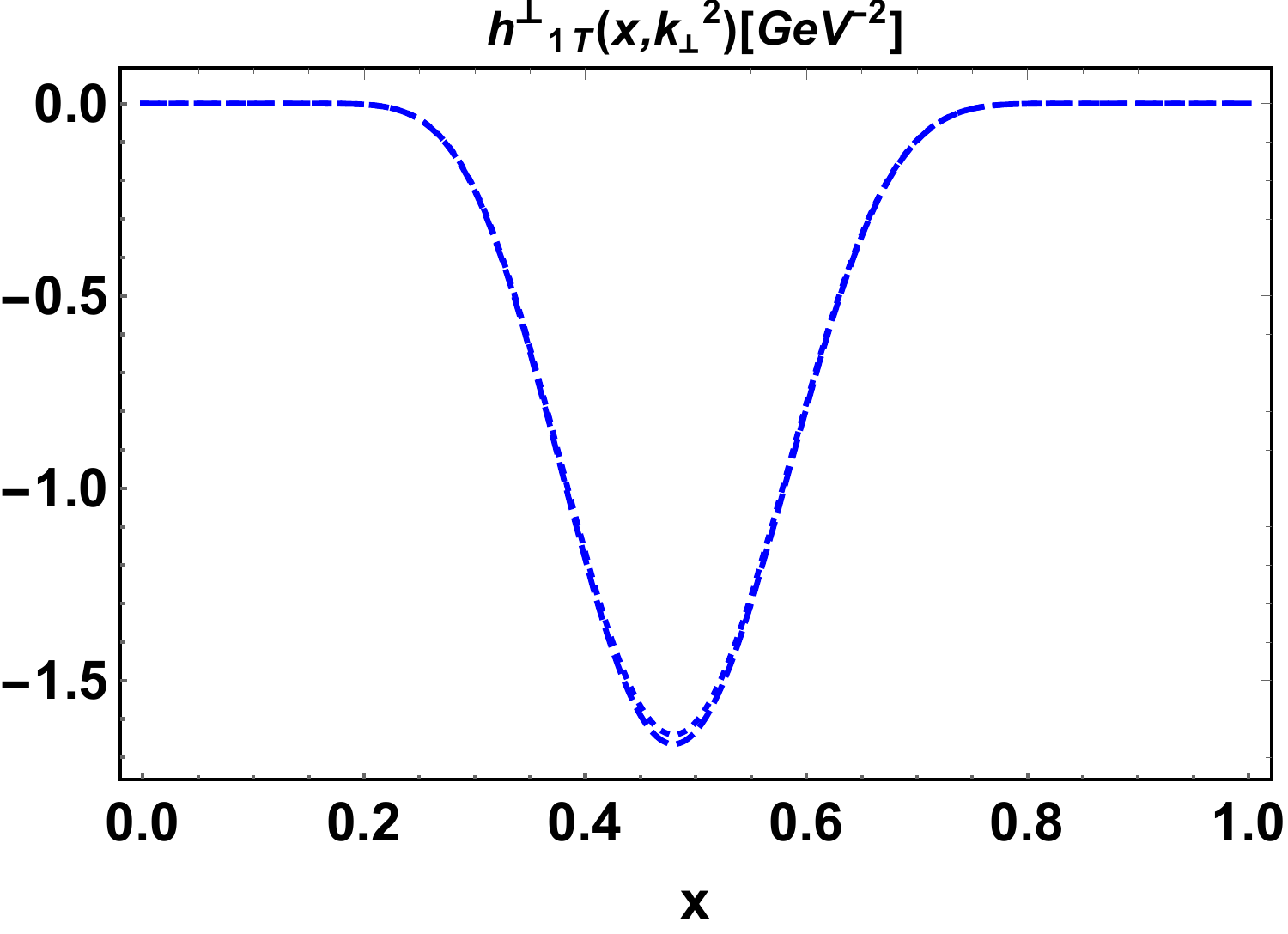}\end{center}
	\end{minipage}
	\caption{(Color online) $h_1(x,{\bf k}^2_\perp)$, $h^\perp_{1L}(x,{\bf k}^2_\perp)$ and $h^\perp_{1T}(x,{\bf k}^2_\perp)$ TMDs are plotted with respect to $x$ at different values of ${\bf k}^2_\perp$ i.e., ${\bf k}^2_\perp=0.1 {\ \rm GeV}^2$  and ${\bf k}^2_\perp=0.2 {\ \rm GeV}^2$. The solid red thick and thin curves represents the TMDs in LFHM, while blue dot and dashed curves are for LFQM. The TMDs comparison of $J/\Psi$ and $\Upsilon$ meson in the LFHM and LFQM at the model scale $\mu_{\rm LFHM}^2=0.20$ GeV$^2$ and $\mu_{\rm LFQM}^2=0.19$ GeV$^2$ in the left and right panels respectively.}
	\label{tmds3}
\end{figure}

\begin{figure}[ht]
	\begin{minipage}[c]{1\textwidth}\begin{center}
			(a)\includegraphics[width=.45\textwidth]{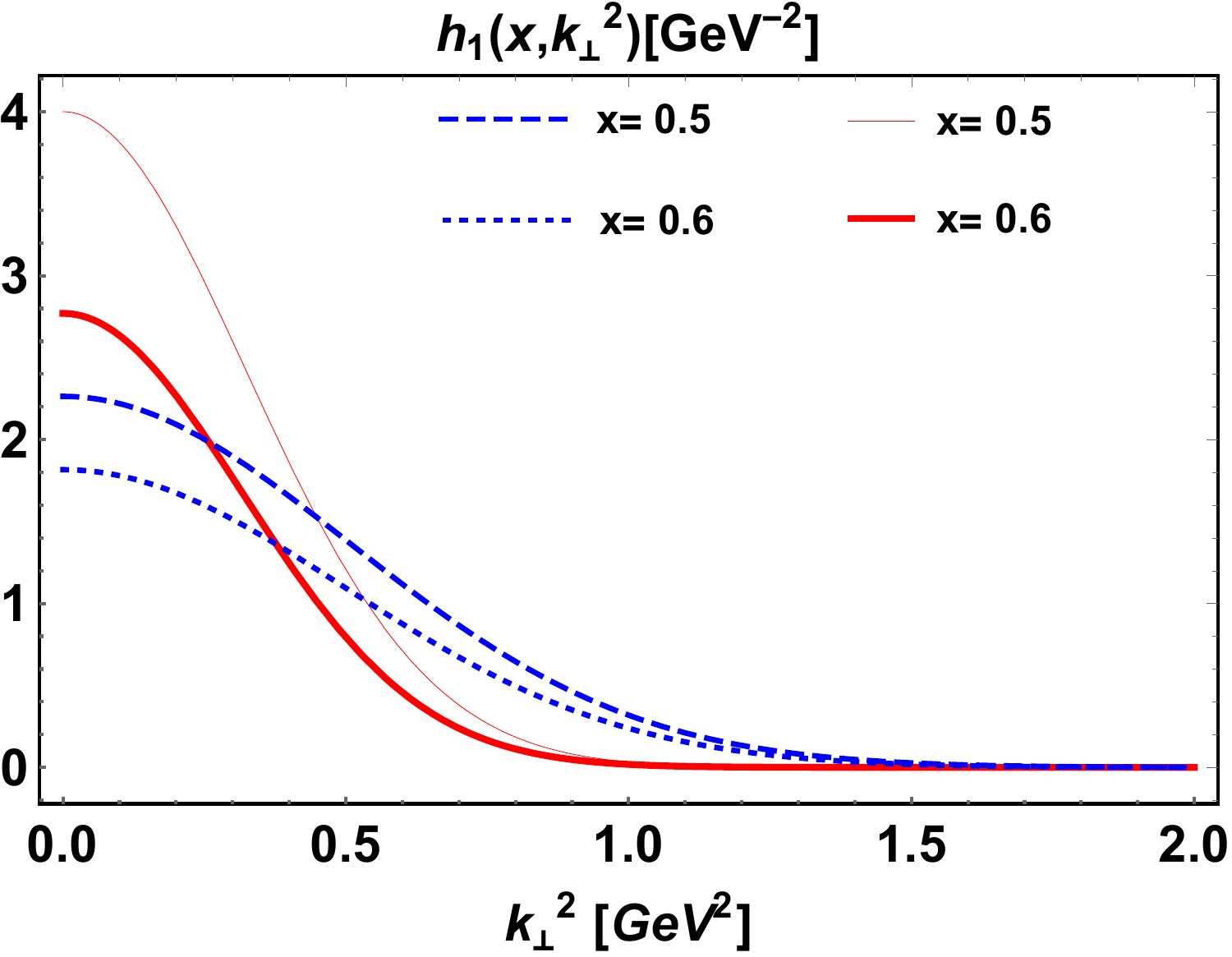}
			(b)\includegraphics[width=.47\textwidth]{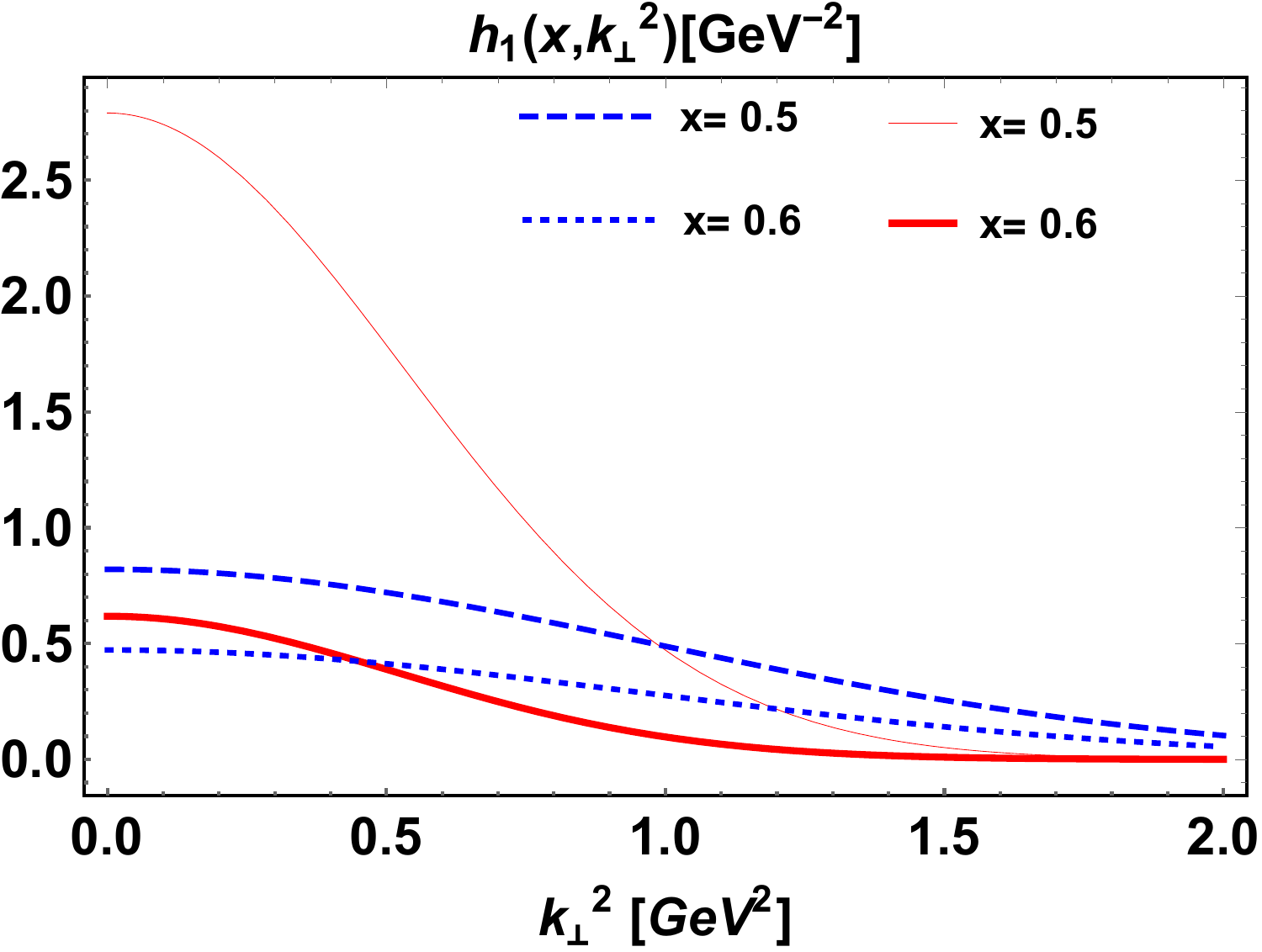}\end{center}
	\end{minipage}
	\begin{minipage}[c]{1\textwidth}\begin{center}
			(c)\includegraphics[width=.455\textwidth]{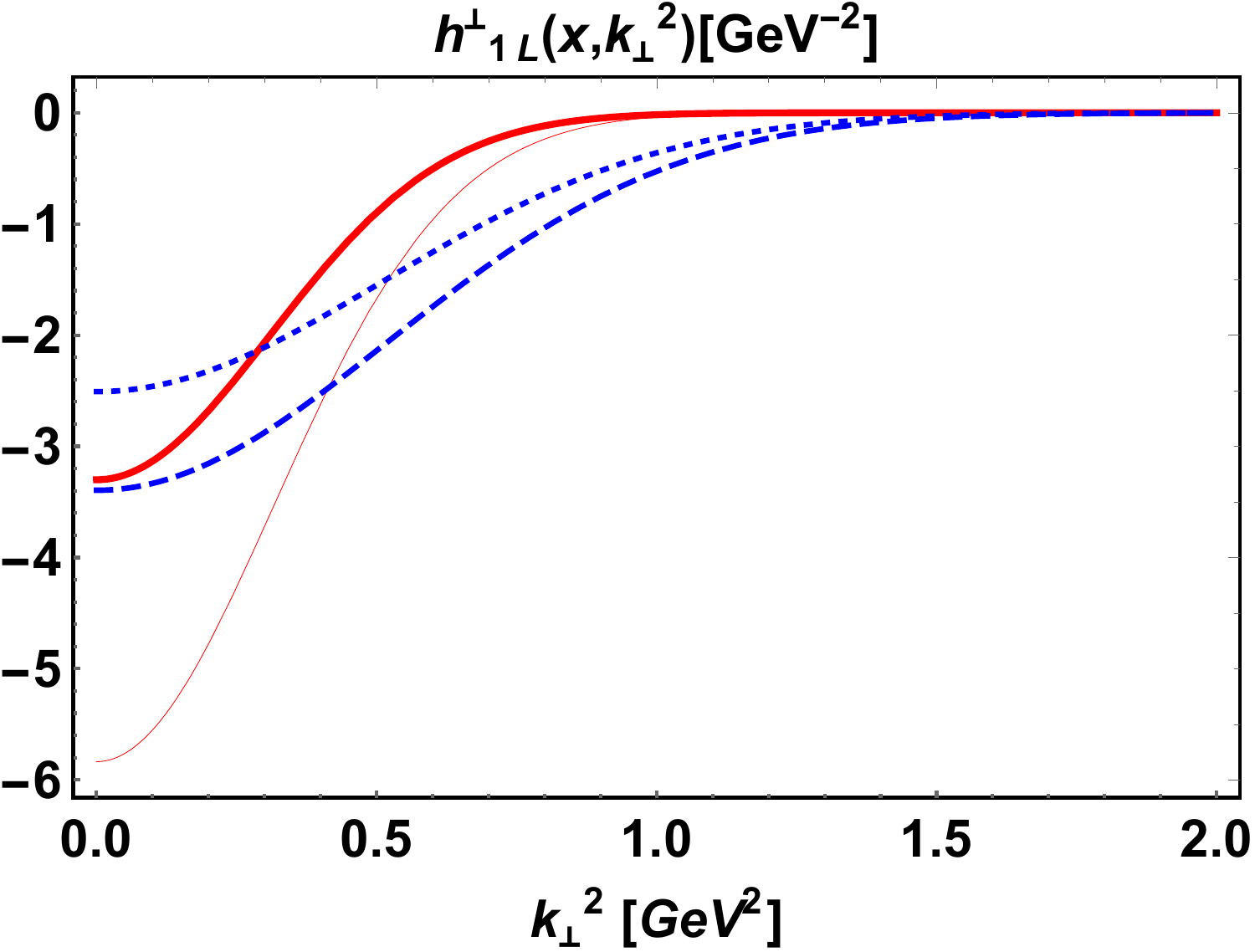}
			(d)\includegraphics[width=.46\textwidth]{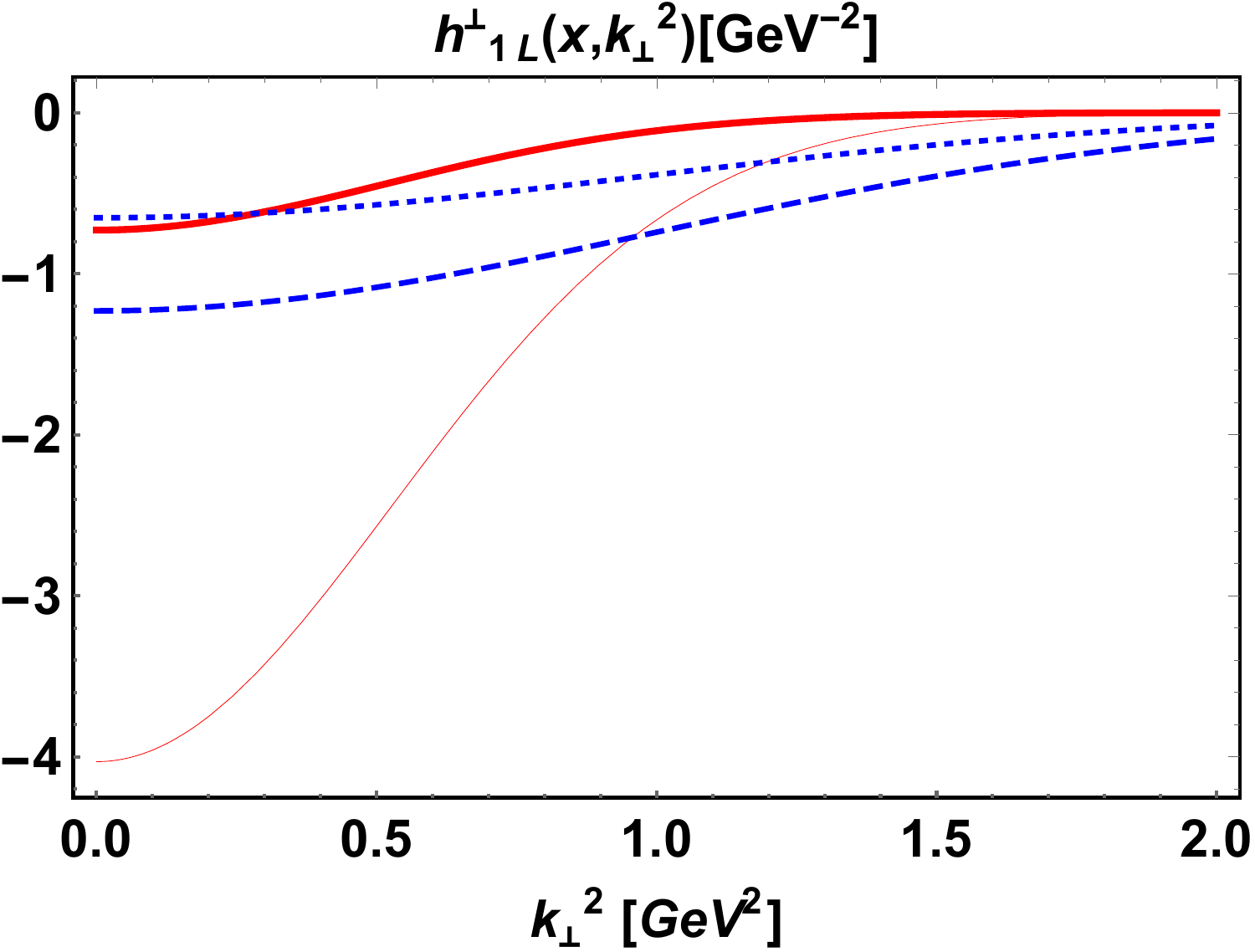}\end{center}
	\end{minipage}
	\begin{minipage}[c]{1\textwidth}\begin{center}
			(e)\includegraphics[width=.455\textwidth]{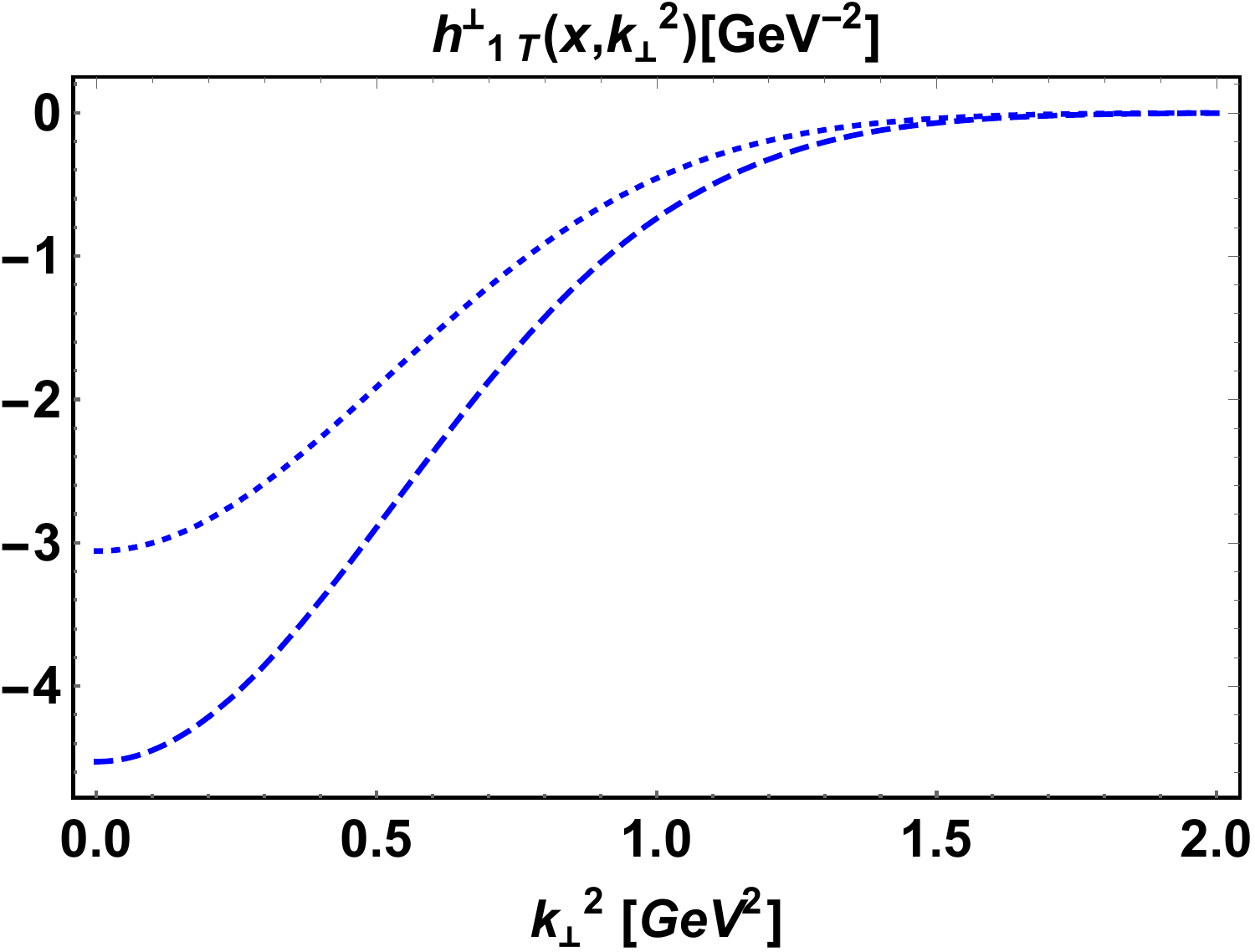}
                (f)\includegraphics[width=.472\textwidth]{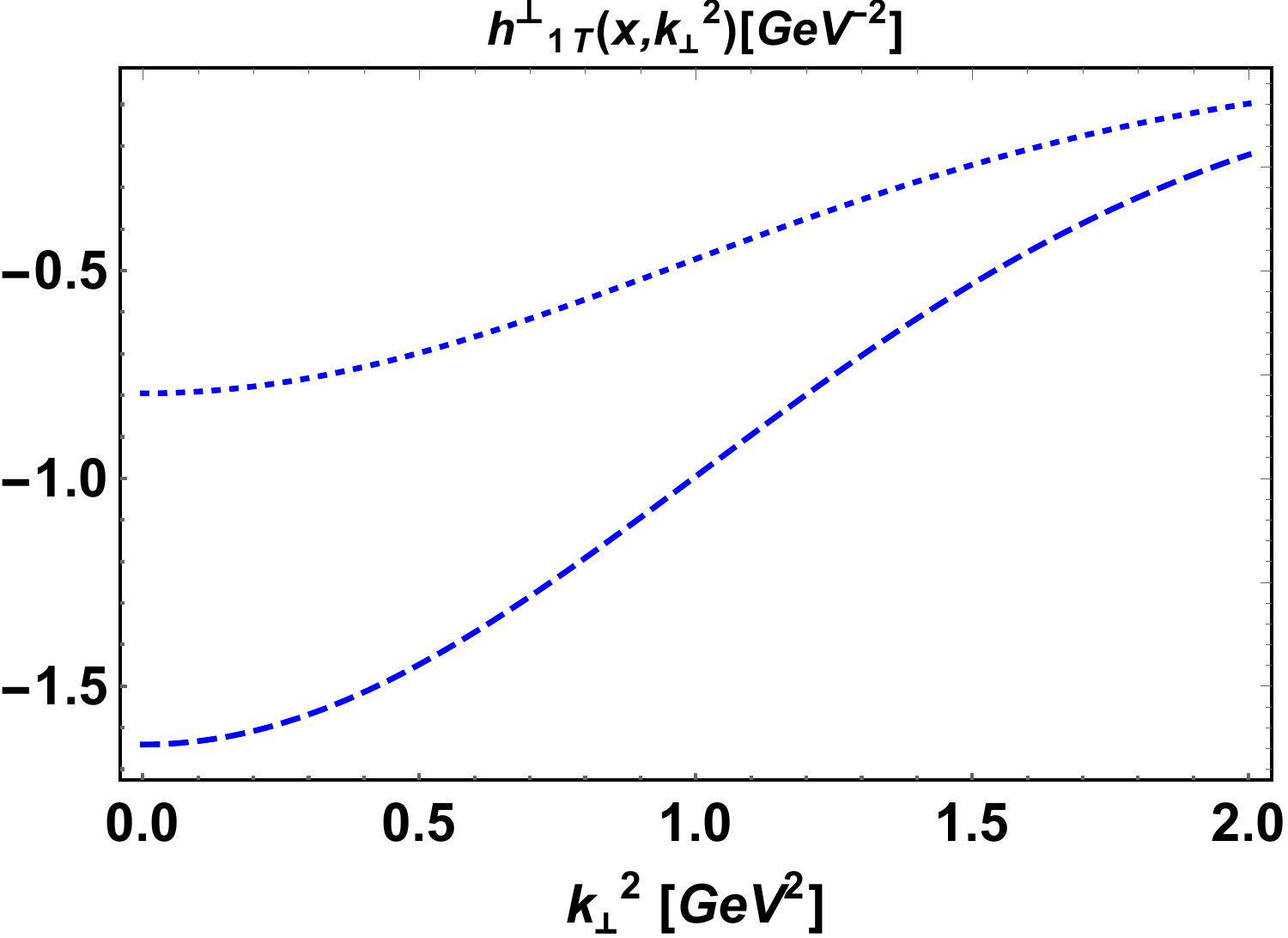}\end{center}
	\end{minipage}
	\caption{(Color online) $h_1(x,{\bf k}^2_\perp)$, $h^\perp_{1L}(x,{\bf k}^2_\perp)$ and $h^\perp_{1T}(x,{\bf k}^2_\perp)$ TMDs are plotted with respect to ${\bf k}^2_\perp$ at different values of $x$ i.e., $x=0.5$  and $x=0.6$. The solid red thick and thin curves represents the TMDs in LFHM, while blue dot and dashed curves are for LFQM. The TMDs comparison of $J/\Psi$ and $\Upsilon$ meson in the LFHM and LFQM at the model scale $\mu_{\rm LFHM}^2=0.20$ GeV$^2$ and $\mu_{\rm LFQM}^2=0.19$ GeV$^2$ in the left and right panels respectively.}
	\label{tmds4}
\end{figure}

\begin{figure}[ht]
	\begin{minipage}[c]{1\textwidth}\begin{center}
			(a)\includegraphics[width=.45\textwidth]{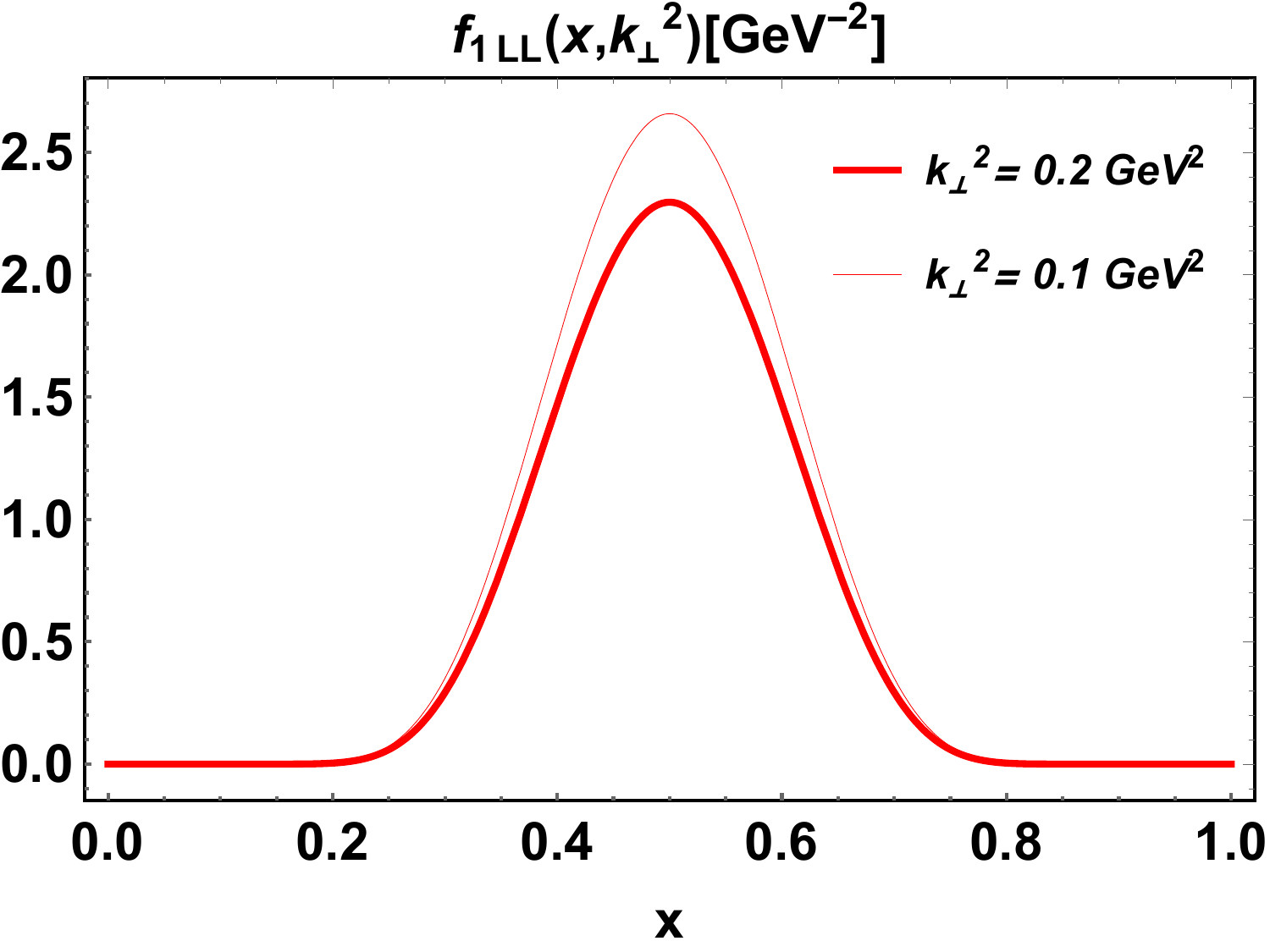}
			(b)\includegraphics[width=.45\textwidth]{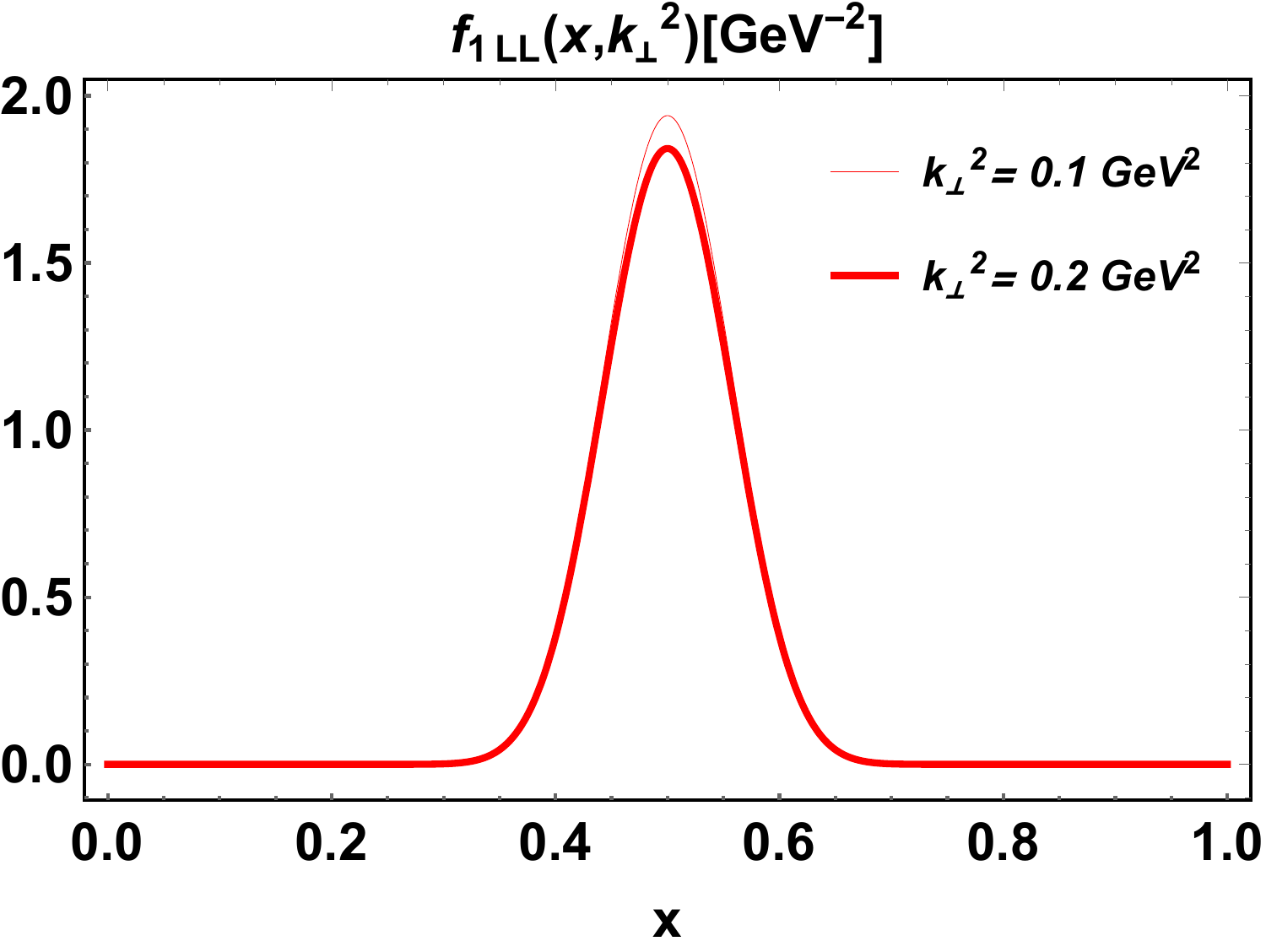}\end{center}
	\end{minipage}
	\begin{minipage}[c]{1\textwidth}\begin{center}
			(c)\includegraphics[width=.455\textwidth]{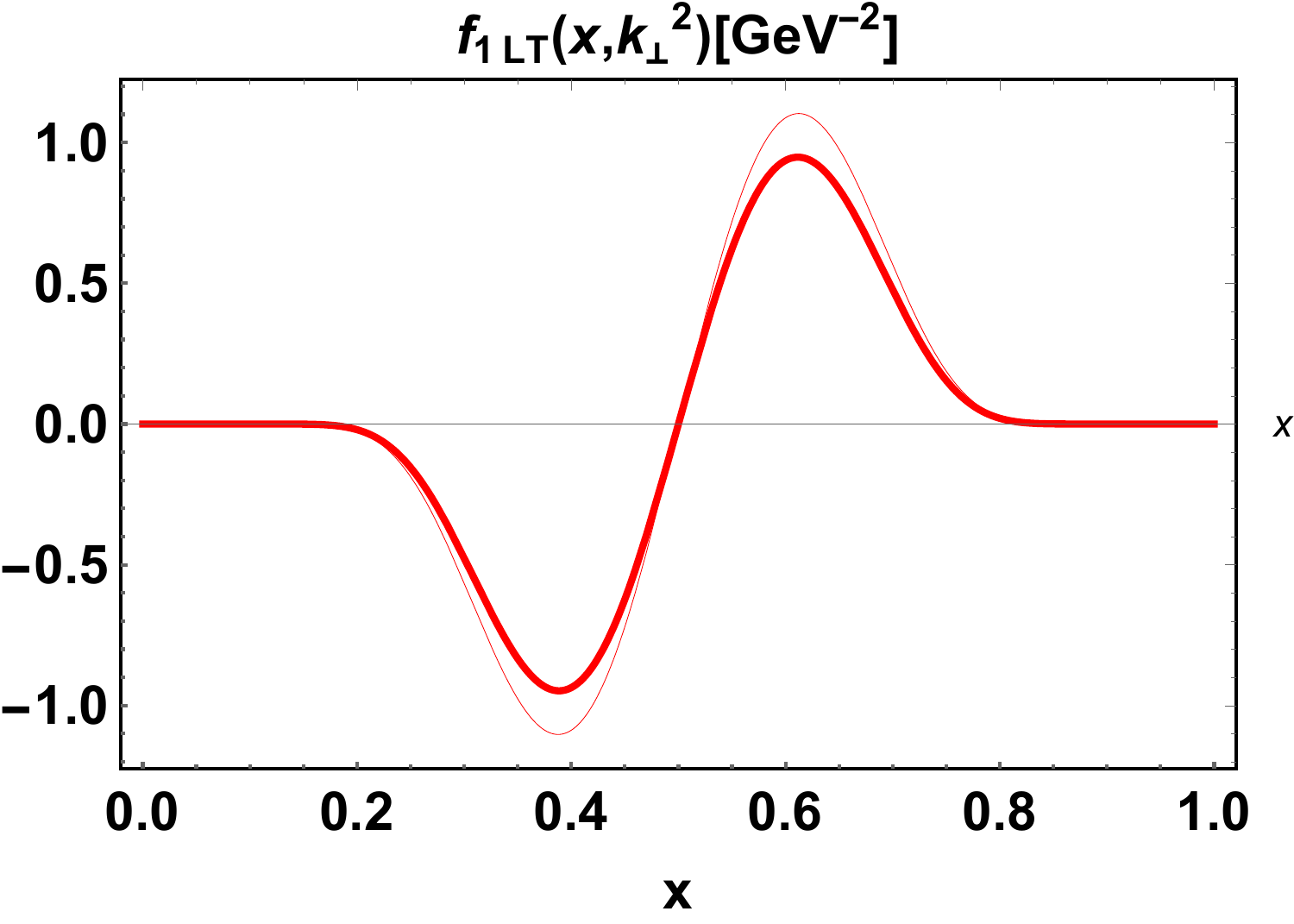}
			(d)\includegraphics[width=.46\textwidth]{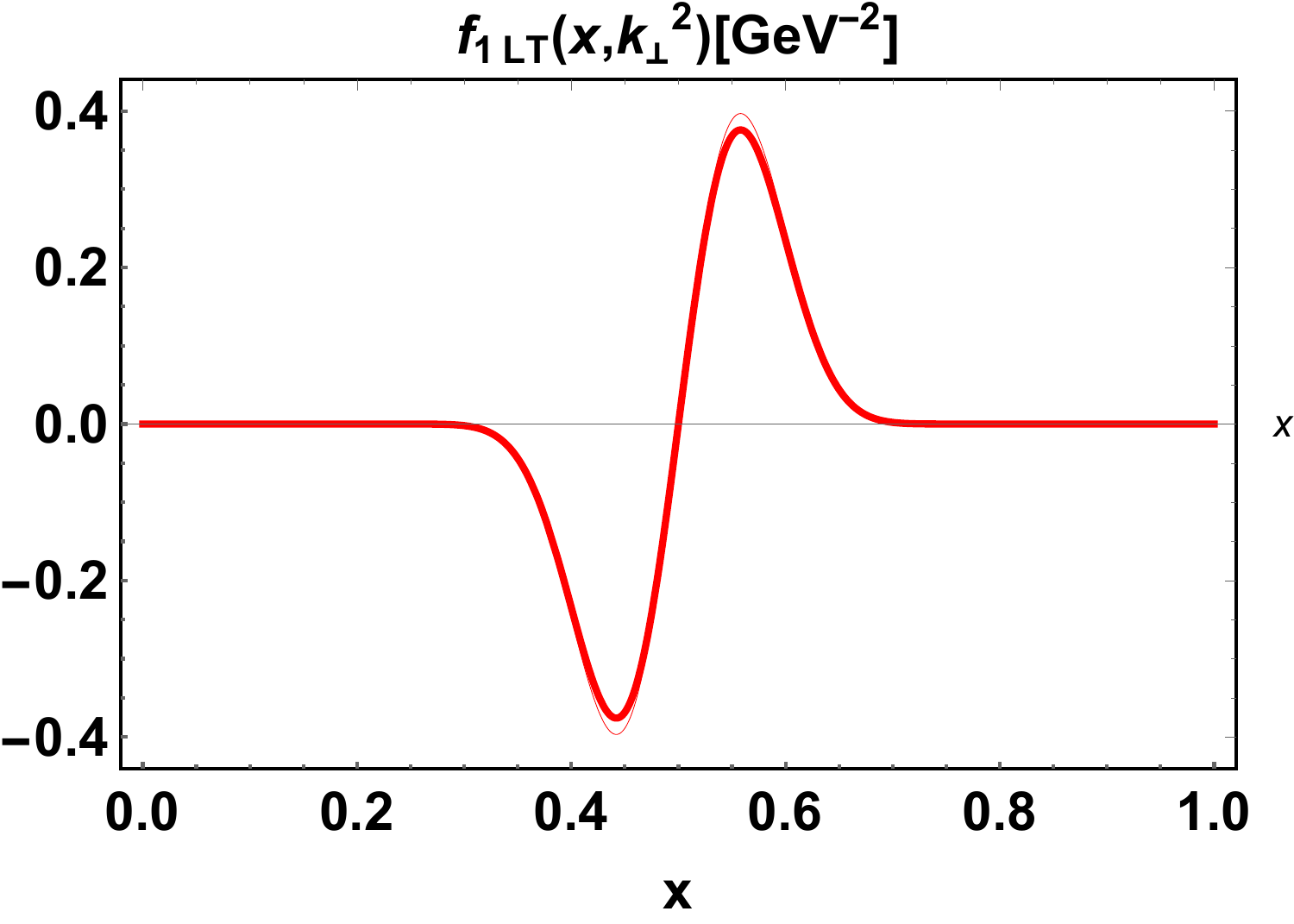}\end{center}
	\end{minipage}
	\begin{minipage}[c]{1\textwidth}\begin{center}
			(e)\includegraphics[width=.47\textwidth]{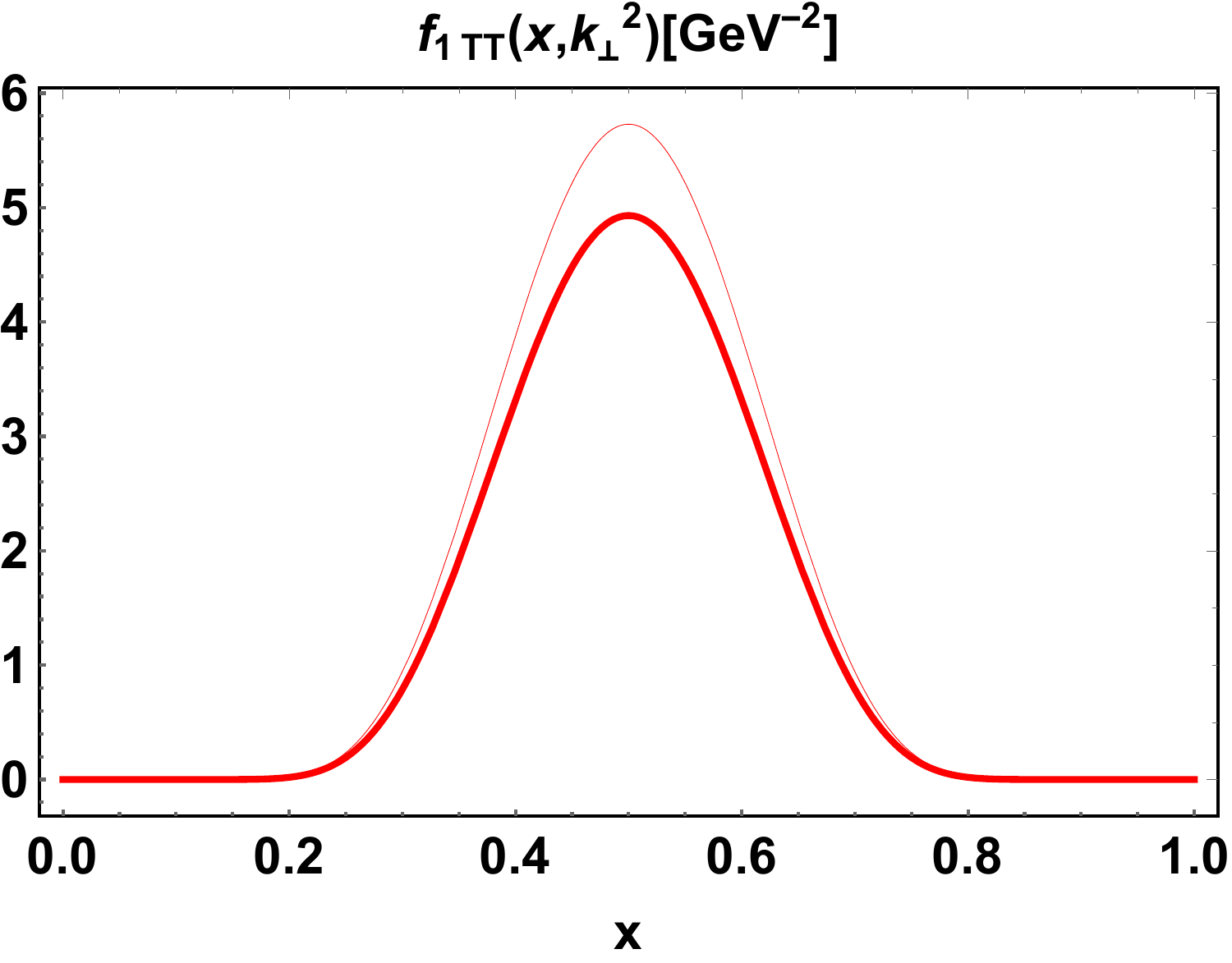}
			(f)\includegraphics[width=.472\textwidth]{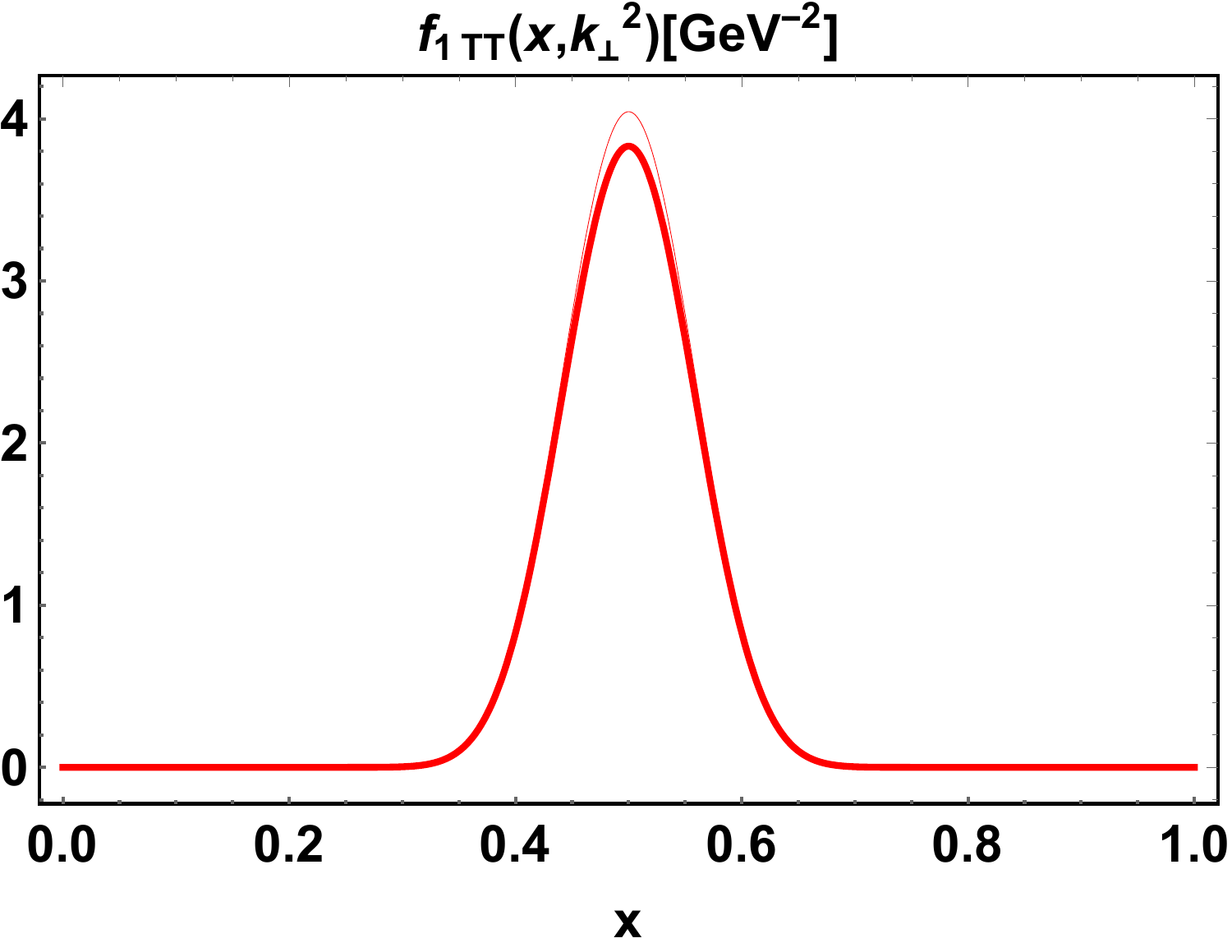}\end{center}
	\end{minipage}
	\caption{(Color online)  $f_{1LL}(x,{\bf k}^2_\perp)$, $f_{1LT}(x,{\bf k}^2_\perp)$ and $f_{1TT}(x,{\bf k}^2_\perp)$ TMDs are plotted with respect to $x$ at different values of ${\bf k}^2_\perp$ i.e., ${\bf k}^2_\perp=0.1 {\ \rm GeV}^2$ (solid thin curves) and ${\bf k}^2_\perp=0.2 {\ \rm GeV}^2$ (solid thick curves) for LFHM at model scale $\mu_{\rm LFHM}^2=0.20$ GeV$^2$ for $J/\psi$ (left panel) and $\Upsilon$ (right panel) meson respectively. }
	\label{tmds5}
\end{figure}

\begin{figure}[ht]
	\begin{minipage}[c]{1\textwidth}\begin{center}
			(a)\includegraphics[width=.45\textwidth]{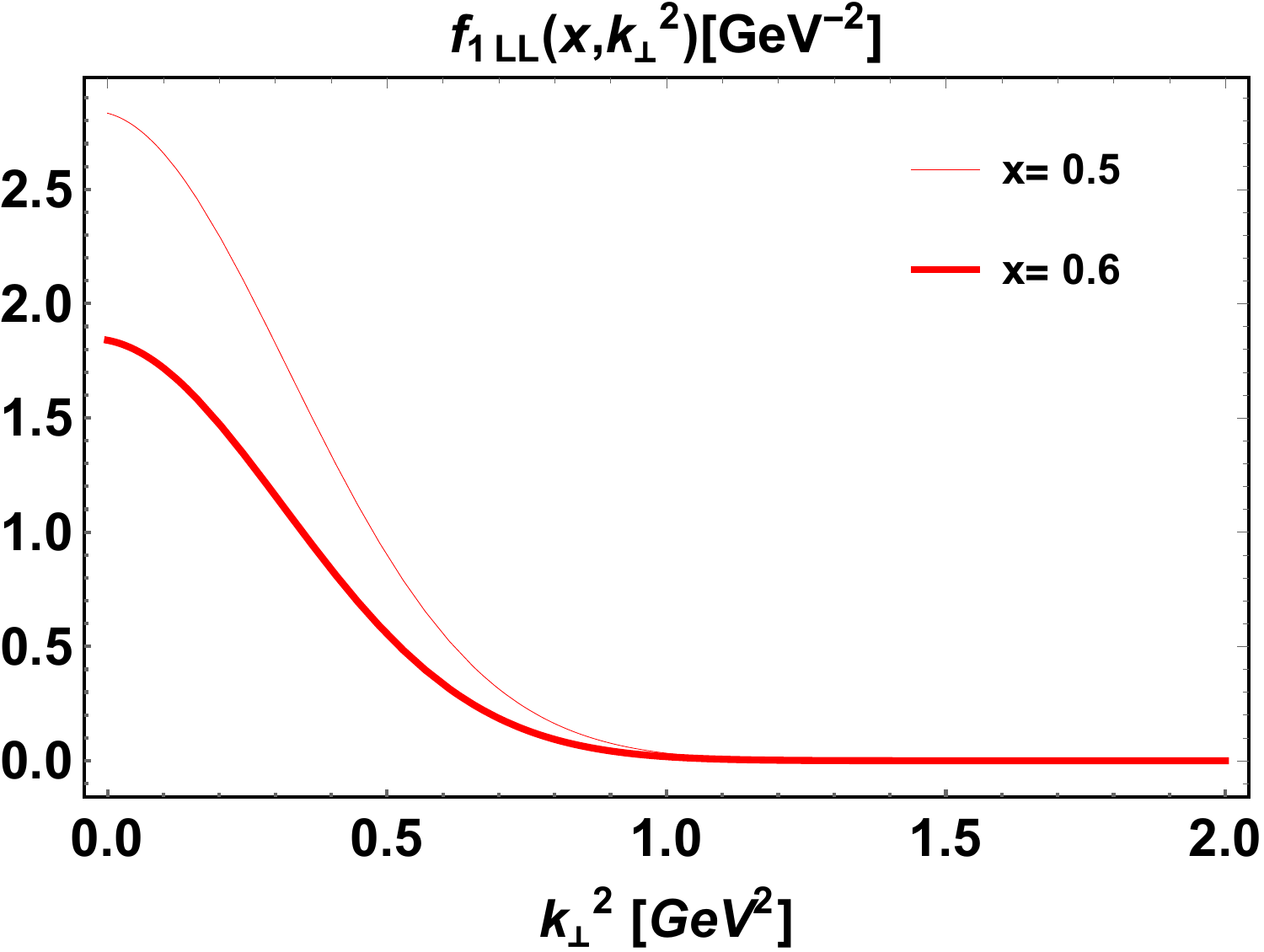}
			(b)\includegraphics[width=.45\textwidth]{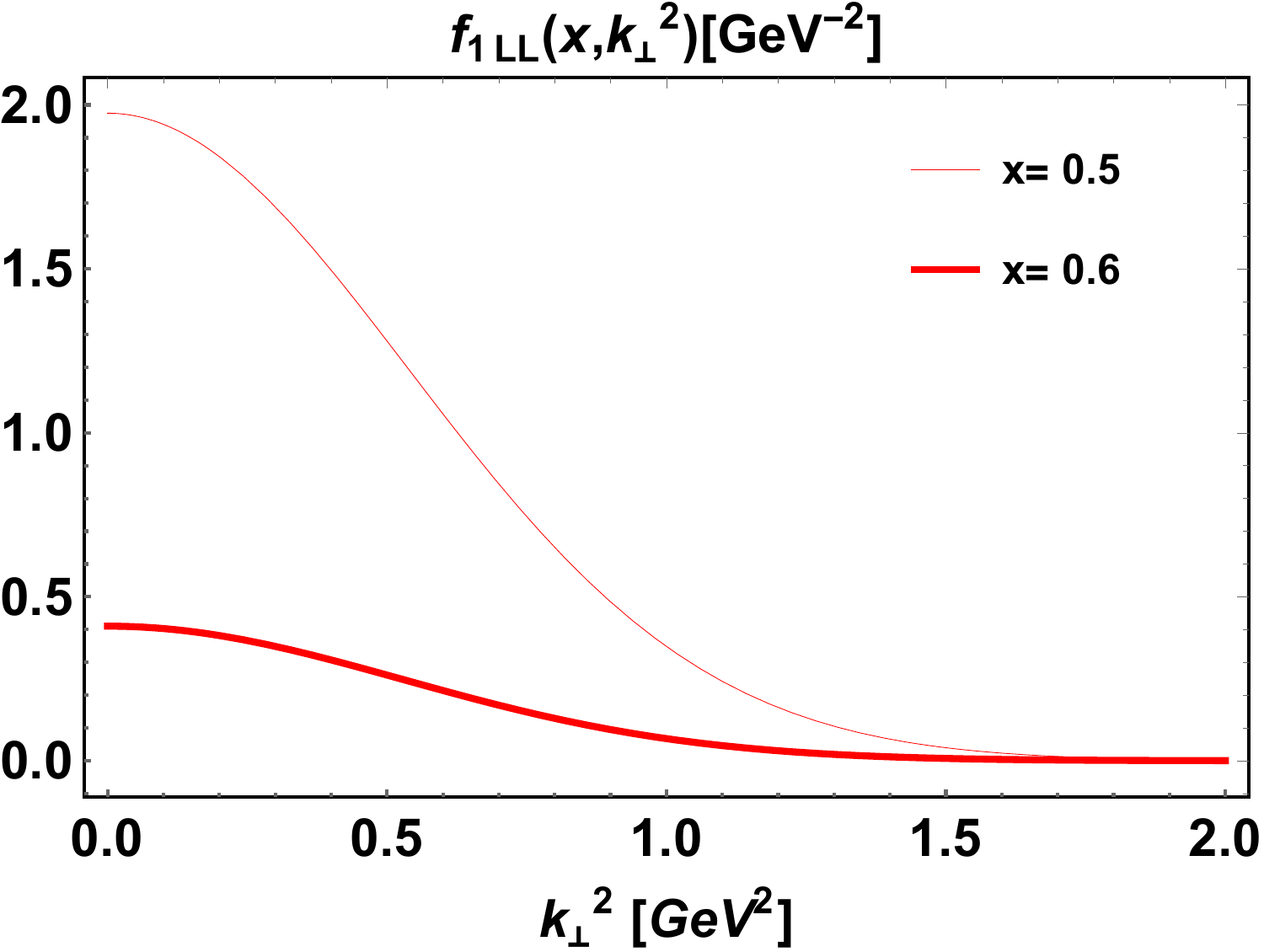}\end{center}
	\end{minipage}
	\begin{minipage}[c]{1\textwidth}\begin{center}
			(c)\includegraphics[width=.455\textwidth]{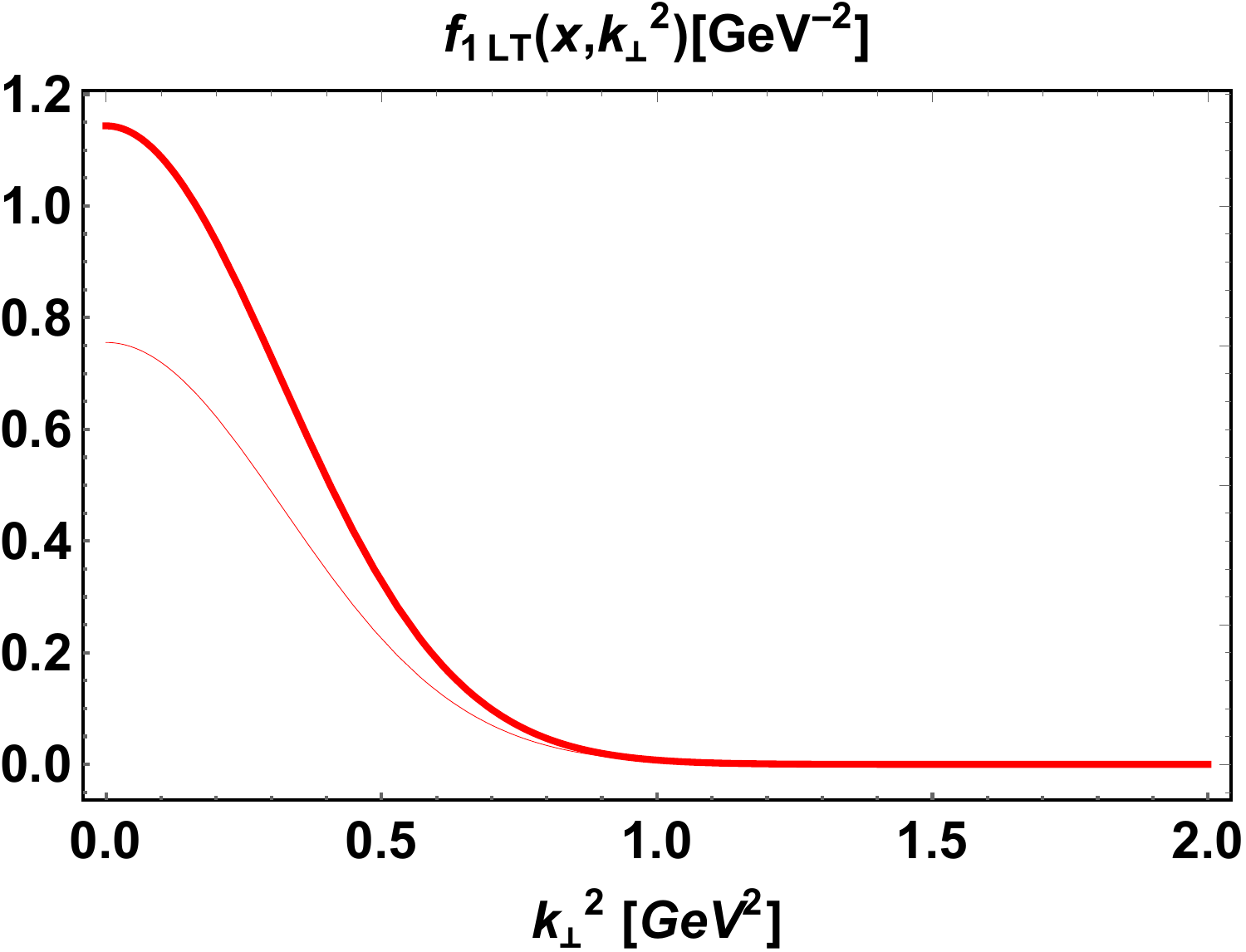}
			(d)\includegraphics[width=.46\textwidth]{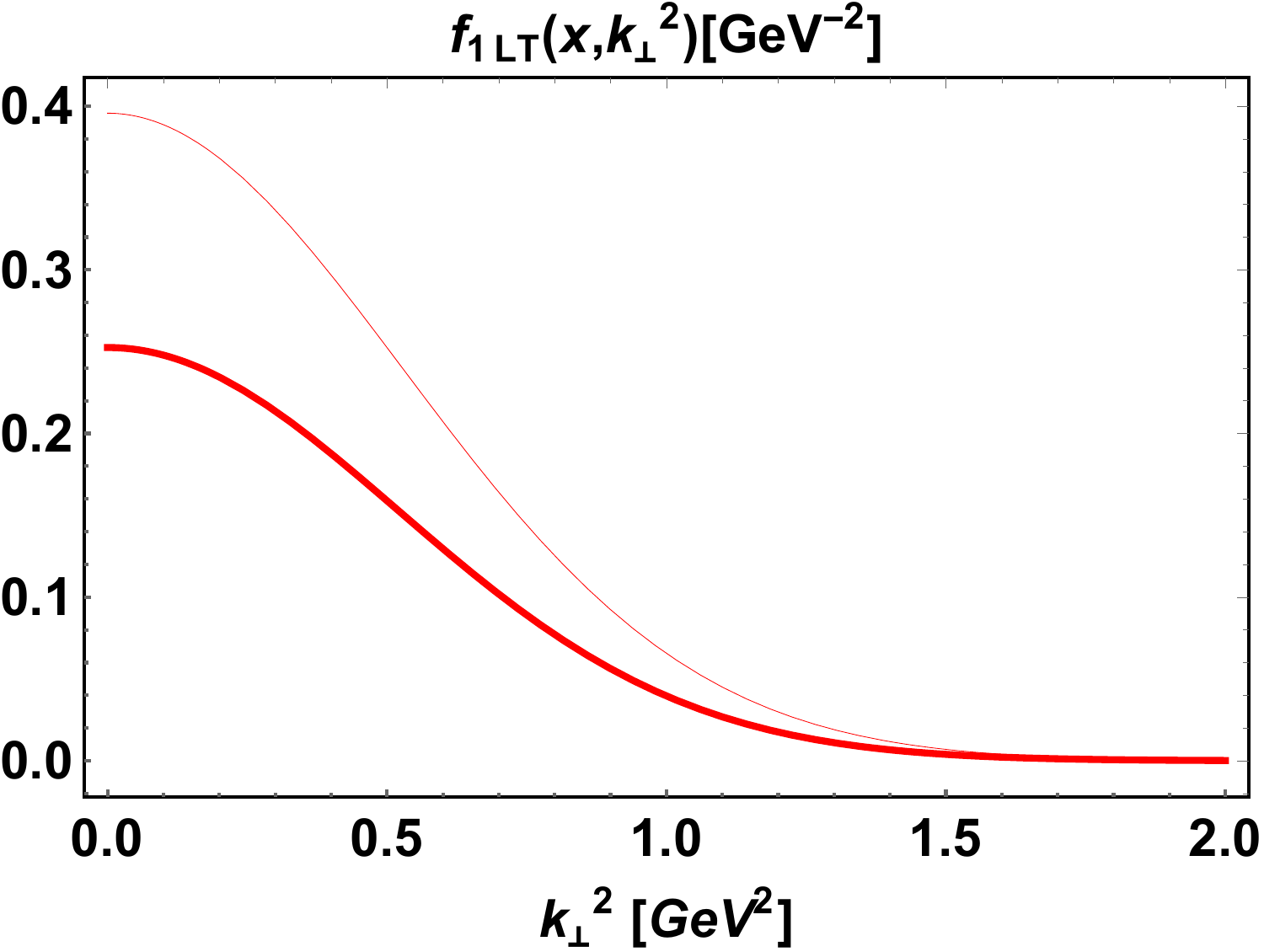}\end{center}
	\end{minipage}
	\begin{minipage}[c]{1\textwidth}\begin{center}
			(e)\includegraphics[width=.47\textwidth]{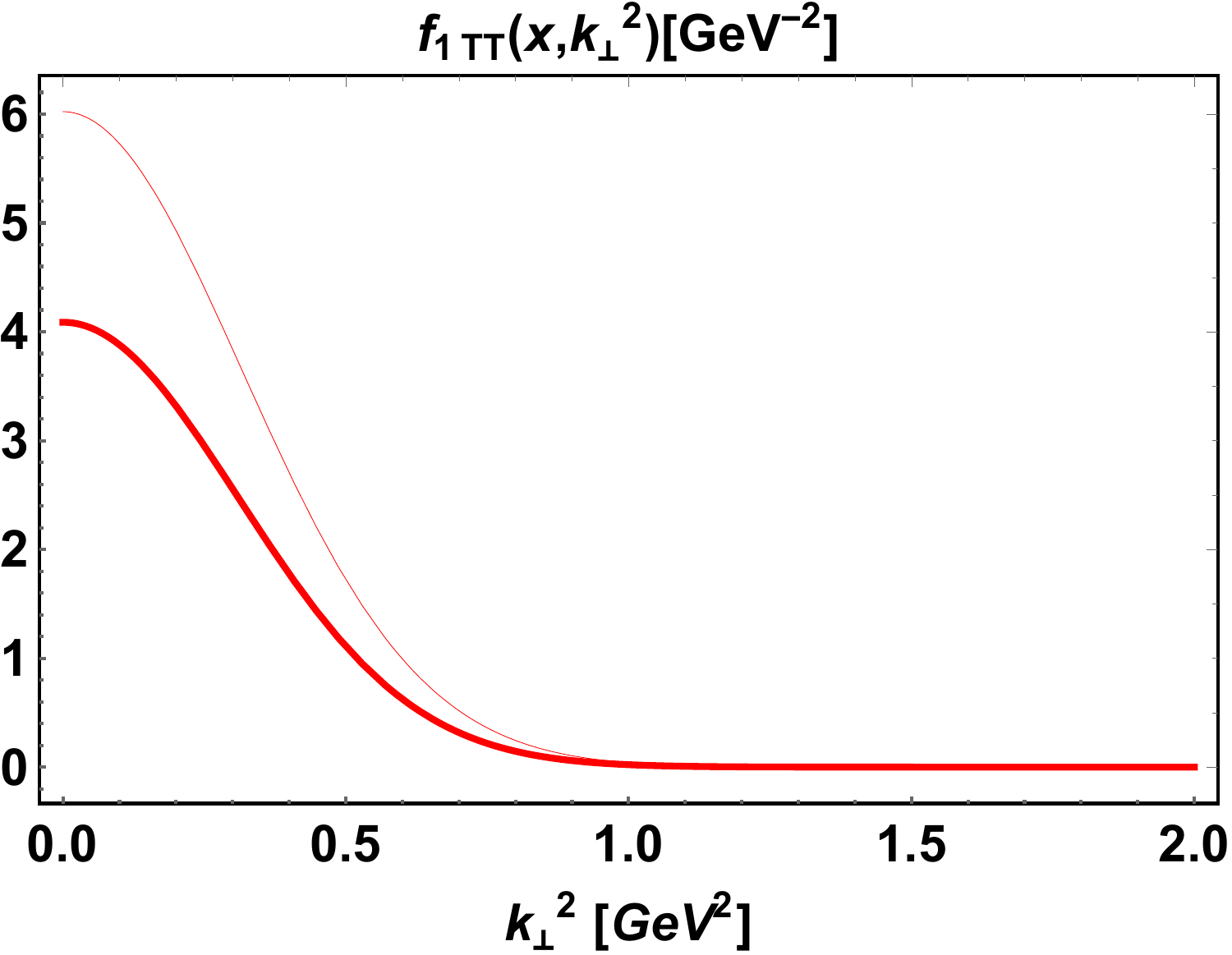}
			(f)\includegraphics[width=.472\textwidth]{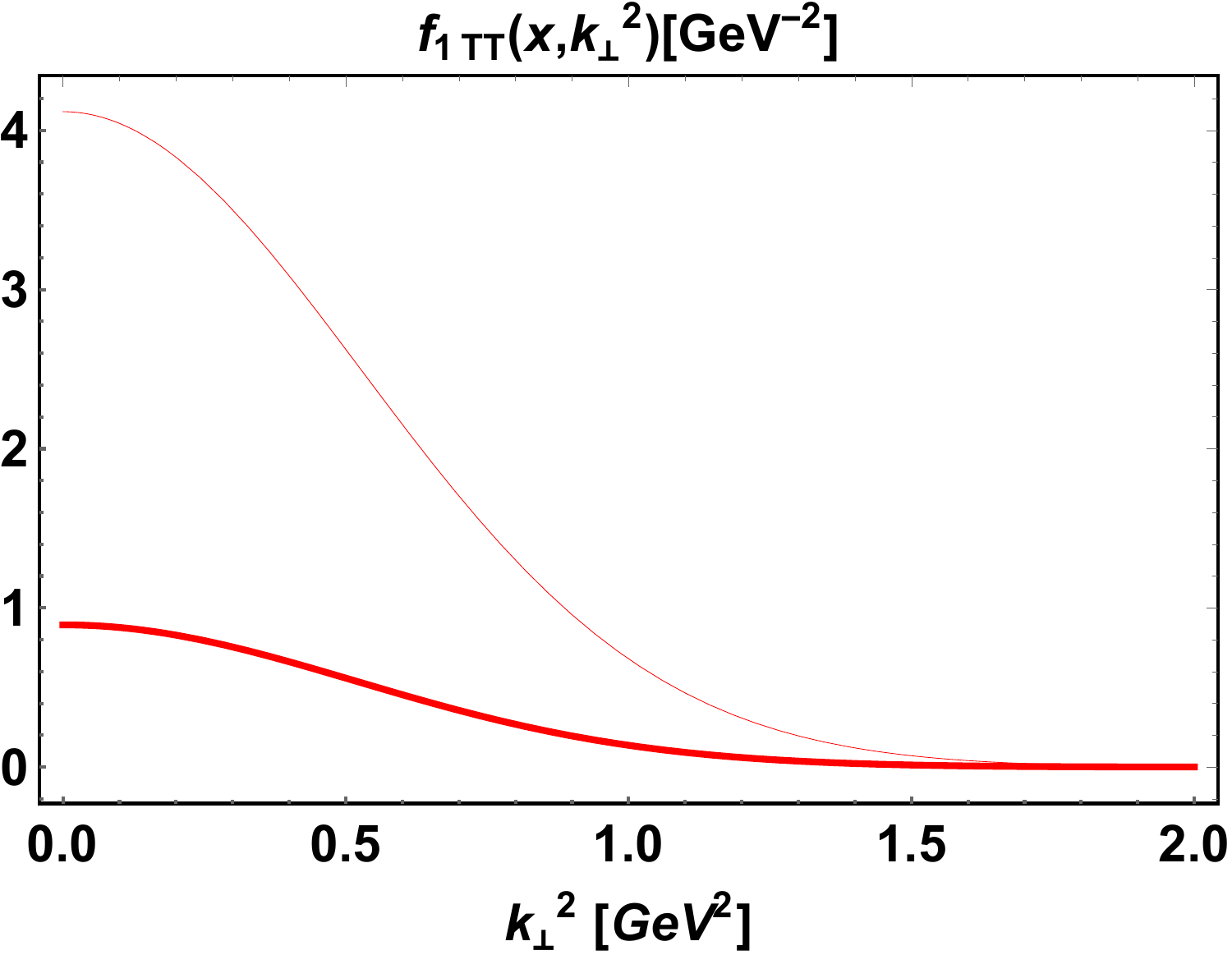}\end{center}
	\end{minipage}
	\caption{(Color online)  $f_{1LL}(x,{\bf k}^2_\perp)$, $f_{1LT}(x,{\bf k}^2_\perp)$ and $f_{1TT}(x,{\bf k}^2_\perp)$ TMDs are plotted with respect to  ${\bf k}^2_\perp$ at $x=0.5$ (solid thin lines) and $0.6$ (solid thick lines) for LFHM at model scale $\mu_{\rm LFHM}^2=0.20$ GeV$^2$ for $J/\psi$ (left panel) and $\Upsilon$ (right panel) meson respectively.}
	\label{tmds51}
\end{figure}

For the numerical predictions of $J/\psi$ and $\Upsilon$ meson TMDs in the  LFHM, we have used the universal AdS/QCD scale $\kappa=0.894$ GeV with quark mass $m_{c}=1.5$ GeV for the case of $J/\psi$-meson \cite{Swarnkar:2015osa} and $\kappa=1.49$ GeV with quark mass $m_{b}=4.63$ GeV for the case of $\Upsilon$-meson \cite{Nielsen:2018ytt}. Similarly, in the LFQM, we have used $\beta=0.699$ GeV with quark mass $m_{c}=1.68$ GeV for the case of $J/\psi$-meson and $\beta=1.376$ GeV with quark mass $m_{b}=5.10$ GeV for the case of $\Upsilon$-meson \cite{ll1}. In Figs. \ref{tmds}, \ref{tmds2}, \ref{tmds3}, \ref{tmds4} and \ref{tmds5}, we illustrate the $J/\psi$ and $\Upsilon$ meson TMDs in the LFHM results at model scale  $\mu_{\rm LFHM}^2=0.20$ GeV$^2$ and compare them with the LFQM at model scale $\mu_{\rm LFQM}^2=0.19$ GeV$^2$. The model scale has been taken from the Ref. \cite{rho}.
On the left side of the Figs. \ref{tmds}, \ref{tmds2}, \ref{tmds3}, \ref{tmds4}, \ref{tmds5} and \ref{tmds51}, we have presented the $2$D $J/\psi$ meson TMD's comparison in both LFHM and LFQM. While on the right side, $\Upsilon$ meson TMD's comparison with respect to $x$ and $\textbf{k}_{\perp}^2$ has been presented for both the models. In all the figures, we have used red (thick, thin) and blue (dashed,dotted) lines for LFHM and LFQM respectively.

In Fig. \ref{tmds}, we have presented the unpolarized quark TMD, $f_{1}(x,\textbf{k}_{\perp}^2)$ as well as the longitudinally polarized quark TMDs: $g_{1L}(x,{\bf k}^2_\perp)$ and $g_{1T}(x,{\bf k}^2_\perp)$ for both the particles in both the models. All the TMDs in Fig. \ref{tmds} are functions of $x$ at different fixed values of $\textbf{k}_{\perp}^2$. The qualitative behavior of $f_{1}(x,\textbf{k}_{\perp}^2)$, $g_{1L}(x,{\bf k}^2_\perp)$ and $g_{1T}(x,{\bf k}^2_\perp)$ in the LFHM  are consistent with those of the LFQM. This is true for both the $J/\psi$ and $\Upsilon$ mesons. Even though the behavior of all the plots look similar, however they have different peak values. In the case of LFHM, the particles have higher peak values compared to the LFQM. At $\bfk^2=0.2$ GeV$^2$, both $J/\psi$ and $\Upsilon$ have almost same peak values for $g_{1L}(x,{\bf k}^2_\perp)$ TMD in Fig. \ref{tmds}(c). The $g_{1T}(x,{\bf k}^2_\perp)$ TMD has higher peak value than the other TMDs and carry highest transverse momentum in LFHM. While in the case of LFQM, $h_{1T}(x,{\bf k}^2_\perp)$ has higher transverse momentum for both the particles. As we move towards higher quark masses values, the quark TMDs become narrower with $x$ and broader with $\textbf{k}_{\perp}^2$. As the $\Upsilon$ meson is less relativistic than $J/\psi$ meson, the TMDs become narrower  for $\Upsilon$ and broader for $J/\psi$. All the quark TMDs are centered at $x=0.5$ and low $\textbf{k}_{\perp}^2$. In Fig. \ref{tmds2}, we demonstrate   $f_{1}(x,\textbf{k}_{\perp}^2)$, $g_{1L}(x,{\bf k}^2_\perp)$ and $g_{1T}(x,{\bf k}^2_\perp)$ TMDs as a function of $\textbf{k}_{\perp}^2$ at different fixed values of $x$ in LFHM (left panel) and LFQM (right panel). These TMDs decrease monotonically with respect to $\textbf{k}_{\perp}^2$ indicating heavy quark and anti-quark have low relative momentum. All the figures have a similar trend in both the models. In Fig. \ref{tmds2}, the $J/\psi$ meson TMDs are found to be zero for ${\bf k}^2_{\perp} \ge 2$ GeV$^2$ for both the models. The heavy $\Upsilon$ carries higher transverse momentum than the $J/\Psi$-meson. The TMDs $f_{1}(x,\textbf{k}_{\perp}^2)$, $g_{1T}(x,{\bf k}^2_\perp)$ and  $g_{1L}(x,{\bf k}^2_\perp)$ describe the momentum distributions of the unpolarized quark in the unpolarized meson, the longitudinally polarized quark in the transversely polarized meson and the longitudinally polarized quark in the longitudinally polarized meson respectively. The S-wave is more dominant in the case of heavy vector meson TMDs  as compared to the P-wave and D-wave. Due to this dominance, all the TMDs in Fig. \ref{tmds} follow symmetry under $x \leftrightarrow (1-x)$ in the LFHM as well as in the LFQM. 

In Figs. \ref{tmds3} and \ref{tmds4}, we have discussed the transversely polarized TMDs,  $h_1(x,\textbf{k}_{\perp}^2)$, $h_{1L}^\perp (x,\textbf{k}_{\perp}^2)$ and $h_{1T}^\perp(x,\textbf{k}_{\perp}^2)$ with respect to $x$ and ${\bf k}^2_{\perp}$. In the LFHM, $h_{1T}^\perp(x,\textbf{k}_{\perp}^2)$ TMD is zero \cite{NJL,rho}, therefore the results have been presented only for the case of LFQM in Figs. \ref{tmds3}(e) and  \ref{tmds4}(e). The TMDs $h_1(x,\textbf{k}_{\perp}^2)$ and $h_{1L}^\perp(x,\textbf{k}_{\perp}^2)$ describe the momentum distribution of the transversely polarized quark in a transversely and longitudinally polarized meson respectively. $h_{1T}^\perp(x,\textbf{k}_{\perp}^2)$ TMD describes the momentum distribution when both the meson and quark are transversely polarized and their polarizations are further perpendicular to each other. The TMD $h_1(x,\textbf{k}_{\perp}^2)$ shows symmetry under $x$. There is one unit of OAM transfer occur for $h_{1L}^\perp(x,\textbf{k}_{\perp}^2)$ and $g_{1T}(x,{\bf k}^2_\perp)$ TMDs between final and initial state of hadron. In Fig. \ref{tmds3}(c), the $h_{1L}^\perp(x,\textbf{k}_{\perp}^2)$ TMD shows a negative distribution as is clear from Eq. (\ref{hlp-LFH}) in LFHM where $h_{1L}^\perp(x,\textbf{k}_{\perp}^2)$TMD has a negative ${N}^2_T$ in LFHM and for the LFQM this is due to the negative mass term in Eq. (\ref{h1lq}). Similarly in Fig. \ref{tmds4}, we have plotted
 $h_1(x,\textbf{k}_{\perp}^2)$, $h_{1L}^\perp(x,\textbf{k}_{\perp}^2)$ and $h_{1T}^\perp(x,\textbf{k}_{\perp}^2)$ as a function of ${\bf k}^2_{\perp}$ at fixed $x=0.5, 0.6$ value.

\begin{figure}[ht]
\centering
\begin{minipage}[c]{1\textwidth}
\begin{center}
(a)\includegraphics[width=.42\textwidth]{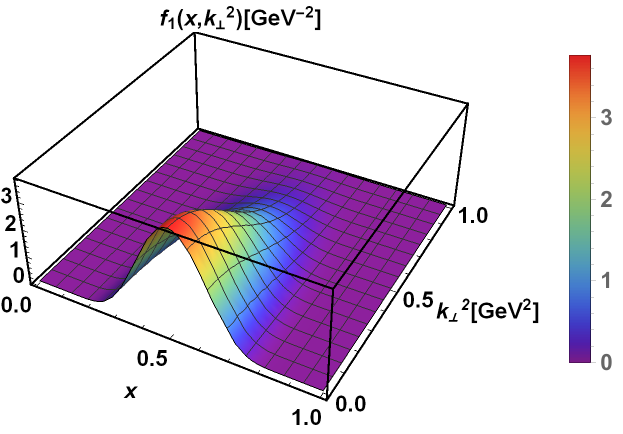}\hspace{1cm}
(b)\includegraphics[width=.45\textwidth]{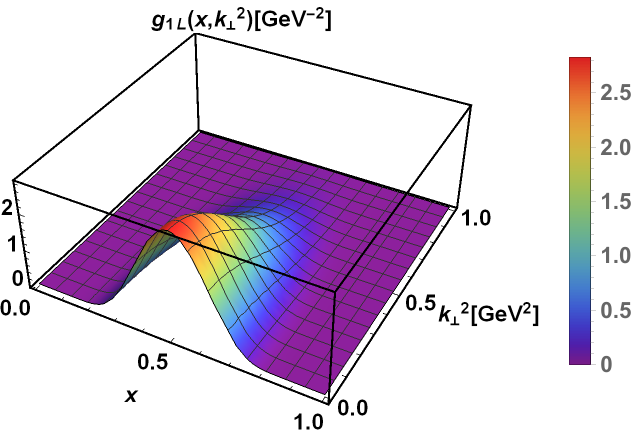}
\end{center}
\end{minipage}
\begin{minipage}[c]{1\textwidth}
\begin{center}
(c)\includegraphics[width=.43\textwidth]{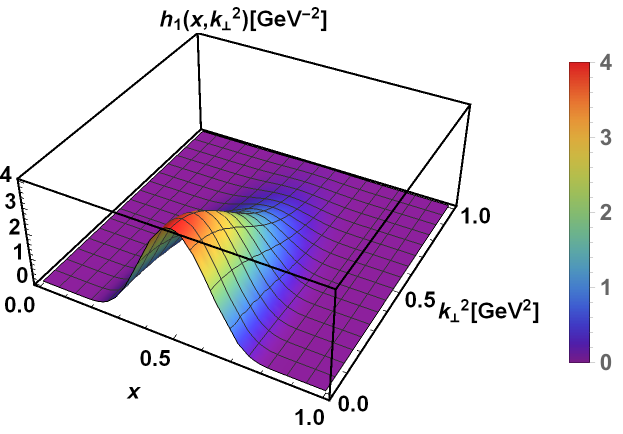}\hspace{1cm}
(d)\includegraphics[width=.44\textwidth]{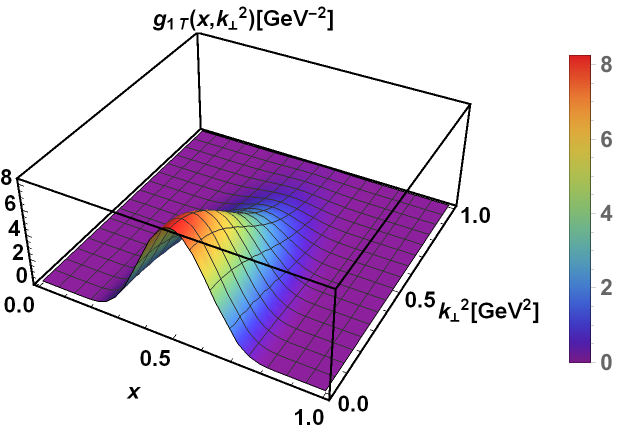}
\end{center}
\end{minipage}
\begin{minipage}[c]{1\textwidth}
\begin{center}
(e)\includegraphics[width=.44\textwidth]{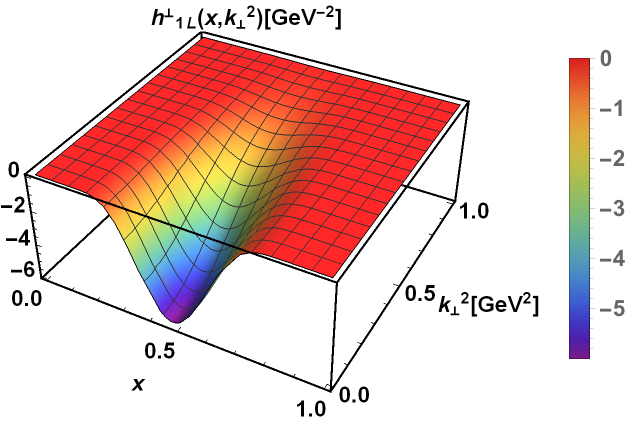}\hspace{1cm}
(f)\includegraphics[width=.44\textwidth]{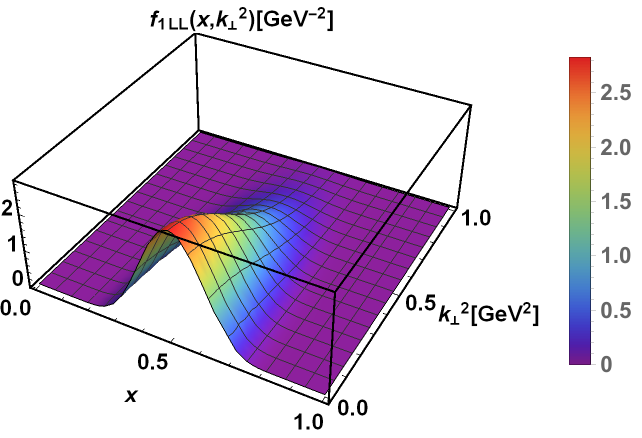}
\end{center}
\end{minipage}
\begin{minipage}[c]{1\textwidth}
\begin{center}
(g)\includegraphics[width=.44\textwidth]{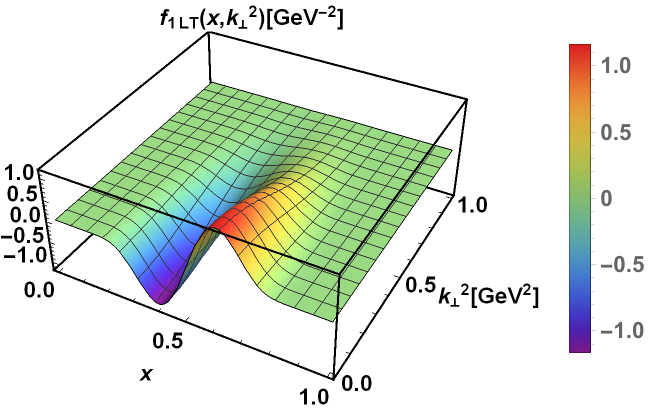}\hspace{1cm}
(h)\includegraphics[width=.43\textwidth]{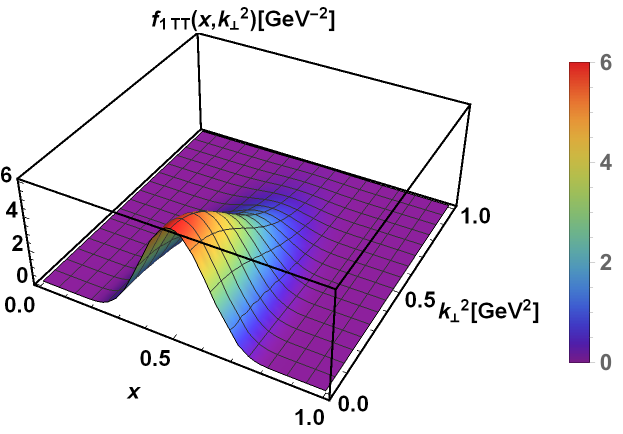}
\end{center}
\end{minipage}
\caption{(Color online) T-even TMDs for the $J/\psi$-meson as a function of $x$ and ${\bf k}^2_\perp$ in the LFHM at the model scale $\mu_{\rm LFHM}^2=0.20$ GeV$^2$.}
\label{3d-tmds}
\end{figure}

\begin{figure}[ht]
\centering
\begin{minipage}[c]{1\textwidth}
\begin{center}
(a)\includegraphics[width=.43\textwidth]{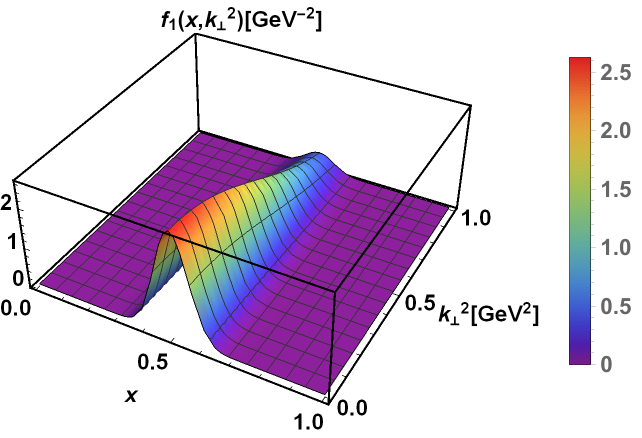}\hspace{1cm}
(b)\includegraphics[width=.44\textwidth]{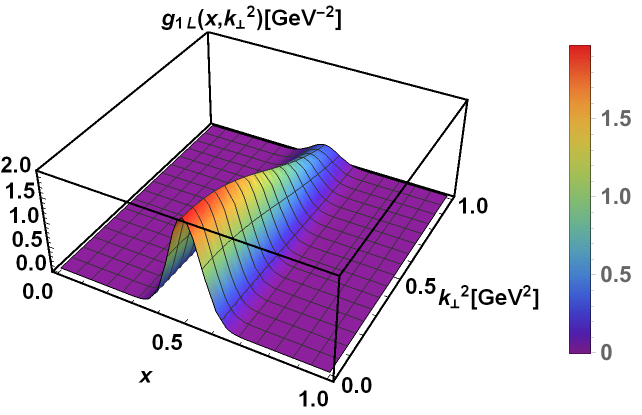}
\end{center}
\end{minipage}
\begin{minipage}[c]{1\textwidth}
\begin{center}
(c)\includegraphics[width=.44\textwidth]{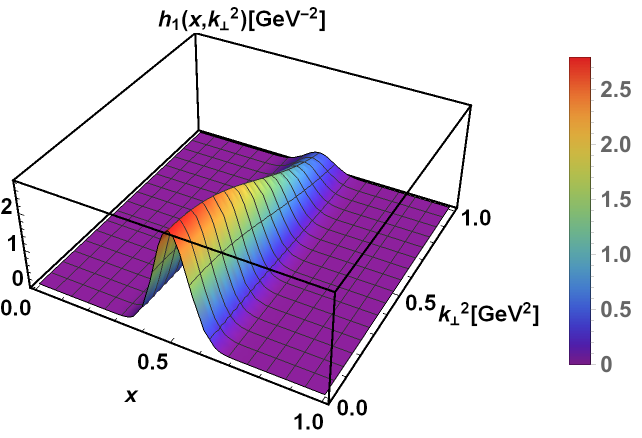}\hspace{1cm}
(d)\includegraphics[width=.44\textwidth]{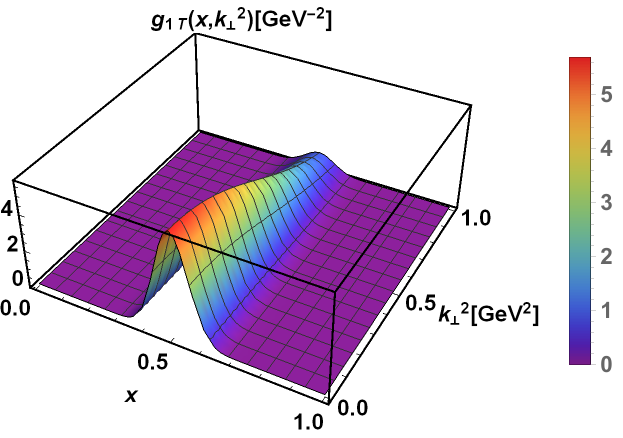}
\end{center}
\end{minipage}
\begin{minipage}[c]{1\textwidth}
\begin{center}
(e)\includegraphics[width=.44\textwidth]{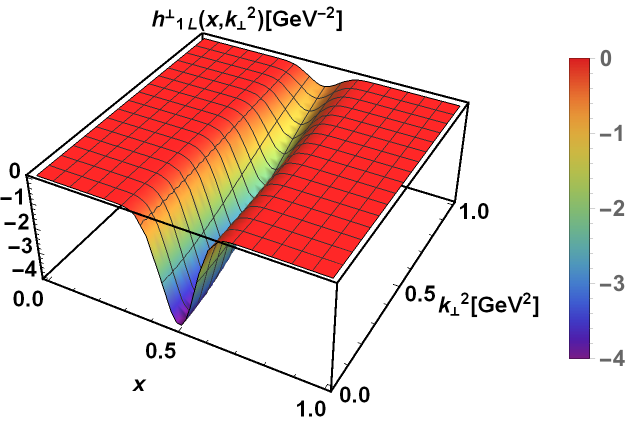}\hspace{1cm}
(f)\includegraphics[width=.44\textwidth]{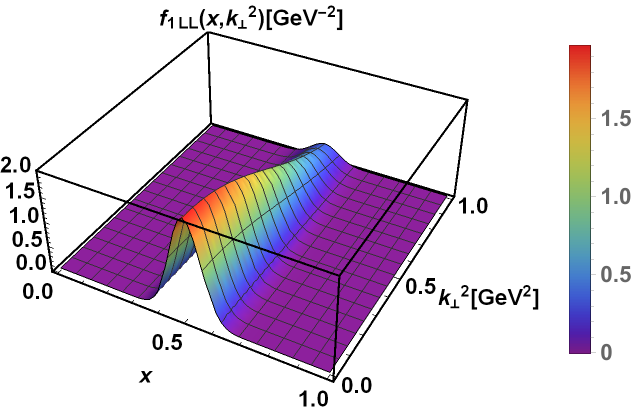}
\end{center}
\end{minipage}
\begin{minipage}[c]{1\textwidth}
\begin{center}
(g)\includegraphics[width=.45\textwidth]{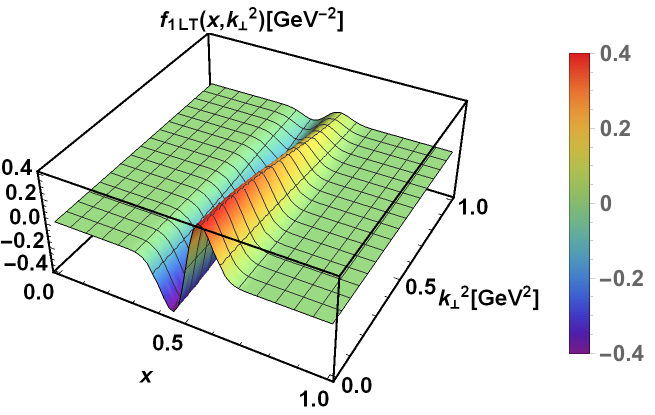}\hspace{1cm}
(h)\includegraphics[width=.41\textwidth]{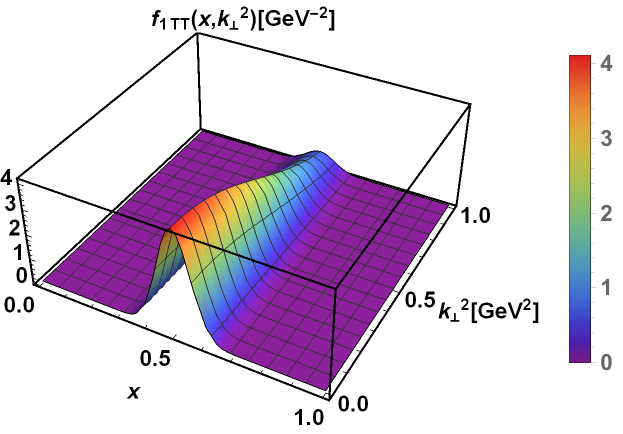}
\end{center}
\end{minipage}
\caption{(Color online) T-even TMDs for $\Upsilon$-meson as a function of $x$ and ${\bf k}^2_\perp$ in the LFHM at the model scale $\mu_{\rm LFHM}^2=0.20$ GeV$^2$.}
\label{3d-tmds2}
\end{figure}

\begin{figure}[ht]
\centering
\begin{minipage}[c]{1\textwidth}
\begin{center}
(a)\includegraphics[width=.43\textwidth]{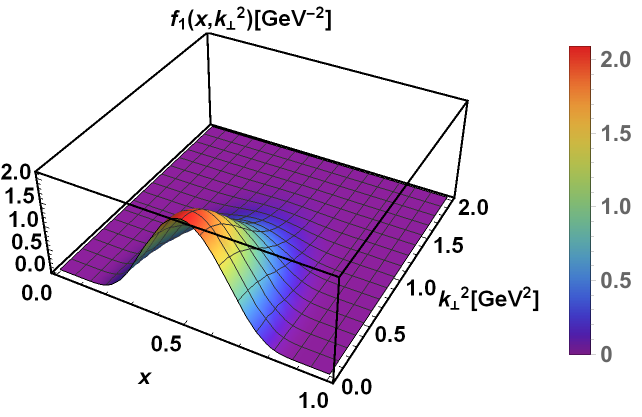}\hspace{1cm}
(b)\includegraphics[width=.45\textwidth]{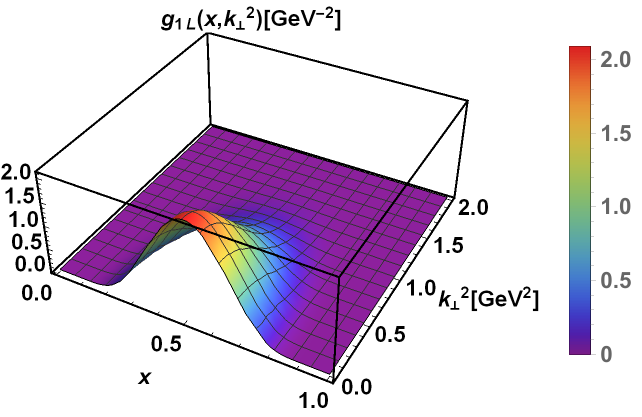}
\end{center}
\end{minipage}
\begin{minipage}[c]{1\textwidth}
\begin{center}
(c)\includegraphics[width=.44\textwidth]{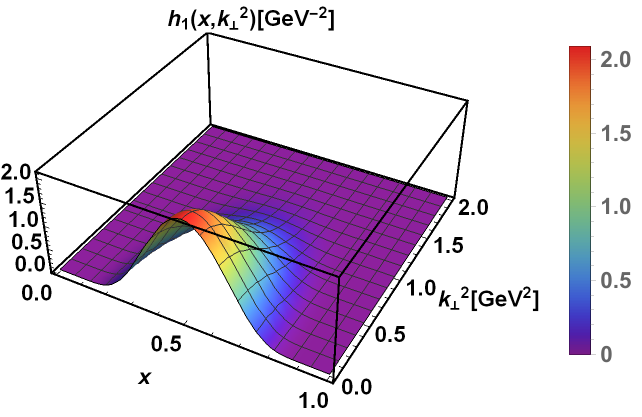}\hspace{1cm}
(d)\includegraphics[width=.44\textwidth]{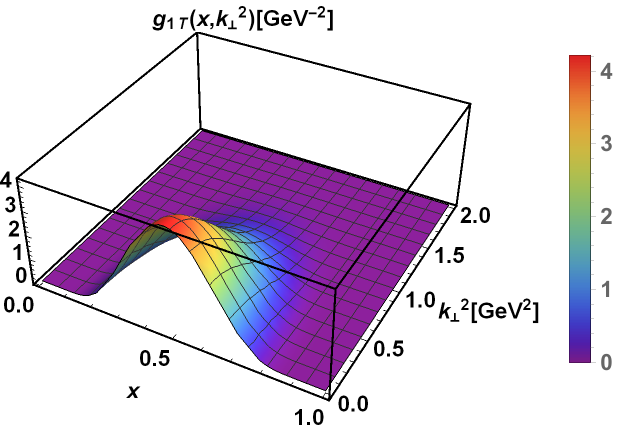}
\end{center}
\end{minipage}
\begin{minipage}[c]{1\textwidth}
\begin{center}
(e)\includegraphics[width=.44\textwidth]{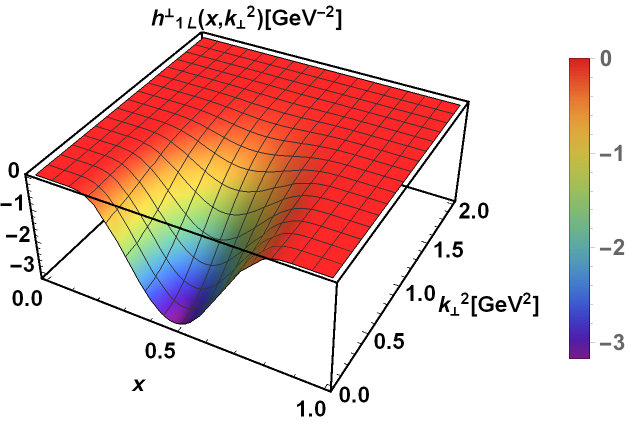}\hspace{1cm}
(f)\includegraphics[width=.44\textwidth]{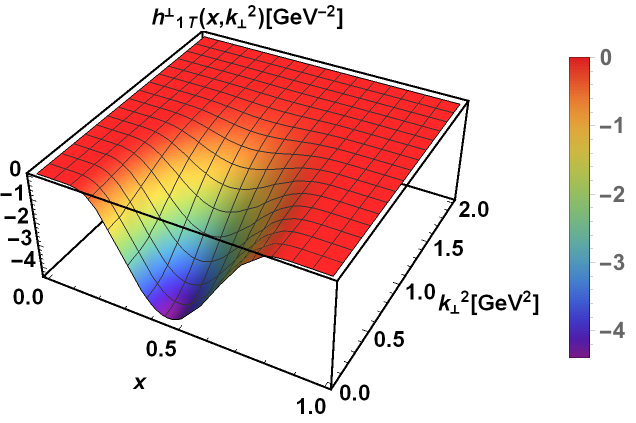}
\end{center}
\end{minipage}
\caption{(Color online) T-even TMDs of for $J/\psi$-meson as a function of $x$ and ${\bf k}^2_\perp$ in the LFQM at the model Scale $\mu_{\rm LFQM}^2=0.19$ GeV$^2$.}
\label{3d-tmds3}
\end{figure}

\begin{figure}[ht]
\centering
\begin{minipage}[c]{1\textwidth}
\begin{center}
(a)\includegraphics[width=.44\textwidth]{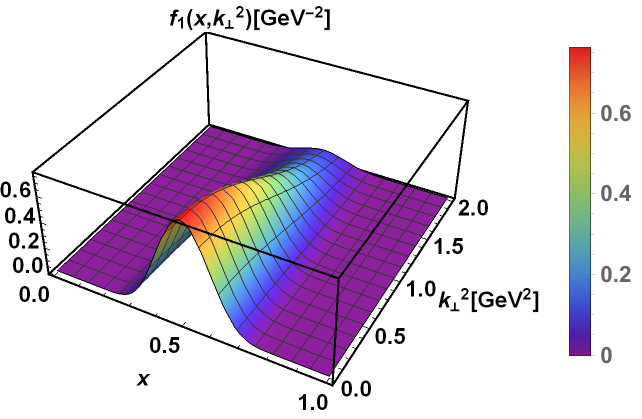}\hspace{1cm}
(b)\includegraphics[width=.44\textwidth]{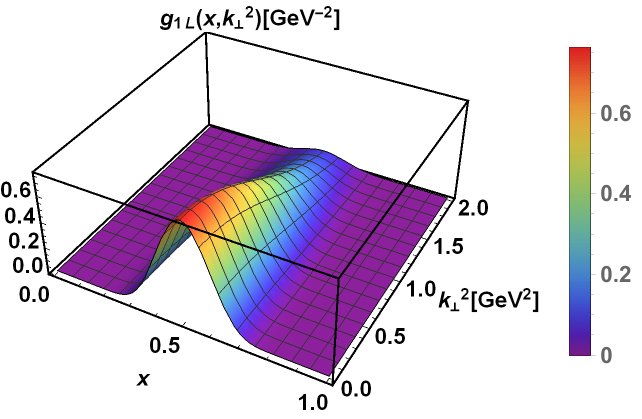}
\end{center}
\end{minipage}
\begin{minipage}[c]{1\textwidth}
\begin{center}
(c)\includegraphics[width=.44\textwidth]{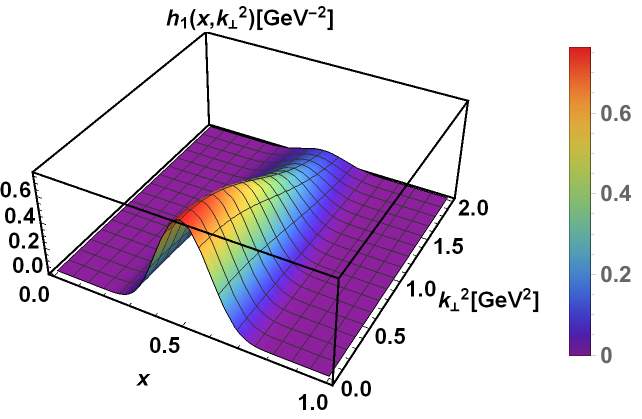}\hspace{1cm}
(d)\includegraphics[width=.44\textwidth]{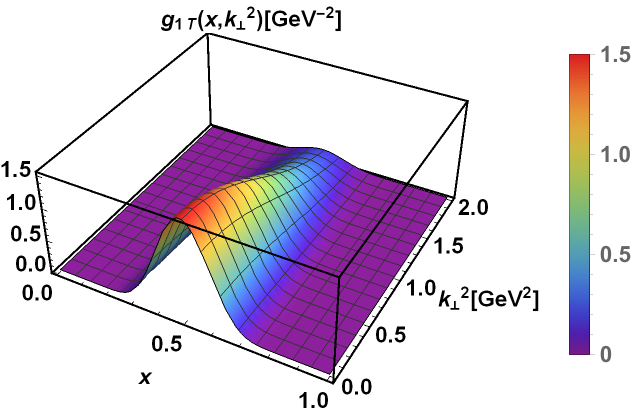}
\end{center}
\end{minipage}
\begin{minipage}[c]{1\textwidth}
\begin{center}
(e)\includegraphics[width=.44\textwidth]{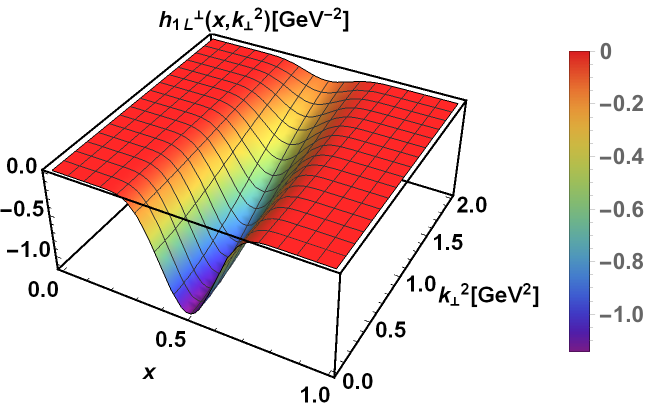}\hspace{1cm}
(f)\includegraphics[width=.44\textwidth]{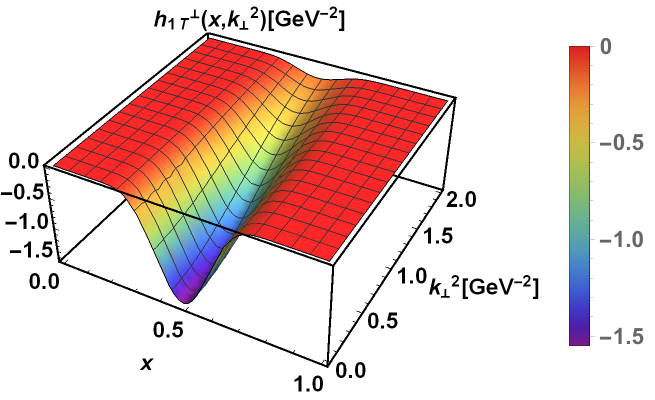}
\end{center}
\end{minipage}
\caption{(Color online)  T-even TMDs of for $\Upsilon$-meson as a function of $x$ and ${\bf k}^2_\perp$ in the LFQM at the model Scale $\mu_{\rm LFQM}^2=0.19$ GeV$^2$.}
\label{3d-tmds4}
\end{figure}

 Now in Fig. \ref{tmds5} and \ref{tmds51}, we have presented the tensor polarized  $f_{1LL}(x,\textbf{k}_{\perp}^2)$, $f_{1LT}(x,\textbf{k}_{\perp}^2)$ and $f_{1TT}(x,\textbf{k}_{\perp}^2)$ TMDs as a function of $x$ at fixed ${\bf k}^2_{\perp}$ and as a function of ${\bf k}^2_{\perp}$ at fixed $x$ in LFHM respectively. For the case of LFQM, the tensor polarized quark TMDs are zero because of the different spin structure of the hadron. In Fig. \ref{tmds5}(a), $f_{1LL}(x,\textbf{k}_{\perp}^2)$ shows a symmetry under $x \leftrightarrow (1-x)$. The TMD  $f_{1LL}(x,\textbf{k}_{\perp}^2)$ has a zero OAM from initial to final state of hadron in Eq. (\ref{f1ll-LFH}). For the case of $\rho$ vector meson, the plot of $f_{1LL}(x,\textbf{k}_{\perp}^2)$ has both positive and negative distribution but as we move to heavy quark masses, there is only positive distribution as the S-wave contribution increases in heavy quark masses. In Fig. \ref{tmds5}, we can clearly see that the $f_{1LT}(x,\textbf{k}_{\perp}^2)$ vanishes at $x=0.5$, exhibits positive distribution for $x>0.5$ and negative distribution for $x<0.5$. The overlap form of $f_{1LT}(x,\textbf{k}_{\perp}^2)$ quark TMD observed due to the transfer of one unit of OAM from initial to final state of hadron. $f_{1LT}$ is anti-symmetric under the $x \leftrightarrow (1-x)$. An interesting fact about $f_{1LL}(x,\textbf{k}_{\perp}^2)$ and $f_{1LT}(x,\textbf{k}_{\perp}^2)$ is that, the $J/\psi$ meson has a higher peak distribution at $k^2_\perp=0.1$ GeV$^2$ than $\Upsilon$ meson when compared to the difference in other TMDs. The last tensor polarized $f_{1TT}(x,\textbf{k}_{\perp}^2)$ TMD is given in Figs. \ref{tmds5}(e)  and \ref{tmds5}(f) which is also symmetric under $x \leftrightarrow (1-x)$. There are two units of OAM transfer from initial to final state of the hadron in this case. An overall factor $1/\sqrt{x(1-x)}$ in LFHM enhance the distributions compared to LFQM TMDs distributions. All the TMDs vanish at the end points  $x\to\{0,1\}$ for any value of ${\bf k}^2_{\perp}$ as mentioned in Refs. \cite{bse,NJL} for heavy vector meson. More details on the study of TMDs with respect to both $x$ and  ${\bf k}^2_{\perp}$ in three-dimensional structure for $J/\psi$-meson have been presented in Fig. \ref{3d-tmds} for LFHM and in Fig. \ref{3d-tmds3} for LFQM. Similarly, for the case of $\Upsilon$-meson results have been presented in Fig. \ref{3d-tmds2} and in Fig. \ref{3d-tmds4} for LFHM and LFQM respectively. Both the models show similar plots when compared to the BSE model \cite{bse} with an exception in the case of $J/\psi$ meson where the BSE model has two different peaks for $f_{1TT}(x,\textbf{k}_{\perp}^2)$ TMD as compared to a single peak in our models. 

 As there is no experimental data available for the case of TMDs to compare our predictions, we have computed average quark transverse momenta ${\langle{{k}_\perp}\rangle}$ for TMDs in both the models for $J/\psi$ and $\Upsilon$-meson. The average quark transverse momenta ${\langle{{k}_\perp}\rangle}$ is expressed as
 \begin{eqnarray}
\langle k_{\perp} \rangle_{\rm TMD} \equiv \frac{{\int} {\rm d}x ~{\rm d}^2{\bf k}_\perp | {\bf k}_\perp|  {\rm TMD}(x,{\bf k}^2_\perp)}{\int {\rm d}x ~{\rm d}^2{\bf k}_\perp {\rm TMD}(x,{\bf k}^2_\perp)}.
\label{moments}
\end{eqnarray}
In Table \ref{table-moment}, we have presented our results from LFHM and LFQM and  have compared them with the only available theoretical prediction from BSE model \cite{bse}. We observe that our results in LFQM for both $J/\psi$ and $\Upsilon$-meson are quite similar to the BSE model results. Meanwhile LFHM under estimate the results and are very less as compared to BSE model as well as LFQM. This is true for the case of already computed $\rho$-meson TMDs as well where the LFHM also gives lower values when compared to NJL model \cite{NJL} and BSE model \cite{bse}. The model input parameters we are choosing for LFHM may be the reason for having lower ${\langle{{k}_\perp}\rangle}$ value as compared to the other models. Since the tensor polarized TMDs for LFQM as well as $h^{\perp}_{1T}(x,\textbf{k}_{\perp}^2)$ TMD in LFHM are zero, we can not compute ${\langle{{k}_\perp}\rangle}$ for these TMDs.
\begin{table}
\centering
\begin{tabular}{|c|c|c|c|c|c|c|}
\hline 
 & \multicolumn{2}{c|}{LFQM} &\multicolumn{2}{c|}{LFHM} & \multicolumn{2}{c|}{BSE model}\\
 TMDs & \multicolumn{2}{c|}{(This work)} &\multicolumn{2}{c|}{(This work)} & \multicolumn{2}{c|}{\cite{bse}} \\
\cline{2-7}
& ${\langle{{k}_\perp}\rangle}^{J/\psi}$ & ${\langle{{k}_\perp}\rangle}^\Upsilon$ & ${\langle{{k}_\perp}\rangle}^{J/\psi}$ & ${\langle{{k}_\perp}\rangle}^\Upsilon$ & ${\langle{{k}_\perp}\rangle}^{J/\psi}$ &  ${\langle{{k}_\perp}\rangle}^\Upsilon$ \\
\hline
$f_1(x,\textbf{k}_{\perp}^2)$ & 0.620 & 1.202 & 0.408 & 0.667 & 0.623 & 1.020\\
$g_{1L}(x,\textbf{k}_{\perp}^2)$ & 0.600 & 1.183 & 0.388 & 0.656 & 0.589 & 1.003 \\
$g_{1T}(x,\textbf{k}_{\perp}^2)$ & 0.619 & 1.201 & 0.411 & 0.668 & 0.615 & 1.020 \\
$h_1(x,\textbf{k}_{\perp}^2)$ & 0.610 & 1.192 & 0.403 & 0.664 & 0.608 & 1.012 \\
$h^\perp_{1L}(x,\textbf{k}_{\perp}^2)$ & 0.622 & 1.204 & 0.395 & 0.660 & 0.608 & 1.017 \\
$h^\perp_{1T}(x,\textbf{k}_{\perp}^2)$ & 0.628 & 1.210 & - & - & 0.602 & 1.017 \\
$f_{1LL}(x,\textbf{k}_{\perp}^2)$ & - & - & 0.441 & 0.685 & - & -\\
$f_{1LT}(x,\textbf{k}_{\perp}^2)$ & - & - & - & - & - & -\\
$f_{1TT}(x,\textbf{k}_{\perp}^2)$ & - & - & 0.403 & 0.664 & 0.764 & 1.063 \\
\hline
\end{tabular}
\caption{The momenta $\langle k_\perp \rangle$ of the LFQM and LFHM compared with the BSE model \cite{bse}. The constituent quark masses $m_{c} = 1.5$ GeV with the AdS/QCD scale $\kappa = 0.894$ GeV for $J/\psi$-meson and $m_{b} = 4.63$ GeV with the AdS/QCD scale $\kappa = 1.49$ GeV for $\Upsilon$-meson in the LFHM respectively. Similarly, in the LFQM, the values of the parameters are $m_{c} = 1.68$ GeV with $\beta = 0.699$ GeV for $J/\psi$-meson and $m_{b} = 5.10$ GeV with $\beta = 1.376$ GeV for $\Upsilon$-meson respectively.}
\label{table-moment}
\end{table}

\section{Parton distribution functions}\label{pdF}
PDFs are the distribution functions used to study the one-dimensional internal structure of hadrons \cite{pdf,pdf1,pdf2,pdf3}. These functions describes the probability of finding the quarks inside the hadrons as a function of longitudinal momentum fraction $x$. The description of longitudinal momentum and polarization carried by the quarks inside a hadron is encoded by the PDFs. Even though the PDFs do not carry any information about the transverse momentum distribution, they have a direct connection with the TMDs and can be easily accessed through the DIS experiments \cite{dis}. There are total $4$ PDFs for the case of spin-$1$ hadrons, which are  $f_1(x)$, $g_{1}(x)$, $h_1(x)$ and tensor PDF $f_{1LL}(x)$ \cite{tmd13}. 

There are two ways to calculate these PDFs. First way is to derive these collinear PDFs by integrating Eq. (\ref{f1}) to Eq. (\ref{f3}) over transverse momentum $k_{\perp}$. We have
\begin{eqnarray}
\langle \gamma^+ \rangle_{\boldsymbol{ \mathcal{S}}}^{(\Lambda)} (x) 
&\equiv f_1(x) + \mathcal{S}_{LL} \,
f_{1LL}(x) \,,
\label{form1-pdf} 
\end{eqnarray}
\begin{eqnarray}
\langle\gamma^{+} \gamma_{5}\rangle^{(\Lambda)}_{\boldsymbol{ \mathcal{S}}} (x) 
&\equiv  \mathcal{S}_L\,g_{1}(x) \,,
\label{form2-pdf} 
\end{eqnarray}
\begin{eqnarray}
\langle\gamma^+\gamma^i\gamma_{5}\rangle^{(\Lambda)}_{\boldsymbol{ \mathcal{S}}} (x) 
& \equiv \mathcal{S}_\perp^i h_1(x)\,.
\label{form3-pdf}
\end{eqnarray}
The PDFs can also be derived from the TMDs using
\begin{equation}  
\mathcal{H}(x) = \int d\mathbf{k}_{\perp}^2 \mathcal{H}(x, \mathbf{k}_{\perp}^2),
\end{equation}
with ${\cal H} =f_1(x), g_{1}(x), h_1(x)$ and  $f_{1LL}(x)$. 

\begin{figure}[ht]
	\centering
	\begin{minipage}[c]{1\textwidth}
		\begin{center}
			(a)\includegraphics[width=.40\textwidth]{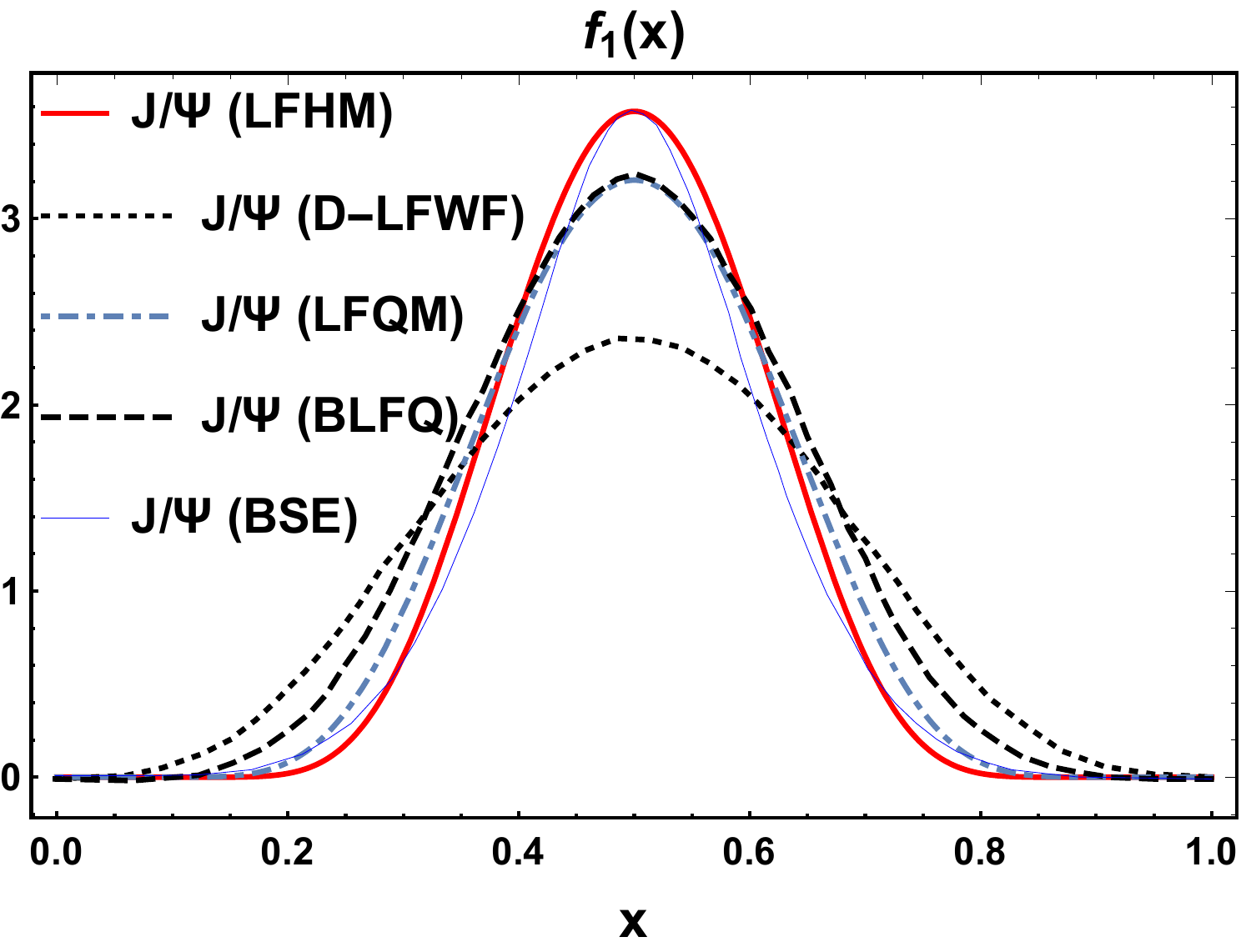}\hspace{1cm}
			(b)\includegraphics[width=.40\textwidth]{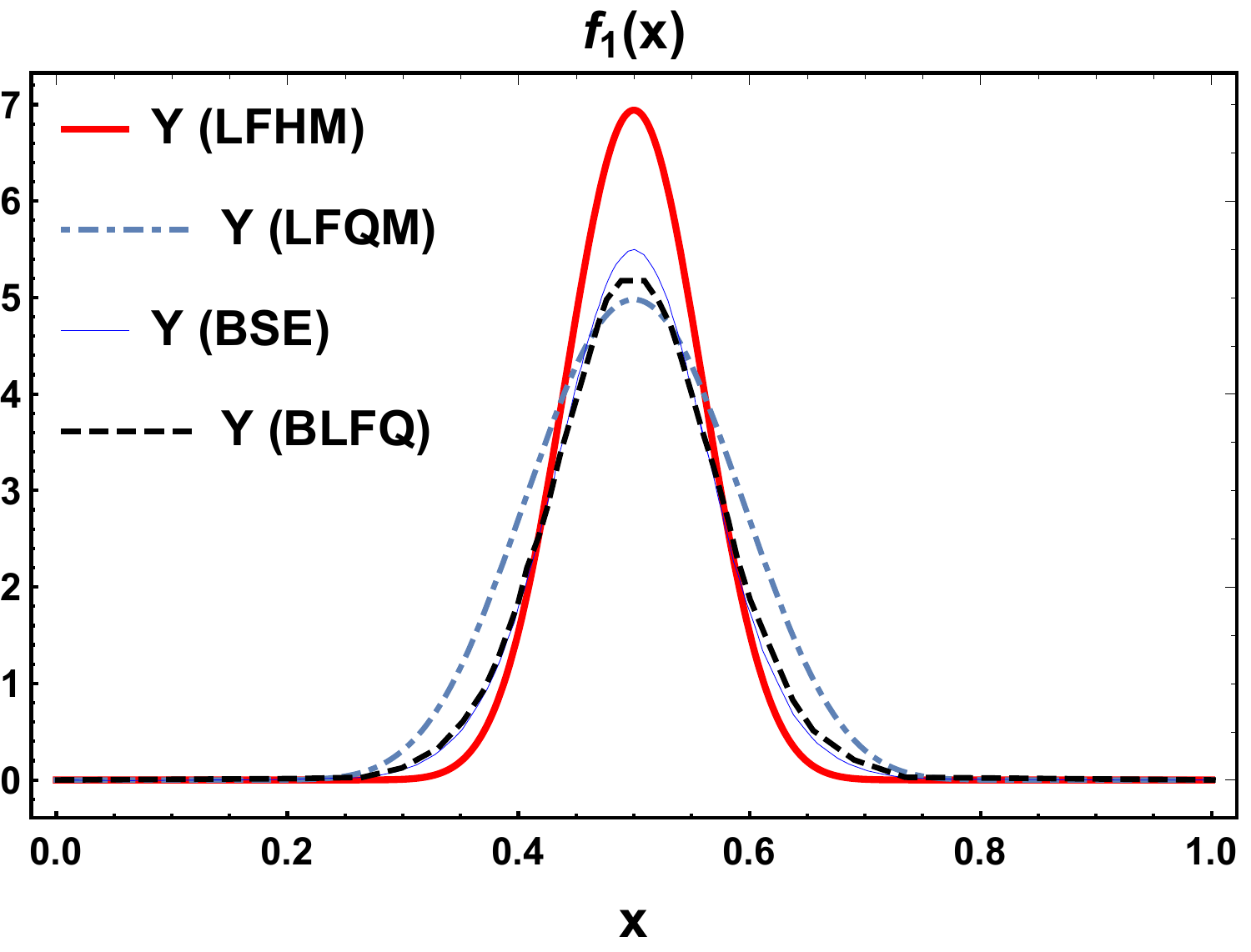}
		\end{center}
	\end{minipage}
	\begin{minipage}[c]{1\textwidth}
		\begin{center}
			(c)\includegraphics[width=.41\textwidth]{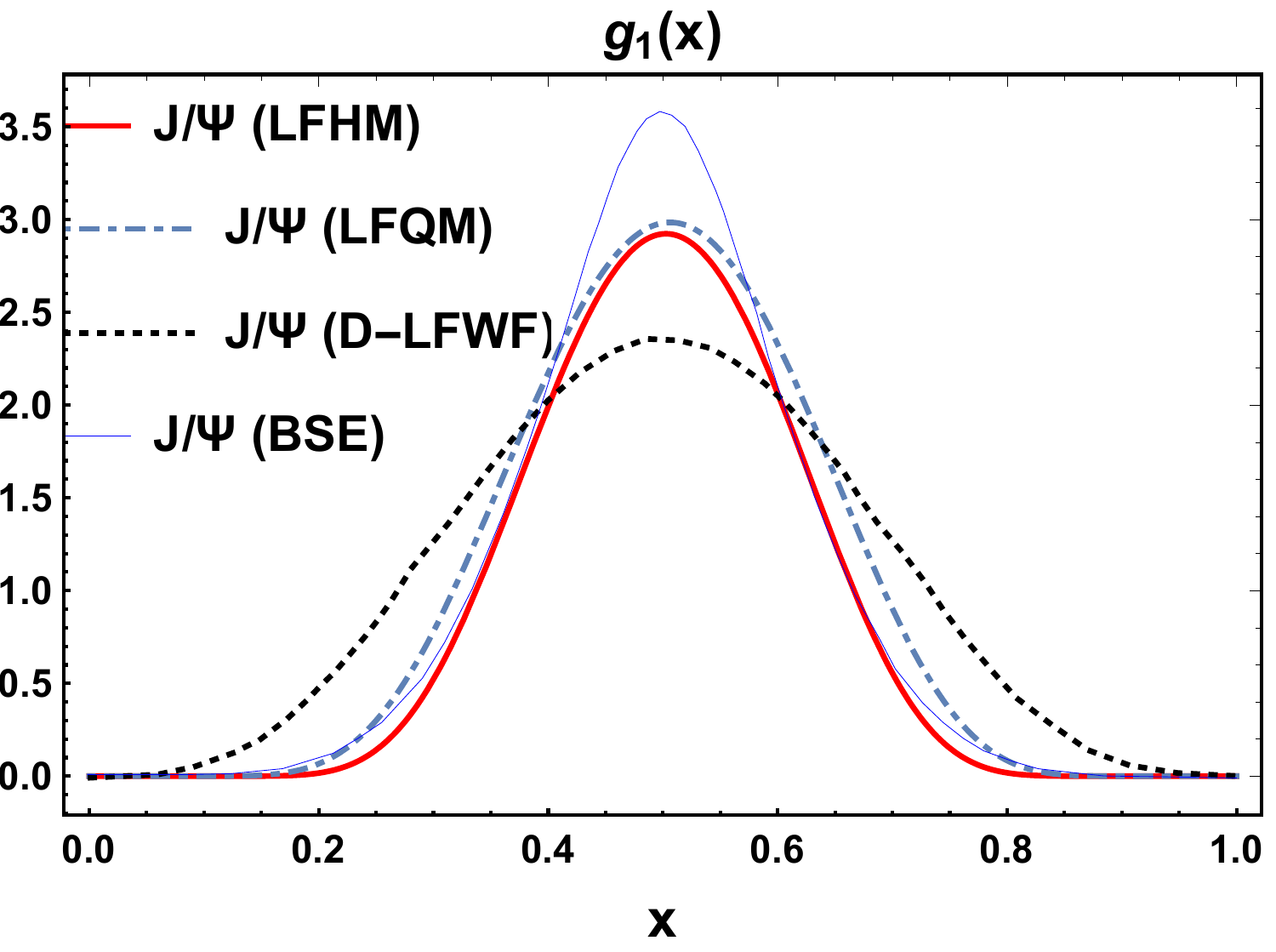}\hspace{1cm}
			(d)\includegraphics[width=.40\textwidth]{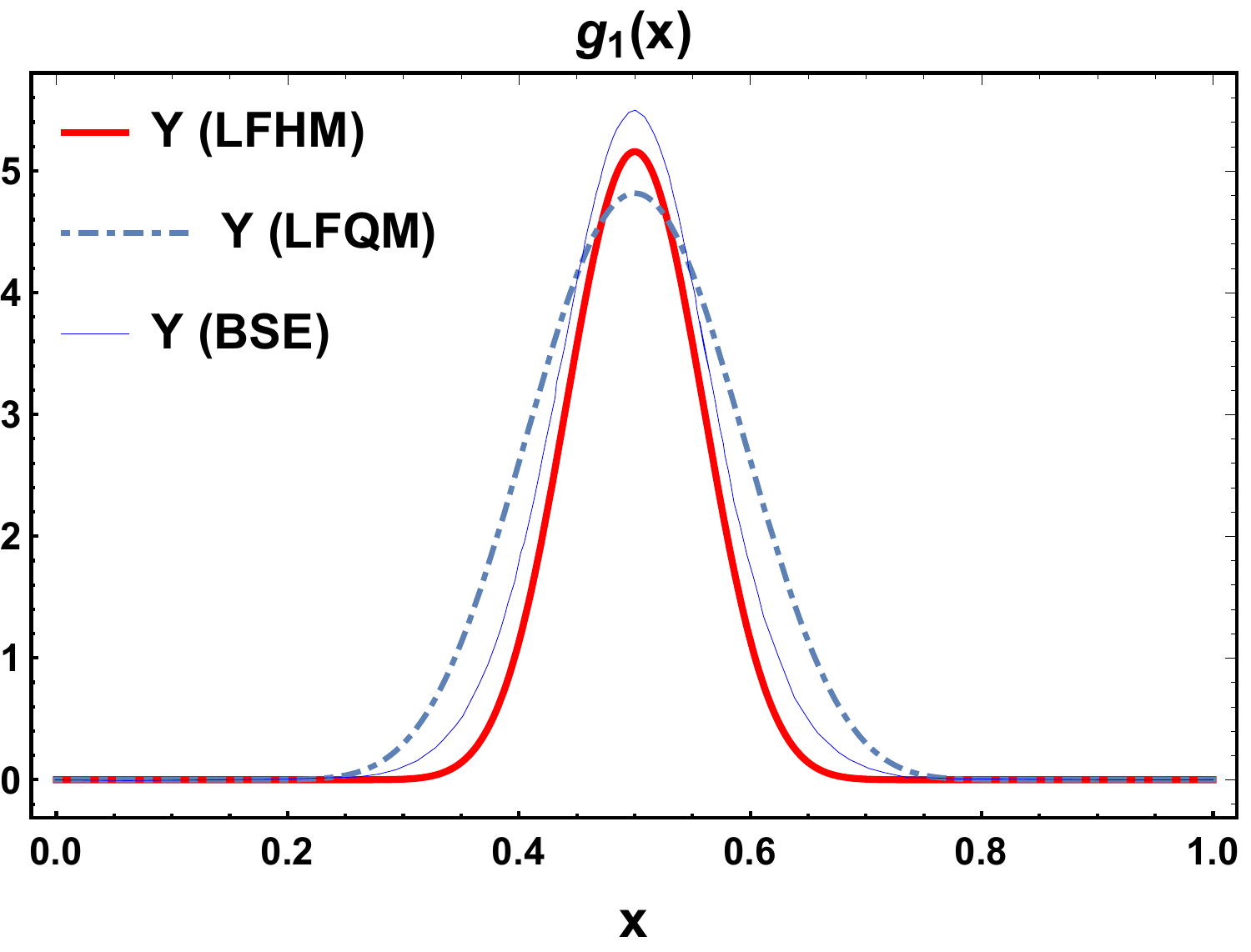}
		\end{center}
	\end{minipage}
	\begin{minipage}[c]{1\textwidth}
		\begin{center}
			(e)\includegraphics[width=.40\textwidth]{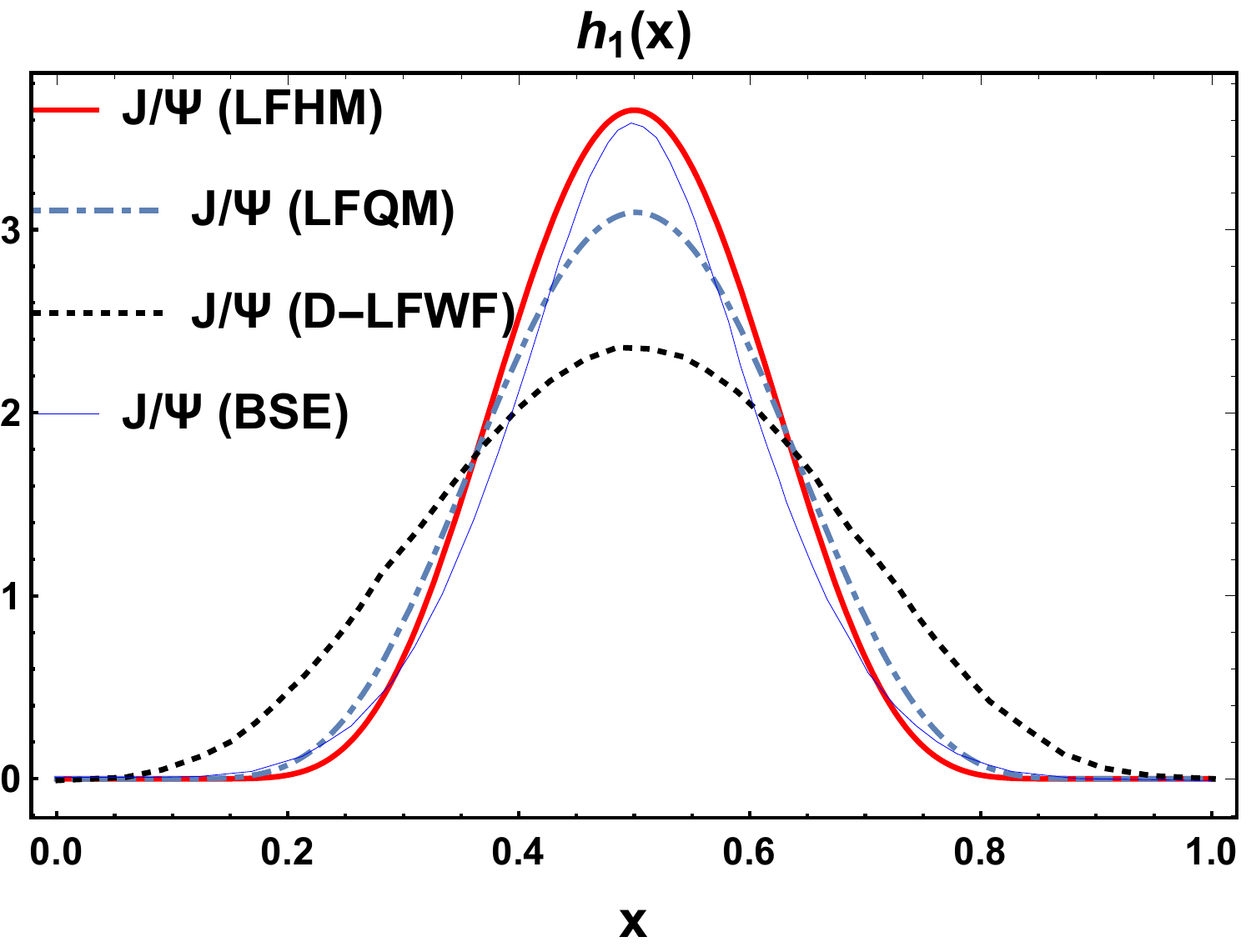}\hspace{1cm}
			(f)\includegraphics[width=.40\textwidth]{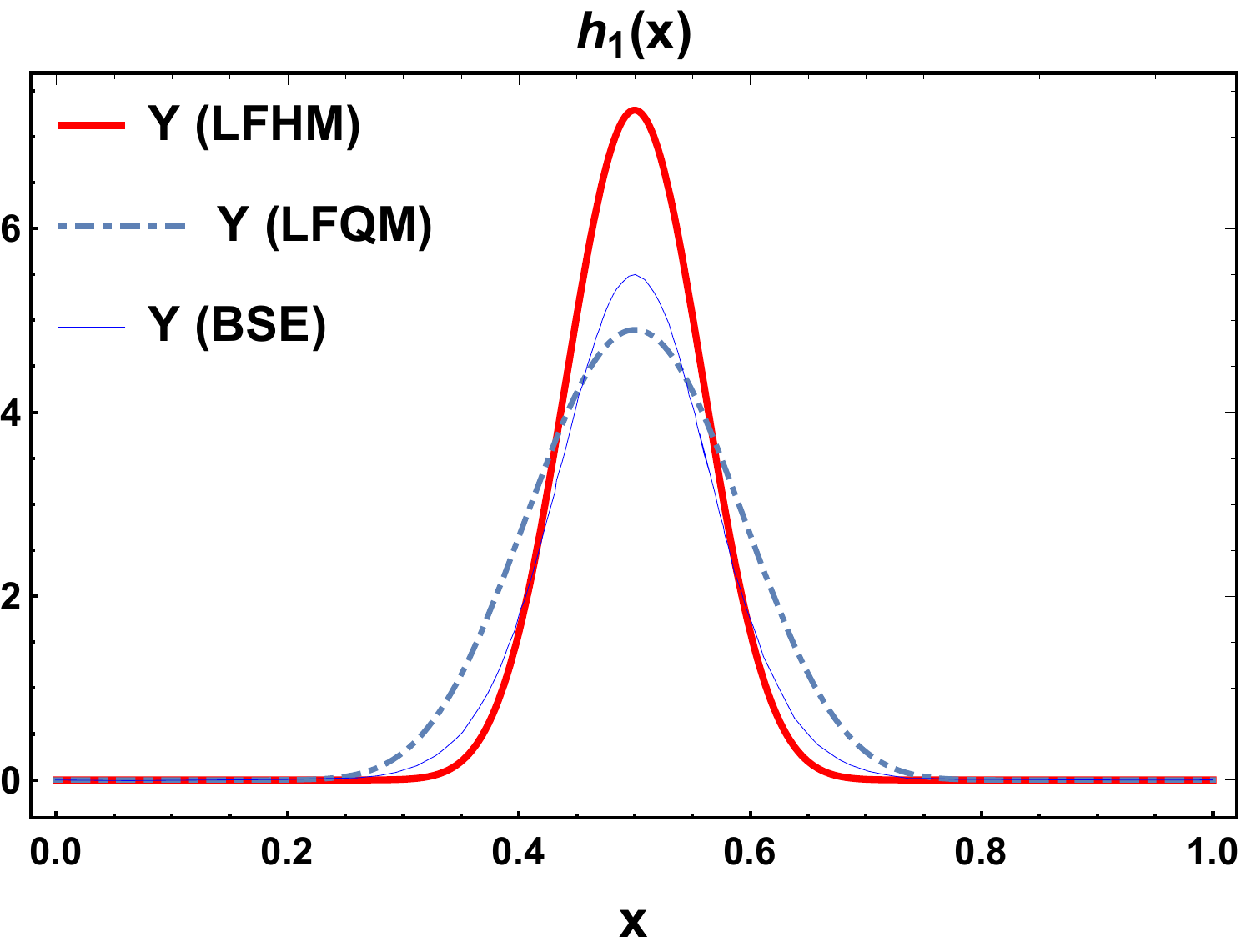}
		\end{center}
	\end{minipage}
	\begin{minipage}[c]{1\textwidth}
		\begin{center}
			(g)\includegraphics[width=.40\textwidth]{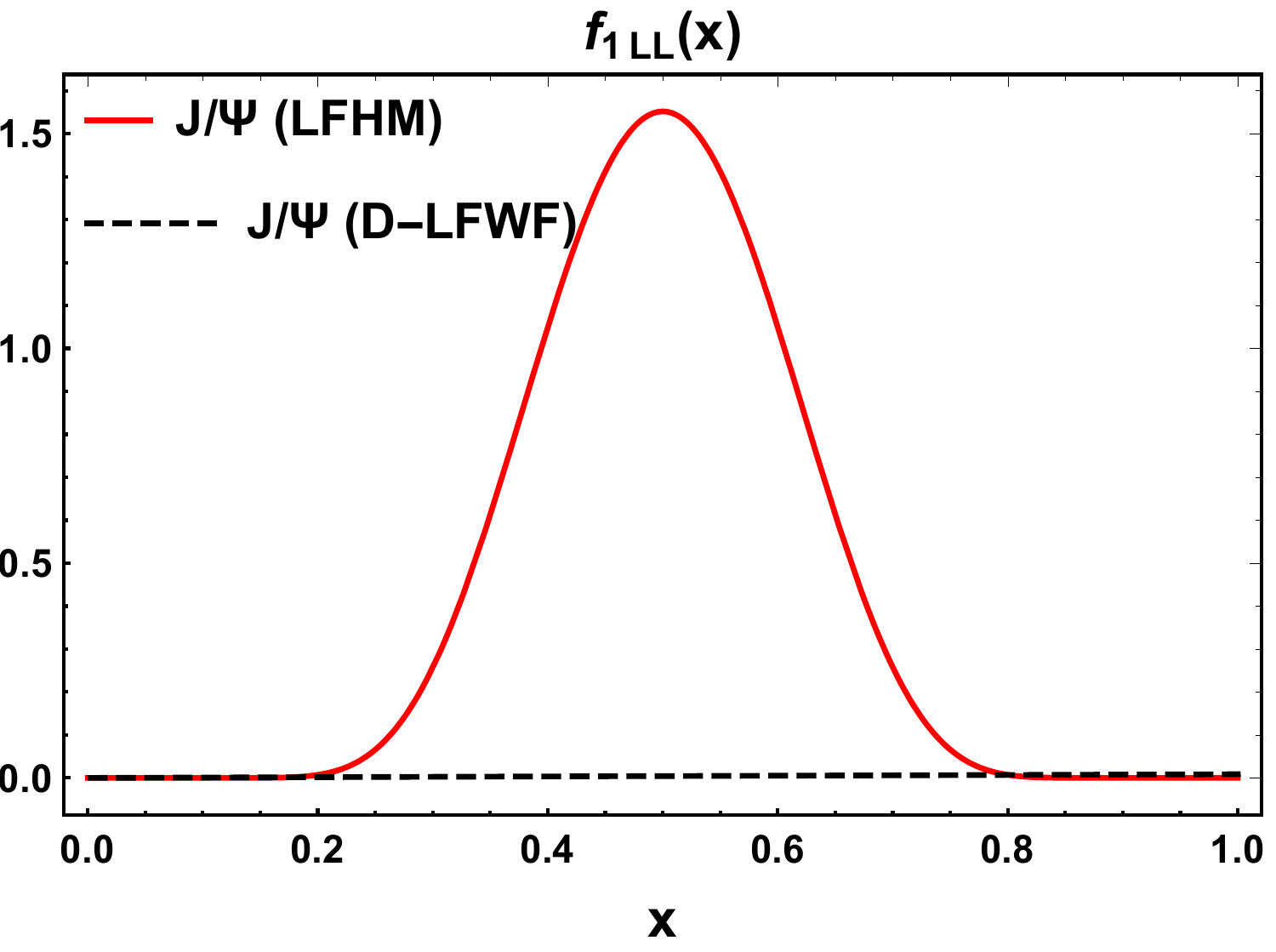}\hspace{1cm}
		(f)\includegraphics[width=.40\textwidth]{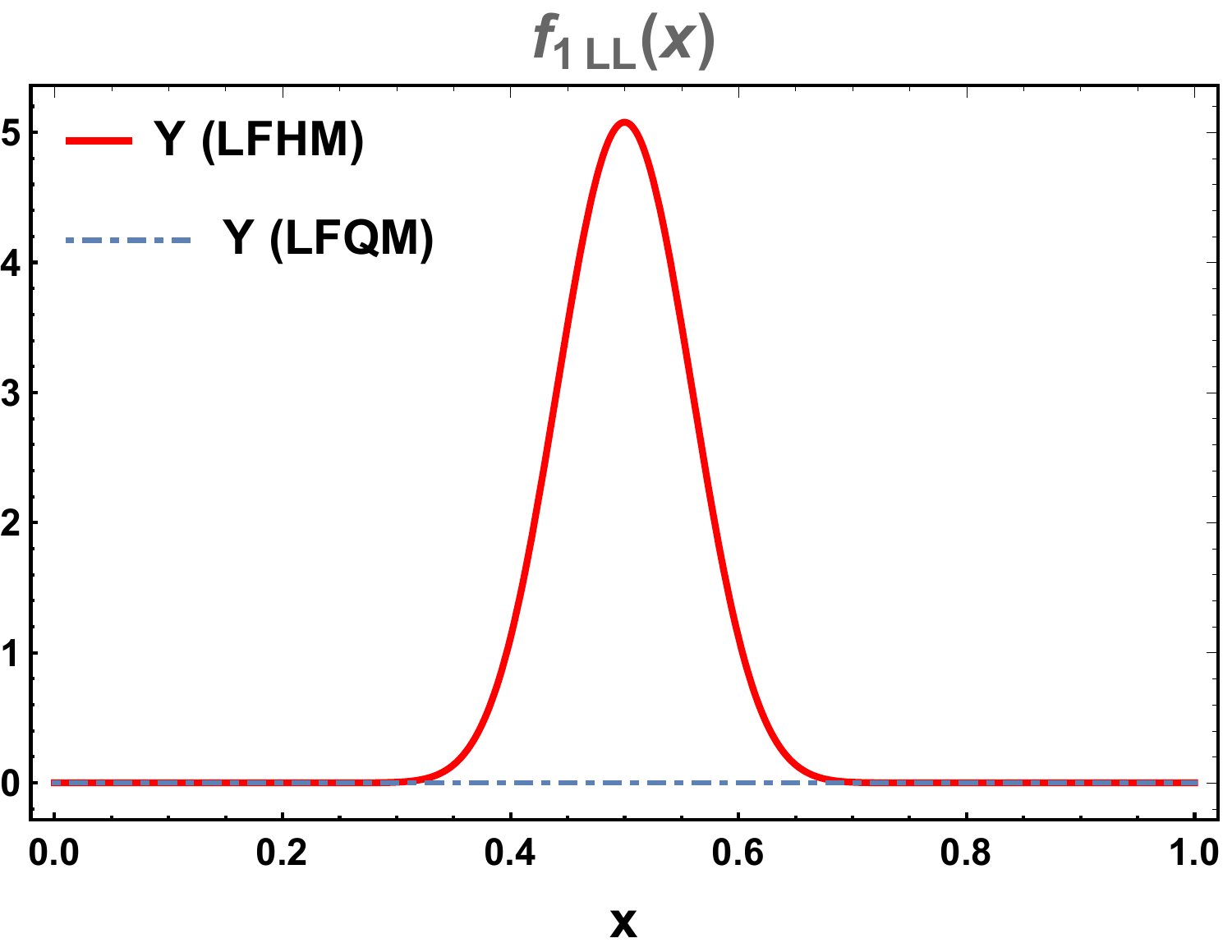}
		\end{center}
	\end{minipage}
	\caption{(Color line) $f_1(x), g_{1}(x)$, $h_1(x)$ and $f_{1LL}(x)$PDFs are plotted with respect to $x$ in different models. The solid thick red and dot-dashed curves in the  left and right panels represent the PDFs for $J/\psi$ and $\Upsilon$ mesons in the LFHM (our work) and LFQM (our work) respectively. The black dotted curves are the results from designed-LFWF \cite{lfwf} for $J/\psi$-meson. The dashed black curves in the left panel are from BLFQ \cite{blfqpdf}. The blue thin curves are from BSE model \cite{bse}.}
	\label{ppp}
\end{figure}

In Fig. \ref{ppp}, we have plotted the trend of the unpolarized distribution $f_{1}(x)$, the transversity distributon $h_{1}(x)$, the helicity distribution $g_{1}(x)$ and the tensor $f_{1LL}(x)$ PDFs with respect to the longitudinal momentum fraction $x$ carried by the quark. We have compared our LFHM and LFQM results with the BSE model \cite{bse}, D-LFWF \cite{lfwf}, BLFQ \cite{blfqpdf} results for both the mesons. Overall, the qualitative behavior of our PDFs is consistent with the predictions of the D-LFWF \cite{lfwf}, BLFQ \cite{blfqpdf} and BSE model \cite{bse}. The heavy meson PDFs are narrow around $x$ and centered at $x=0.5$ for all the models.

The PDFs of our models satisfy the quark sum rules \cite{rho, NJL}
\begin{equation}
\int_0^1 {\rm d}x~f_1(x)=1 \,,
\end{equation}
\begin{equation}
\int_0^1 {\rm d}x~x\,f_1(x) + \int_0^1 {\rm d}x~(1-x)\,f_1(x)=1\,,
\end{equation}
for both the particles as shown in Table \ref{table-quark}. The quark spin sum values 
 $\langle x \rangle= \int d x \mathcal{H}(x)$ for all the PDFs are given in Table \ref{table-quark}. $f_1(x), h_1(x)$  and $g_{1}(x)$ are all come to close to unity for both the mesons. However, due to relativistic effects, the quark spin sum value for $\Upsilon$-meson is greater than $J/\psi$-meson. On the other hand, the quark spin sum for the tensor PDF $f_{1LL}(x)$ comes out to be quite less as compared to other PDFs. This may be due to the higher Fock-state contributions playing an important role in the tensor PDF. A similar observation has been made in the BSE model \cite{bse}. The positivity constrains on spin-$1$ TMDs and PDFs \cite{rho,NJL} are also satisfied in our models.
\begin{table}
\centering
\begin{tabular}{|c|c|c|c|c|c|c|}
\hline
 & \multicolumn{2}{c|}{LFHM} & \multicolumn{2}{c|}{LFQM} &\multicolumn{2}{c|}{BSE model}\\
 PDFs & \multicolumn{2}{c|}{(This work)}  &\multicolumn{2}{c|}{(This work)}  &  \multicolumn{2}{c|}{\cite{bse}} \\
 \cline{2-7}
 & ${\langle{{x}}\rangle}^{J/\psi}$ & ${\langle{{x}}\rangle}^\Upsilon$ &  ${\langle{{x}}\rangle}^{J/\psi}$ & ${\langle{{x}}\rangle}^\Upsilon$ &${\langle{{x}}\rangle}^{J/\psi}$ &  ${\langle{{x}}\rangle}^\Upsilon$ \\
\hline
$f_1(x)$  & 0.97 & 1.00& 1.00& 1.00& 1.00 & 1.00\\
$g_{1}(x)$  & 0.81 & 0.74 &0.93 &0.97 & 0.92 & 0.98 \\
$h_1(x)$  & 1.00 & 1.00 &0.97 & 0.98& 0.96 & 0.99 \\
$f_{1LL}(x)$ & 0.42 & 0.73 &- &- & - & -\\
\hline
\end{tabular}
\caption{The quark spin sum $\langle x \rangle$ of all the PDFs for both the particles in both LFHM and LFQM compared with the BSE model \cite{bse}.}
\label{table-quark}
\end{table}

\section{conclusion}\label{conclude}
In this work, we have presented the $J/\psi$ and $\Upsilon$-meson T-even TMDs in the light-from holographic model and the light-front quark model. Both the models have been introduced with an addition of the dynamic spin effects in their LFWFs leading to different spin structure for both the models. We have calculated the TMDs using the helicity amplitudes which are basically the overlap form of LFWFs. The relativistic effect being less for the heavy mesons as compared to light mesons, the distribution function are narrower in longitudinal momentum fraction $(x)$ and wider in transverse momentum $(\textbf{k}^2_{\perp})$ of the quarks when compared to the light mesons case. The P and D-wave contributions are quite small compared to the S-wave resulting in the symmetry of most of the TMDs. We have observed that $f_{1}(x,\textbf{k}^2_{\perp})$, $g_{1L}(x,\textbf{k}^2_{\perp})$, $g_{1T}(x,\textbf{k}^2_{\perp})$, $h_{1}(x,\textbf{k}^2_{\perp})$ and $h^{\perp}_{1L}(x,\textbf{k}^2_{\perp})$ TMDs show similar behavior in both the models for both the mesons. The TMD $h_{1T}(x,\textbf{k}^2_{\perp})$ vanishes in LFHM and shows a negative distribution in LFQM. Similarly, the tensor polarized TMDs $f_{1LL}(x,\textbf{k}^2_{\perp})$, $f_{1LT}(x,\textbf{k}^2_{\perp})$ and $f_{1TT}(x,\textbf{k}^2_{\perp})$ are zero for LFQM and non-zero for LFHM. $f_{1LL}(x,\textbf{k}^2_{\perp})$ and $f_{1TT}(x,\textbf{k}^2_{\perp})$ TMDs shows positive distribution, whereas $f_{1LT}(x,\textbf{k}^2_{\perp})$ have both positive and negative distributions. All the TMD are quite narrow around $x=0.5$ in LFHM than the LFQM. The TMDs shows symmetry under $x \leftrightarrow (1-x)$ except $f_{1LT}(x,k^2_{\perp})$ TMD. $f_{1LT}(x,k^2_{\perp})$ TMD shows anti-symmetry under $x\leftrightarrow (1-x)$. Nevertheless,  the predictions of spin-$1$ TMDs in both the models  are consistent with the BSE model \cite{bse}. All the TMDs, in our models also satisfy the necessary positivity constraints \cite{NJL,rho}. 
We have also calculated the first momenta $\langle k_{\perp} \rangle$ for various TMDs and compared with available BSE model \cite{bse} data. We observe that the LFQM results are similar to the BSE model, whereas the LFHM underestimate them which may be due to the interplay of some other subtle non-perturbative effects. Finally, we have calculated the unpolarized $f_{1}(x)$, the transversity $h_{1}(x)$, the helicity $g_{1}(x)$ and the tensor $f_{1LL}(x)$ PDFs for $J/\psi$ and $\Upsilon$-meson in LFHM as well as in LFQM. We have compared these PDFs with the other available theoretical predictions \cite{blfqpdf,bse,lfwf}. The PDFs in this work agree qualitatively with other models except the $f_{1LL}(x)$ PDF which is zero in Ref. \cite{lfwf}. We have calculated quark spin sum of these PDFs and the results are more close to unity as compared to other models.

In conclusion, this work can also be extended in future to compute the GPDs with different form factors (FFs) for spin-$1$ particles and spin asymmetry arising in case of $J/\psi$-meson. Further, adding a nontrival gauge link to calculate the T-odd TMDs for spin-$1$ particles in future would not only complete the study of the associated distributions but will also give a complete structure of hadrons through the future experiments to be conducted in BNL and J-Lab.

\section{Acknowledgement}
H.D. would like to thank the Science and Engineering Research Board, Department of Science and Technology, Government of India through the grant (Ref No.TAR/2021/000157) under TARE scheme for financial support.



\end{document}